\documentclass{JHEP3}
\usepackage[english]{babel}
\usepackage[dvips]{graphicx}
\usepackage{amsmath} 
\usepackage{amssymb}
\usepackage{psfrag}
\usepackage{cite}
\usepackage{esint}

\title{Mueller Navelet jets at LHC -- complete next-to-leading BFKL calculation}

\author{Dimitri Colferai\\
        Dipartimento di Fisica, Universit{\`a} di Firenze, Italy\\
        INFN, Florence, Italy\\
        Email: \email{colferai@fi.infn.it}}
\author{Florian Schwennsen\\
        Deusches Elektronen-Synchrotron DESY, Hamburg, Germany \\
        Email: \email{florian.schwennsen@desy.de}}
\author{Lech Szymanowski\\
        Soltan Institute for Nuclear Studies, Warsaw, Poland {\em \&} \\
        CPHT, {\'E}cole Polytechnique, CNRS, 91128 Palaiseau Cedex, France\\ 
        Email: \email{lech.szymanowski@fuw.edu.pl}}
\author{Samuel Wallon\\ 
        LPT, Universit{\'e} Paris-Sud, CNRS, Orsay, France {\em \&} \\
        UPMC Univ. Paris 06, facult\'e de physique, 4 place Jussieu, 75252 Paris Cedex 05, France\\
        Email: \email{samuel.wallon@th.u-psud.fr}}

\abstract{We calculate cross section and azimuthal decorrellation of Mueller Navelet jets at the LHC in the complete next-lo-leading order BFKL framework, i.e. including next-to-leading corrections to the Green's function as well as next-to-leading corrections to the Mueller Navelet vertices. The obtained results for standard observables proposed for studies of Mueller Navelet jets show that both sources of corrections are of equal, big importance for final magnitude and final behavior of observables. The astonishing conclusion of our analysis is that the observables obtained within the complete next-lo-leading order BFKL framework of the present paper are quite similar to the same observables obtained within next-to-leading DGLAP type treatment. This fact sheds doubts on general belief that the studies of Mueller Navelet jets at the LHC will lead to clear discrimination between the BFKL and the DGLAP dynamics. \\

\textsc{Date}: 17/11/2010\\
\textsc{Pacs}: 12.38.Bx, 12.38.Cy, 13.85.Hd}

\keywords{Hadronic Colliders, Jets, NLO Computations}

\preprint{CPHT-RR139.1209, LPT-ORSAY-09-111, DFF 452/10/2009, DESY 10-031 }

\begin{document}

\newcommand{\veck}{{\bf k}}
\newcommand{\vecki}{{\bf k}_i}
\newcommand{\veckone}{{\bf k}_1}
\newcommand{\vecktwo}{{\bf k}_2}
\newcommand{\veckj}{{\bf k}_{J}}
\newcommand{\veckji}{{\bf k}_{J,i}}
\newcommand{\veckjone}{{\bf k}_{J,1}}
\newcommand{\veckjtwo}{{\bf k}_{J,2}}
\newcommand{\vecqecht}{{\bf q}}
\newcommand{\vecq}{{\bf k}-{\bf k}'}
\newcommand{\vecl}{{\bf l}}
\newcommand{\kmin}{k_{{\rm min}}}
\newcommand{\kmax}{k_{{\rm max}}}
\newcommand{\kminone}{k_{{\rm min},1}}
\newcommand{\kmaxone}{k_{{\rm max},1}}
\newcommand{\kmintwo}{k_{{\rm min},2}}
\newcommand{\kmaxtwo}{k_{{\rm max},2}}
\newcommand{\deins}[1]{{\rm d}#1\,}
\newcommand{\dzwei}[1]{{\rm d}^2#1\,}
\newcommand{\dk}{\dzwei{\veck}}
\newcommand{\dl}{\dzwei{\vecl}}
\newcommand{\dkprime}{\dzwei{\veck'}}
\newcommand{\dkone}{\dzwei{\veckone}}
\newcommand{\dktwo}{\dzwei{\vecktwo}}
\newcommand{\dkjet}{\deins{|\veckj|}}
\newcommand{\dkjetone}{\deins{|\veckjone|}}
\newcommand{\dkjettwo}{\deins{|\veckjtwo|}}
\newcommand{\dsigma}{\deins{\sigma}}
\newcommand{\dsigmahat}{\deins{{\hat\sigma}_{\rm{ab}}}}
\newcommand{\dnu}{\deins{\nu}}
\newcommand{\dz}{\deins{z}}
\newcommand{\dx}{\deins{x}}
\newcommand{\dxone}{\deins{x_1}}
\newcommand{\dxtwo}{\deins{x_2}}
\newcommand{\dyjetone}{\deins{y_{J,1}}}
\newcommand{\dyjettwo}{\deins{y_{J,2}}}
\newcommand{\dphij}{\deins{\phi_{J}}}
\newcommand{\dphijone}{\deins{\phi_{J,1}}}
\newcommand{\dphijtwo}{\deins{\phi_{J,2}}}
\newcommand{\dtwojets}{{\rm d}|\veckjone|\,{\rm d}|\veckjtwo|\,\dyjetone \dyjettwo}
\newcommand{\shat}{{\hat s}}
\newcommand{\non}{\nonumber\\}
\newcommand{\asbar}{{\bar{\alpha}}_s}
\newcommand{\fourint}{\int \dphij \dk \dx f(x)E_{n,\nu}(\veck)\cos(m\phi_J)}
\newcommand{\chihat}{{\omega}}

\setcounter{page}{0}
\newpage
\section{Introduction} 

The large center of mass energy of hadron colliders like the Tevatron and the Large Hadron Collider (LHC) is not only of interest for the production of possible new heavy particles, but also allows to investigate the high energy regime of Quantum Chromodynamics (QCD).
An especially interesting situation appears if two different large scales enter the game.
If the two scales are ordered, large logarithms of the ratio of the two scales compensate the smallness of the coupling and therefore have to be resummed to all orders. One famous example is the case of high energy scattering with fixed momentum transfer. If the center of mass energy $s$ is much larger than the momentum transfer $|t|$ -- the so-called Regge asymptotics of the process -- the gluon exchange in the crossed channel dominates and logarithms of the type $[\alpha_s\ln(s/|t|)]^n$ have to be resummed. This is realized by the leading logarithmic (LL) Balitsky-Fadin-Kuraev-Lipatov (BFKL) \cite{Fadin:1975cb,Kuraev:1976ge,Kuraev:1977fs,Balitsky:1978ic} equation for the gluon Green's function describing the momentum exchange in the $t$-channel.

\begin{figure}[hb!]
\psfrag{p1}{\footnotesize \hspace{-.2cm}$p(p_1)$}
\psfrag{p2}{\footnotesize \hspace{-.2cm}$p(p_2)$}
\psfrag{j1}{\footnotesize\hspace{-1.9cm}$jet_1$ ($\veckjone, \, \phi_{J,1})$}
\psfrag{j2}{\footnotesize\hspace{0cm}$jet_2$ ($\veckjtwo, \, \phi_{J,2})$}
\psfrag{a}{\footnotesize \hspace{-.3cm}$\phi_{J,1}$}
\psfrag{b}{\footnotesize \hspace{-.1cm}$\phi_{J,2}-\pi$}
\psfrag{rp}{\ large $+$ rapidity}
\psfrag{rm}{\ large $-$ rapidity}
\psfrag{r0}{\ zero rapidity}
\psfrag{perp}{$\perp$ plane}
\psfrag{B}{\rotatebox{90}{$\leftarrow$Beam axis$\rightarrow$}}
\vspace{-.2cm}
\centerline{\hspace{-2.5cm}\includegraphics[height=7cm]{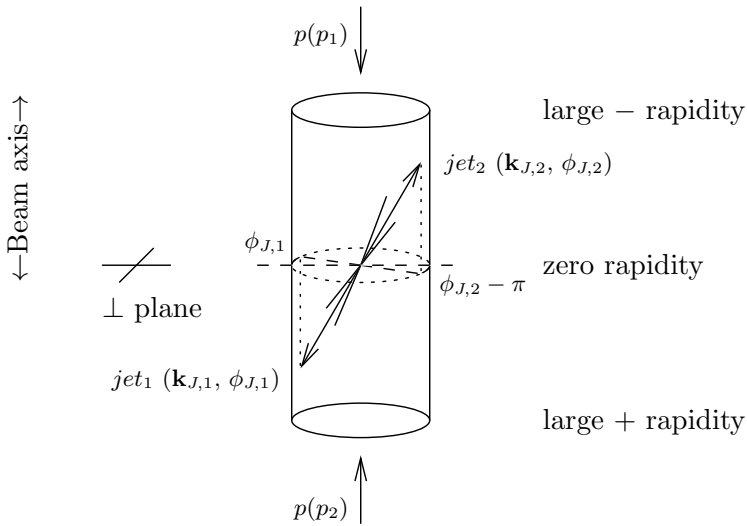}}
\caption{Mueller Navelet jets, illustrated at lowest order.}
\label{fig:jets}
\end{figure}

One of the most famous testing ground for BFKL physics are the Mueller Navelet
jets \cite{Mueller:1986ey}, illustrated in Fig.~\ref{fig:jets}. The predicted
power like rise of the cross section with increasing energy has been observed at
the Tevatron $p\bar{p}$-collider \cite{Abbott:1999ai}, but the measurements
revealed an even stronger rise than predicted by BFKL calculations. Beside the
cross section also a more exclusive observable within this process drew the
attention, namely the azimuthal correlation between these jets.  Considering
hadron-hadron scattering in the common parton model to describe two jet
production at LO, one deals with a back-to-back reaction and expects the
azimuthal angles of the two jets always to be $\pi$ and hence completely
correlated. This corresponds in Fig.~\ref{fig:jets} to
$\phi_{J,1}=\phi_{J,2}-\pi\,$. But when we increase the rapidity difference
between these jets, the phase space allows for more and more emissions leading
to an angular decorrelation between the jets.  In the academical limit of
infinite rapidity, the angles should be completely uncorrelated. In the regime
of large, but realizable rapidity differences the resummation of large
logarithms calls for a description within the BFKL theory. Unfortunately, the
leading logarithmic approximation \cite{DelDuca:1993mn,Stirling:1994zs}
overestimates this decorrelation by far. Improvements have been obtained by
taking into account some corrections of higher order like the running of the
coupling \cite{Orr:1997im,Kwiecinski:2001nh}.  In particular, the effect of
energy-momentum conservation, which is beyond BFKL approximation, was shown to
have an important impact for large rapidity separation of jets
\cite{Orr:1997im}. In our present study this can affect the reliability of the
predictions at the borders of the phase space, as discussed in Sec.~\ref{Ap_emc}.  Some earlier calculations with
the next-to-leading (NLL) BFKL Green's function have been
published in Ref.~\cite{Vera:2007kn,Marquet:2007xx}.

In this paper we present the full NLL BFKL calculation where also the NLL result for the Mueller Navelet vertices~\cite{Bartels:2001ge,Bartels:2002yj} will be taken into account. In Sec.~\ref{sec:LLcalculation} we recall the LL BFKL calculation deriving also the key formulas which are then used in Sec.~\ref{sec:NLLcalculation} where the NLL calculation is presented. In Sec.~\ref{sec:Results} we present results and give a summary in Sec.~\ref{sec:Summary}. Technical details of the numerical implementation are given in an appendix.

\section{LL calculation}
\label{sec:LLcalculation}

\subsection{Kinematics}
\label{sec:kinematics}

\begin{figure}
 \centering
\includegraphics[height=11cm]{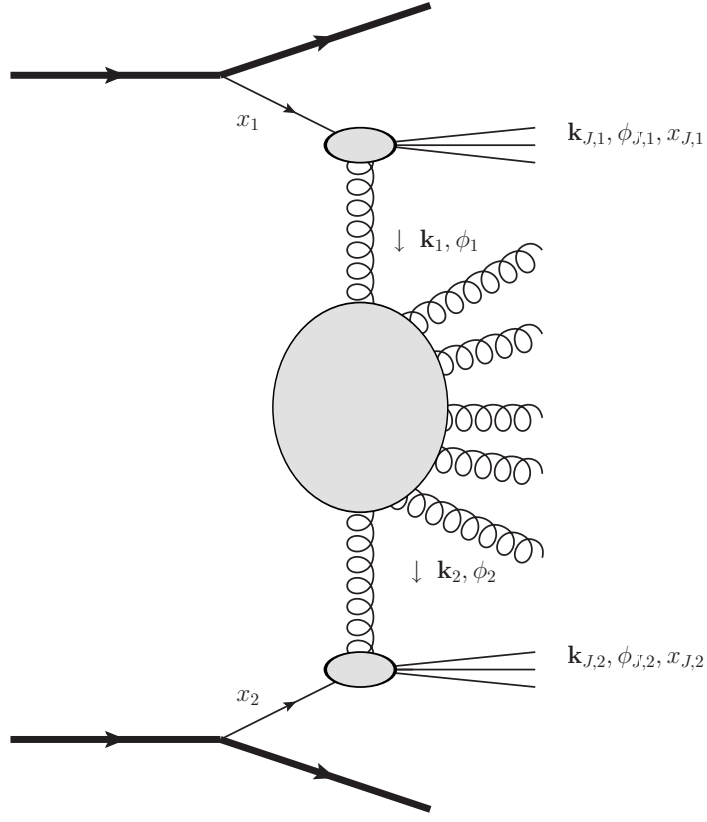}
\caption{Schematical illustration of the kinematics as described in Sec.~\ref{sec:kinematics}.}\label{fig:kinematics}
\end{figure}

The kinematic setup is schematically shown in Fig.~\ref{fig:kinematics}. The two hadrons collide at a center of mass energy $s$ producing two very forward jets, the transverse momenta of the jets are labeled by Euclidean two dimensional vectors $\veckjone$ and $\veckjtwo$, while their azimuthal angles are noted as $\phi_{J,1}$ and $\phi_{J,2}$. We will denote the rapidities of the jets by $y_{J,1}$ and $y_{J,2}$ which are related to the longitudinal momentum fractions of the jets via $x_J = \frac{|\veckj|}{\sqrt{s}}e^{y_J}$.

At any real experiment transverse momenta as well as rapidities are measured within certain intervals. A proper theoretical calculation should take this into account and integrate $|\veckji|$ and $y_{J,i}$ over the according interval. 
However, since at the LHC the binning in rapidity and in transverse momentum will be quite narrow \cite{Cerci:2008xv}, we consider the case of fixed rapidities and transverse momenta. 

\subsection{LL BFKL calculation}

Due to the large longitudinal momentum fractions $x_{J,1}$ and $x_{J,2}$
of the forward jets, collinear factorization holds and the differential cross section can be written as
\begin{equation}
  \frac{\dsigma}{\dtwojets} = \sum_{{\rm a},{\rm b}} \int_0^1 \dxone \int_0^1 \dxtwo f_{\rm a}(x_1) f_{\rm b}(x_2) \frac{\dsigmahat}{\dtwojets},
\end{equation}
where $f_{\rm a,b}$ are the standard parton distribution functions~(PDFs) of a parton a (b) in the according proton. They depend furthermore on the renormalization scale $\mu_R$ and the factorization scale $\mu_F$.

The partonic cross section at lowest order in the collinear factorization approach would just be described by simple two-to-two scattering processes as they are discussed in standard text books. However, the necessary resummation of logarithmically enhanced contributions calls for a description of the partonic cross section in $k_T$-factorization:
\begin{equation}
  \frac{\dsigmahat}{\dtwojets} = \int \dphijone\dphijtwo\int\dkone\dktwo V_{\rm a}(-\veckone,x_1)G(\veckone,\vecktwo,\shat)V_{\rm b}(\vecktwo,x_2),\label{eq:bfklpartonic}
\end{equation}
where $G$ is the BFKL Green's function depending on $\shat=x_1 x_2 s$, and the jet vertex $V$ at lowest order reads~\cite{Bartels:2001ge,Bartels:2002yj}:
\begin{align}
  V_{\rm a}^{(0)}(\veck,x) =& h_{\rm a}^{(0)}(\veck)\mathcal{S}_J^{(2)}(\veck;x) , & h_{\rm a}^{(0)}(\veck) =& \frac{\alpha_s}{\sqrt{2}}\frac{C_{A/F}}{\veck^2} , \label{def:V0}\\
 & & \mathcal{S}_J^{(2)}(\veck;x) =& \delta\left(1-\frac{x_J}{x}\right)|\veckj|\delta^{(2)}(\veck-\veckj).
\end{align}
In the definition of $h_{\rm a}^{(0)}$, $C_A=N_c=3$ is to be used for initial gluon and $C_F=(N_c^2-1)/(2N_c)=4/3$ for initial quark. Following the notation of Ref.~\cite{Bartels:2001ge,Bartels:2002yj}, the dependence of $V$ on the jet variables is implicit. 

Combining the PDFs with the jet vertices we now write
\begin{multline}
  \frac{\dsigma}{\dtwojets} = \\
= \int \dphijone\dphijtwo\int\dkone\dktwo \Phi(\veckjone,x_{J,1},-\veckone)G(\veckone,\vecktwo,\shat)\Phi(\veckjtwo,x_{J,2},\vecktwo) ,
\end{multline}
where
\begin{align}
  \Phi(\veckjtwo,x_{J,2},\vecktwo) =& \int \dxtwo f(x_2) V(\vecktwo,x_2).
\end{align}
These $\Phi$ are no longer impact factors in the classical sense as they depend, after the convolution in $x$ with the PDF, on the total energy $s$. In the `pure' BFKL formula of Eq.~\eqref{eq:bfklpartonic} the longitudinal momentum fractions $x_i$ were just some external parameter and the vertices $V$ would not depend on $\shat$ nor on $s$.

In view of the azimuthal decorrelation we want to investigate later, it is useful to define the following coefficients:
\begin{multline}
  \label{def:mathcalc}
  \mathcal{C}_m \equiv \int \dphijone\dphijtwo\cos\big(m(\phi_{J,1}-\phi_{J,2}-\pi)\big)\\
\times\int\dkone\dktwo \Phi(\veckjone,x_{J,1},-\veckone)G(\veckone,\vecktwo,\shat)\Phi(\veckjtwo,x_{J,2},\vecktwo) .
\end{multline}
Knowing these coefficients, one can easily obtain the differential cross section
\begin{equation}
  \frac{\dsigma}{\dtwojets} = \mathcal{C}_0,\label{def:dsigma}
\end{equation}
and the following measure of azimuthal decorrelation
\begin{equation}
  \langle\cos(m\varphi)\rangle \equiv \langle\cos\big(m(\phi_{J,1}-\phi_{J,2}-\pi)\big)\rangle = \frac{\mathcal{C}_m}{\mathcal{C}_0} .\label{def:decor}
\end{equation}

By decomposing $\Phi$ in terms of the LL-BFKL eigenfunctions
\begin{equation}
  E_{n,\nu}(\veckone) = \frac{1}{\pi\sqrt{2}}\left(\veckone^2\right)^{i\nu-\frac{1}{2}}e^{in\phi_1},
\label{def:eigenfunction}
\end{equation}
we can reduce the number of final integrations. To this purpose we define the intermediate coefficients
\begin{align}
   \hat{C}^{(1)}_{n_1,\nu_1}(\veckjone,x_{J,1}) =& \int\dkone\Phi(\veckjone,x_{J,1},-\veckone)E_{n_1,\nu_1}(\veckone) , \non
=& (-1)^{n_1} \int\dkprime \Phi(\veckjone,x_{J,1}\veck')E_{n_1,\nu_1}(\veck')\\
\hat{C}^{(2)}_{n_2,\nu_2}(\veckjtwo,x_{J,2}) =& \int\dktwo\Phi(\veckjtwo,x_{J,2},\vecktwo)E_{n_2,\nu_2}^*(\vecktwo),\\
\intertext{and make use of the following relations between different representations of the BFKL Green's function introducing the -- at LL arbitrary -- scale $s_0$:}
G(\veckone,\vecktwo,\shat) =& \int\frac{\deins{\omega} }{2\pi i}G_\omega(\veckone,\vecktwo) \left(\frac{\shat}{s_0}\right)^\omega , \label{eq:Gomega}\\
G_{n_1,n_2,\nu_1,\nu_2;\omega} =& \int\dkone\int\dktwo E^*_{n_1,\nu_1}(\veckone) G_\omega(\veckone,\vecktwo) E_{n_2,\nu_2}(\vecktwo) \non
  =& \frac{1}{\omega-\chihat(n_1,\nu_1)}\delta_{n_1,n_2}\delta(\nu_1-\nu_2),\label{eq:Gnnu}
\end{align}
where $\chihat(n_1,\nu_1)$ is given by the LL eigenvalue of the BFKL equation, namely
\begin{align}
  \chihat(n,\nu) =& \asbar\chi_0\left(|n|,\frac{1}{2}+i\nu\right) ,\\
\chi_0(n,\gamma) =& 2\Psi(1)-\Psi\left(\gamma+\frac{n}{2}\right)-\Psi\left(1-\gamma+\frac{n}{2}\right),
\end{align}
with $\Psi(x) = \Gamma'(x) /\Gamma(x)$, and $\asbar = N_c\alpha_s/\pi$.

With these new definitions we can write Eq.~\eqref{def:mathcalc} as:
\begin{align}
  \mathcal{C}_m \equiv& \sum_n \int \dnu \int \dphijone\dphijtwo\hat{C}^{(1)}_{n,\nu}(\veckjone,x_{J,1}) \left(\frac{\shat}{s_0}\right)^{\chihat(n,\nu)}
\hat{C}^{(2)}_{n,\nu}(\veckjtwo,x_{J,2})\cos(m\varphi) \non
=& (-1)^{m}\sum_n \int \dnu \left(\frac{\shat}{s_0}\right)^{\chihat(n,\nu)}\non
&\times\Bigg[
\left(\int \dphijone\cos(m\phi_{J,1})\hat{C}^{(1)}_{n,\nu}(\veckjone,x_{J,1})\right)
    \left(\int \dphijtwo\cos(m\phi_{J,2})\hat{C}^{(2)}_{n,\nu}(\veckjtwo,x_{J,2})\right) \non
&\hphantom{\times\Bigg[}+
\left(\int \dphijone\sin(m\phi_{J,1})\hat{C}^{(1)}_{n,\nu}(\veckjone,x_{J,1})\right)
    \left(\int \dphijtwo\sin(m\phi_{J,2})\hat{C}^{(2)}_{n,\nu}(\veckjtwo,x_{J,2})\right) \Bigg] .
\label{eq:cm}
\end{align}

After a little bit of simple algebra we end up with
\begin{equation}
  \mathcal{C}_m = (4-3\delta_{m,0})\int \dnu C_{m,\nu}(|\veckjone|,x_{J,1})C^*_{m,\nu}(|\veckjtwo|,x_{J,2})\left(\frac{\shat}{s_0}\right)^{\chihat(m,\nu)}.
\label{eq:cm2}
\end{equation}
Here we have defined
\begin{align}
  C_{m,\nu}(|\veckj|,x_{J})
=& \int\dphij\dk \dx f(x) V(\veck,x)E_{m,\nu}(\veck)\cos(m\phi_J) . \label{eq:mastercnnu}
\end{align}
The origin of the factor $(4-3\delta_{m,0})$ in Eq.~\eqref{eq:cm2} is twofold. Firstly the integration over $\phi_J$ leads to a $\delta_{m,|n|}$. Secondly when using the addition formula for $\cos(m\varphi)$ to disentangle $\phi_{J,1}$ and $\phi_{J,2}$ also coefficients with sine instead of cosine are generated. While for $m= 0$ they vanish, for $m\ne 0$ they give the same contribution as those with the cosine.

Inserting Eq.~\eqref{def:V0} into Eq.~\eqref{eq:mastercnnu}, we obtain for the LL Mueller Navelet jet vertices in conformal space
\begin{equation}
  C_{m,\nu}^{\rm (LL)}(|\veckj|,x_{J}) = \frac{\alpha_s C_{A/F}}{2}\left(\veckj^2\right)^{i\nu-1}x_Jf_{\rm a}(x_J)(1+\delta_{m,0}).
\label{eq:cnnuLO}
\end{equation}

It is worth to note, that $C_{m,\nu}^{\rm (LL)}$ depends on $m$ only in a trivial way ($1+\delta_{m,0}$) such that the azimuthal correlations \eqref{def:decor} do not depend on the PDFs at all. In the following section we will see, that this changes when one takes into account the NLL corrections to the jet vertices.

\section{NLL calculation}
\label{sec:NLLcalculation}

The master formulae of the LL calculation~(\ref{eq:cm2}, \ref{eq:mastercnnu}) will also be used for the NLL calculation. Even though the vertices do not simplify as drastically as in the LL case, we gain the possibility to calculate for a limited number of $m$ the coefficients $C_{m,\nu}$ as universal grids in $\nu$. In transverse momentum space one would need a two dimensional grid. Moreover, at NLL there are some contributions with an additional transverse momentum integration, such that some contributions would be analytic functions in {\it e.g.} $\veckone$ while other would be proportional to distributions like $\delta^{(2)}(\veckone-\veckjone)$. 

\subsection{Strong coupling, renormalization scheme and PDFs at NLL}

Based on the $\overline{\rm MS}$ renormalization scheme, we use the MSTW 2008 PDFs \cite{Martin:2009iq} and the two-loop strong coupling in the following form:
\begin{equation}
  \label{eq:runningcoupling}
  \alpha_s(\mu_R^2) = \frac{1}{b_0 L}\left(1+\frac{b_1}{b_0^2}\frac{\ln L}{L}\right),
\end{equation}
with $L=\ln \mu_R^2/\Lambda_{\rm QCD}^2$, and
\begin{align}
  b_0 =& \frac{33-2N_f}{12\pi} ,  & 
  b_1 =& \frac{153-19N_f}{24\pi^2} .
\end{align}
In the following $\alpha_s$ or $\asbar$ without argument is to be understood as $\alpha_s(\mu_R^2)$ or $\asbar(\mu_R^2)$ respectively. Since in the MSTW 2008 PDFs $\mu_R$ and $\mu_F$ are set to be equal, for a consistent calculation we are forced to perform this identification throughout the whole calculation as well.

\subsection{Jet vertices at NLL}

To calculate the coefficients $C_{m,\nu}$ \eqref{eq:mastercnnu} at next
to leading order level, we take for $V_{\rm a}(\veck,x)$ instead of just the leading order result $V^{(0)}(\veck,x)$ \eqref{def:V0} the full NLL vertex
\begin{equation}
 \label{eq:VLO+NLO}
  V_{\rm a}(\veck,x) = V^{(0)}_{\rm a}(\veck,x) + \alpha_s V^{(1)}_{\rm a}(\veck,x).
\end{equation}
The matrix elements needed to calculate the Mueller Navelet jet vertex at next
to leading order -- namely the partonic $2\to 3$ process at tree level and the
partonic $2\to 2$ process at one loop level -- are known for a long time.
The separation of collinear singularities (to be absorbed by renormalized PDFs)
from the BFKL large logarithms in $s$ was performed by Bartels, Vacca and one of
us~\cite{Bartels:2001ge,Bartels:2002yj} in terms of a generic and infrared-safe
jet algorithm. In this paper, we shall apply such procedure to a concrete jet
algorithm, namely the cone algorithm, as will be explained in Sec.~\ref{sec:jd}.

We will build on the results obtained in
Ref.~\cite{Bartels:2001ge,Bartels:2002yj} using their notation as well, but we correct an inconsistency in the treatment of the collinear cutoff parameter $\Lambda$ which later is identified with the factorization scale $\mu_F$. In the `real' $C_F$ term the authors rescale the transverse momentum which is integrated over but do not adapt the cutoff parameter $\Lambda$.
The correction of this point does not change the singular terms, and all the
discussion of the arrangement of divergences and subtractions remains unchanged.
However the finite part of the subtraction changes such that beside the cutoff
functions also the `virtual' part of the vertex changes, {\it e.g.} the term
proportional to $\left(\frac{\ln(1-z)}{1-z}\right)_+$ vanishes completely.
\footnote{
We note a misprint in equation~(105) of Ref.~\cite{Bartels:2001ge}: in the
`real' $C_A$ term the expression $\vecqecht-\veck$ must be replaced by
$\vecqecht-z \veck$ both in numerator and in the denominator. Just after it,
$+-$ is to be interpreted as $-$.}

The final expressions for the NLL correction to the vertices read:
\begin{eqnarray}
&&V_{\rm g}^{(1)}(\veck,x)\non
 &=& \left[\left(\frac{11}{6}\frac{C_A}{\pi}-\frac{1}{3}\frac{N_f}{\pi}\right)\ln\frac{\veck^2}{\Lambda^2}+\left(\frac{\pi^2}{4}-\frac{67}{36}\right)\frac{C_A}{\pi}+\frac{13}{36}\frac{N_f}{\pi}-b_0\ln\frac{\veck^2}{\mu^2}\right]V_{\rm g}^{(0)}(\veck,x)\non
&& +\int\dz\frac{N_f}{\pi}\frac{C_F}{C_A}z(1-z)V_{\rm g}^{(0)}(\veck,xz)\non
&& +\frac{N_f}{\pi}\int\frac{\dkprime}{\pi}\int_0^1\dz P_{\rm qg}(z)\Bigg[\frac{h_{\rm q}^{(0)}(\veck')}{(\vecq)^2+\veck'^2}\mathcal{S}_J^{(3)}(\veck',\vecq,xz;x)\non
&&\hphantom{+\frac{N_f}{\pi}\int\frac{\dkprime}{\pi}\int_0^1\dz P_{\rm qg}(z)\Bigg[}-\frac{1}{\veck'^2}\Theta(\Lambda^2-\veck'^2)V_{\rm q}^{(0)}(\veck,xz) \Bigg]\non
&& +\frac{N_f}{2\pi}\int\frac{\dkprime}{\pi}\int_0^1\dz P_{\rm qg}(z)\frac{\mathcal{N}C_A}{\big((1-z)\veck-\veck'\big)^2}\Bigg[
z(1-z)\frac{(\vecq)\cdot\veck'}{(\vecq)^2\veck'^2}\mathcal{S}_J^{(3)}(\veck',\vecq,xz;x)\non
&&\hphantom{+\frac{N_f}{2\pi}\int\frac{\dkprime}{\pi}\int_0^1\dz P_{\rm qg}(z)\frac{\mathcal{N}C_A}{\big((1-z)\veck-\veck'\big)^2}\Bigg[}
\hspace{-3mm}
-\frac{1}{\veck^2}\Theta\left(\Lambda^2-\big((1-z)\veck-\veck'\big)^2\right)\mathcal{S}_J^{(2)}(\veck,x)\Bigg]\non
&&+\frac{C_A}{\pi}\int_0^1 \frac{\dz}{1-z} \left[(1-z)P(1-z)\right]\int\frac{\dl}{\pi\vecl^2}\non
&&\hspace{1cm}\times\Bigg\{\frac{\mathcal{N}C_A}{\vecl^2+(\vecl-\veck)^2}\Big[
\mathcal{S}_J^{(3)}(z\veck+(1-z)\vecl,(1-z)(\veck-\vecl),x(1-z);x)\non
&&\hspace{1cm}\hphantom{\times\Bigg\{\frac{\mathcal{N}C_A}{\vecl^2+(\vecl-\veck)^2}\Big[}
+\mathcal{S}_J^{(3)}(\veck-(1-z)\vecl,(1-z)\vecl,x(1-z);x)\Big]\non
&&\hspace{1cm}\hphantom{\times\Bigg\{}-\Theta\left(\frac{\Lambda^2}{(1-z)^2}-\vecl^2\right)\left[V_{\rm g}^{(0)}(\veck,x) +V_{\rm g}^{(0)}(\veck,x z) \right]\Bigg\}\non
&&-\frac{2C_A}{\pi}\int_0^1 \frac{\dz}{1-z} \int\frac{\dl}{\pi\vecl^2}\Bigg[\frac{\mathcal{N}C_A}{\vecl^2+(\vecl-\veck)^2}S_J^{(2)}(\veck,x)-\Theta\left(\frac{\Lambda^2}{(1-z)^2}-\vecl^2\right)V_{\rm g}^{(0)}(\veck,x)\Bigg]\non 
&&+\frac{C_A}{\pi}\int\frac{\dkprime}{\pi}\int_0^1\dz\Bigg[ P(z) \bigg(
(1-z)\frac{(\vecq)\cdot\big((1-z)\veck-\veck'\big)}{(\vecq)^2\big((1-z)\veck-\veck'\big)^2} h_{\rm g}^{(0)}(\veck')\non
&&\hphantom{+\frac{C_A}{\pi}\int\frac{\dkprime}{\pi}\int_0^1\dz\Bigg[}\times\mathcal{S}_J^{(3)}(\veck',\vecq,xz;x)
-\frac{1}{\veck'^2}\Theta(\Lambda^2-\veck'^2)V_{\rm g}^{(0)}(\veck,xz)\bigg)\non
&&\hphantom{+\frac{C_A}{\pi}\int\frac{\dkprime}{\pi}\int_0^1\dz\Bigg[ }
-\frac{1}{z(\vecq)^2}\Theta\big(|\vecq|-z(|\vecq|+|\veck'|)\big)V_{\rm g}^{(0)}(\veck',x)\Bigg]\label{eq:Vgright}.
\end{eqnarray}
\begin{eqnarray}
&& V_{\rm q}^{(1)}(\veck,x) \non
&=&  \left[\left(\frac{3}{2}\ln\frac{\veck^2}{\Lambda^2}-\frac{15}{4}\right)\frac{C_F}{\pi}
  +\left(\frac{85}{36}+\frac{\pi^2}{4}\right)\frac{C_A}{\pi}
  -\frac{5}{18}\frac{N_f}{\pi}-b_0\ln\frac{\veck^2}{\mu^2}\right]V_{\rm q}^{(0)}(\veck,x)\non
&& +\int\dz\left(\frac{C_F}{\pi}\frac{1-z}{2}+\frac{C_A}{\pi}\frac{z}{2}\right)V_{\rm q}^{(0)}(\veck,x z)\non
&&+\frac{C_A}{\pi}\int\frac{\dkprime}{\pi}\int\dz \Bigg[\frac{1+(1-z)^2}{2z}\non
&&\hspace{2cm}\times\Bigg((1-z)\frac{(\vecq) \cdot \big((1-z)\veck-\veck'\big)}{(\vecq)^2\big((1-z)\veck-\veck'\big)^2}h_{\rm q}^{(0)}(\veck') \mathcal{S}_J^{(3)}(\veck',\vecq,xz;x)\non
&&\hspace{2cm}\hphantom{\Bigg(}
-\frac{1}{\veck'^2}\Theta(\Lambda^2-\veck'^2)V_{\rm q}^{(0)}(\veck,xz)\Bigg)\non
&&\hphantom{+\frac{C_A}{\pi}\int\frac{\dkprime}{\pi}\int\dz \Bigg[}-\frac{1}{z(\vecq)^2}\Theta\big(|\vecq|-z(|\vecq|+|\veck'|)\big)V_{\rm q}^{(0)}(\veck',x)\Bigg]\non
&&+\frac{C_F}{2\pi}\int\dz \frac{1+z^2}{1-z}\int\frac{\dl}{\pi\vecl^2}\non
&&\hspace{1cm}\times\Bigg[\frac{\mathcal{N}C_F}{\vecl^2+(\vecl-\veck)^2}\Big(\mathcal{S}_J^{(3)}(z\veck+(1-z)\vecl,(1-z)(\veck-\vecl),x(1-z);x)\non
&&\hspace{1cm}\hphantom{\times\Bigg[\frac{\mathcal{N}C_F}{\vecl^2+(\vecl-\veck)^2}\bigg(}+\mathcal{S}_J^{(3)}(\veck-(1-z)\vecl,(1-z)\vecl,x(1-z);x)\Big)\non
&&\hspace{1cm}\hphantom{\times\Bigg[}-\Theta\left(\frac{\Lambda^2}{(1-z)^2}-\vecl^2\right)\left(V_{\rm q}^{(0)}(\veck,x)+V_{\rm q}^{(0)}(\veck,xz)\right)\Bigg]\non
&&-\frac{2C_F}{\pi}\int\dz \left(\frac{1}{1-z}\right)\int\frac{\dl}{\pi\vecl^2}\Bigg[\frac{\mathcal{N}C_F}{\vecl^2+(\vecl-\veck)^2}S_J^{(2)}(\veck,x)\non
&&\hphantom{-\frac{2C_F}{\pi}\int\dz \left(\frac{1}{1-z}\right)\int\frac{\dl}{\pi\vecl^2}\Bigg[}-\Theta\left(\frac{\Lambda^2}{(1-z)^2}-\vecl^2\right)V_{\rm q}^{(0)}(\veck,x)\Bigg]\label{eq:Vqright},
\end{eqnarray}
Here $N_f$ denotes the number of active quark flavors, $b_0=(11N_c-2N_f)/(12\pi)$, and $\mathcal{N}=\alpha_s/\sqrt{2}$. A priori, the factorization scale $\mu_F=\Lambda$ and the renormalization scale $\mu_R=\mu$ are independent of each other even though in the end we will set them equal. 

\subsubsection{Jet definition\label{sec:jd}}

For a concrete calculation of Mueller Navelet jet production one also has to
choose a concrete jet algorithm obeying the property of infra-red safety, as
required by the general procedure of Refs.~\cite{Bartels:2001ge,Bartels:2002yj}.
Two of the most common ones are the cone algorithm
\begin{align}
 \mathcal{S}_J^{(3,{\rm cone})}(\veck',\vecq,xz;x) =& 
\hphantom{+}   \mathcal{S}_J^{(2)}(\veck,x)\;\Theta\left(\left[\frac{|\vecq|+|\veck'|}{\max(|\vecq|,|\veck'|)}R_{\rm cone}\right]^2-\left[\Delta y^2+\Delta\phi^2\right]\right) \non
  & \hspace{-2cm}+ \mathcal{S}_J^{(2)}(\vecq,xz)\;\Theta\left(\left[\Delta y^2+\Delta\phi^2\right]-\left[\frac{|\vecq|+|\veck'|}{\max(|\vecq|,|\veck'|)}R_{\rm cone}\right]^2\right) \non
  & \hspace{-2cm}+ \mathcal{S}_J^{(2)}(\veck',x(1-z))\;\Theta\left(\left[\Delta y^2+\Delta\phi^2\right]-\left[\frac{|\vecq|+|\veck'|}{\max(|\vecq|,|\veck'|)}R_{\rm cone}\right]^2\right), \label{eq:jetdefcone}\\
\intertext{as it has been adapted for NLL calculation in Ref.~\cite{Ellis:1989vm}, and the $k_T$ algorithm}
 \mathcal{S}_J^{(3,k_T)}(\veck',\vecq,xz;x) =& 
\hphantom{+}   \mathcal{S}_J^{(2)}(\veck,x)\;\Theta\big(R_{k_T}^2-\left[\Delta y^2+\Delta\phi^2\right]\big) \non
  & + \mathcal{S}_J^{(2)}(\vecq,xz)\;\Theta\big(\left[\Delta y^2+\Delta\phi^2\right]-R_{k_T}^2\big) \non
  &  + \mathcal{S}_J^{(2)}(\veck',x(1-z))\;\Theta\big(\left[\Delta y^2+\Delta\phi^2\right]-R_{k_T}^2\big)\label{eq:jetdefkt} ,
\end{align}
where
\begin{align}
  \Delta y=& \log\left(\frac{1-z}{z}\frac{|\vecq|}{|\veck'|}\right) , &
  \Delta\phi =& \arccos\frac{\veck'(\vecq)}{\sqrt{{\veck'}^2(\vecq)^2}} . 
\end{align}

In our study we will use the cone algorithm with a cone size of $R_{\rm cone}=0.5$ as it probably will be used in a CMS analysis at the LHC~\cite{Cerci:2008xv}.

\subsubsection{LL subtraction and $s_0$}

The requirement of a BFKL calculation that the two scattering objects have a similar hard scale is reflected by the fact that in this standard situation of BFKL physics the energy scale $s_0$ can be written as a product of two energy scales each assigned to one of these scattering objects.
\begin{equation}
  s_0 =\sqrt{s_{0,1} s_{0,2}} .
\end{equation}
In Ref.~\cite{Bartels:2001ge,Bartels:2002yj} the energy scale $s_{0,i}$ (assigned to the Mueller Navelet jet) was chosen as $(|\veckj|+|\veckj-\veck|)^2$. 
While $\veck$ is integrated over, it is preferable to let $s_0$ depend only on external scales. Also $\shat=x_1 x_2 s$ is in fact not an external scale since the longitudinal momentum fractions $x_1$ and $x_2$ are integrated over as well. Therefore, we want to change to a new $s_0'$:
\begin{align}
  s_{0,1}= (|\veckjone|+|\veckjone-\veckone|)^2 \quad\rightarrow\quad& s_{0,1}'= \frac{x_{1}^2}{x_{J,1}^2}\veckjone^2 , \\
  s_{0,2}= (|\veckjtwo|+|\veckjtwo-\vecktwo|)^2 \quad\rightarrow\quad& s_{0,2}'= \frac{x_{2}^2}{x_{J,2}^2}\veckjtwo^2 , \\
  \frac{\shat}{s_0}\quad\rightarrow\quad& \frac{\shat}{s_0'}=\frac{x_{J,1}x_{J_2}s}{|\veckjone|\cdot|\veckjtwo|} = e^{y_{J,1}-y_{J,2}} \equiv e^Y, \label{eq:s0}
\end{align}
where we introduced the relative rapidity $Y=y_{J,1}-y_{J,2}\,.$

The energy scale $s_0$ is a free parameter in the calculation. However, like for the renormalization scale at NLL level a change of it does not go without consequences. In fact, a change of $s_0\to s_0'$ in the Green's function has to be accompanied by an according correction term to the impact factors \cite{Fadin:1998sh,Fadin:1999qc}:
\begin{equation}
  \Phi_{\rm NLL} (\veck_i;s_{0,i}') = 
  \Phi_{\rm NLL} (\veck_i;s_{0,i})
+\int\dkprime \Phi_{\rm LL} (\veck_i')\mathcal{K}_{\rm LL}(\veck_i',\veck_i)\frac{1}{2}\ln\frac{s_{0,i}'}{s_{0,i}},\label{eq:s0changeinphi}
\end{equation}
with $\mathcal{K}_{\rm LL}$ being the LL BFKL kernel. Due to the Dirac delta distribution $\delta(1-x_{J,i}/x_i)$ in the jet algorithm inside $\Phi$ the ratio of longitudinal momentum fractions in $s_{0,i}'$ reduces to 1 and hence the logarithm in Eq.~\eqref{eq:s0changeinphi} vanishes for $\veck_i'=\veck_i$ such that only the real part of the kernel contributes.

To study the role of $s_0$, we will investigate the effect when changing it. A subsequent change of $s_{0,i}$ by just a factor $\lambda$ can be easily performed at the very end because of the use of BFKL eigenfunctions:
\begin{align}
& \hspace{-.6cm}C_{m,\nu}(|\veckj|,x_{J};s_0''=\lambda s_0')- C_{m,\nu}(|\veckj|,x_{J};s_0')\non
 =& \int\dphij\dk\int\dkprime \dx f(x) V^{(0)}(\veck',x)\mathcal{K}(\veck',\veck)E_{m,\nu}(\veck)\cos(m\phi_J)\frac{1}{2}\ln\frac{s_0''}{s_0'}\non
=& \int\dphij\int\dkprime \dx f(x) V^{(0)}(\veck',x) \asbar\chi_{0}\left(m,\frac{1}{2}+i\nu\right)E_{m,\nu}(\veck')\cos(m\phi_J)\frac{1}{2}\ln\lambda\non
=& \asbar\chi_{0}\left(m,\frac{1}{2}+i\nu\right) C_{m,\nu}^{\rm (LL)}(|\veckj|,x_{J})\frac{1}{2}\ln\lambda .
\end{align}

The LL subtraction, {\it i.e.} the terms multiplied by $\Theta(|\vecq|-z(|\vecq|+|\veck'|))$ in Eqs.~(\ref{eq:Vqright}, \ref{eq:Vgright}), cancels some part in the limit of the additional emission having a big rapidity distance to the jet. In fact, numerically this cancellation works very poorly due to an azimuthal averaging which has been performed for the LL subtraction. A significant improvement can be obtained by omitting this averaging and introducing new LL subtraction terms
\begin{subequations}
  \begin{align}
   V_{\rm q;\; LL\; subtraction}^{(1)} =& -\frac{C_A}{\pi^2}\frac{1}{z(\veck-\veck')^2}\frac{(\veck-\veck')(\veck-\veck'-z \veck') }{(\veck-\veck')^2(\veck-\veck'-z \veck')^2}V_{\rm q}^{(0)}(\veck',x) ,\\
   V_{\rm g;\; LL\; subtraction}^{(1)} =& -\frac{C_A}{\pi^2}\frac{1}{z(\veck-\veck')^2}\frac{(\veck-\veck')(\veck-\veck'-z \veck') }{(\veck-\veck')^2(\veck-\veck'-z \veck')^2}V_{\rm g}^{(0)}(\veck',x) .
 \end{align}
\label{eq:dimitrisoptimization}
\end{subequations}
As a consequence $s_{0,i}$ changes from $s_{0,i}=(|\veckji|+|\veckji-\vecki|)^2$ to $s_{0,i}=(\vecki-2\veckji)^2$.
It is also possible to use 
\begin{subequations}
\begin{align}
  \widetilde V_{\rm q;\; LL\; subtraction}^{(1)} =& -\frac{C_A}{\pi^2}\frac{1}{z(\veck-\veck')^2}\frac{(\veck-\veck')(\veck-\veck'-z \veck) }{(\veck-\veck')^2(\veck-\veck'-z \veck)^2}V_{\rm q}^{(0)}(\veck',x) ,\\ 
  \widetilde V_{\rm g;\; LL\; subtraction}^{(1)} =& -\frac{C_A}{\pi^2}\frac{1}{z(\veck-\veck')^2}\frac{(\veck-\veck')(\veck-\veck'-z \veck) }{(\veck-\veck')^2(\veck-\veck'-z \veck)^2}V_{\rm g}^{(0)}(\veck',x) ,
 \end{align}
\end{subequations}
which are slightly inferior concerning the numerical performance but give a $s_0$ change from $s_{0,i}=(|\veckji|+|\veckji-\vecki|)^2$ to $s_{0,i}=\veckji^2$ which already is close to the final $s_0'$ making a correction term (Eq.~\eqref{eq:s0changeinphi}) needless since the ratio of longitudinal momentum fractions in $s_{0,i}'$ effectively reduces to 1 as described above.

We have checked that all three possible subtraction terms after combining them with the according correction term \eqref{eq:s0changeinphi} lead to the same result. For reasons of numerical performance we have chosen Eqs.~\eqref{eq:dimitrisoptimization} for the final calculation. As we nevertheless aim for the final $s_0'$ defined in Eq.~\eqref{eq:s0} we still have to use the correction term introduced in Eq.~\eqref{eq:s0changeinphi} additionally.

\subsection{BFKL Green's function at NLL}

Last but not least, we also have to take the BFKL Green's function at NLL level. The key to the Green's function is the BFKL kernel at NLL \cite{Fadin:1998py,Ciafaloni:1998gs}. 
While at LL the BFKL equation is conformally invariant, at NLL it is not such that in fact the LL eigenfunctions $E_{n,\nu}$ \eqref{def:eigenfunction} are strictly speaking not eigenfunctions of the NLL kernel.
Nevertheless, the action of the NLL BFKL kernel on the eigenfunctions has been calculated in Ref.~\cite{Kotikov:2000pm}. The status of the $E_{n,\nu}$ being eigenfunctions formally can be saved if one accepts the eigenvalue to become an operator containing a derivative with respect to $\nu$ \cite{Ivanov:2005gn,Vera:2006un,Vera:2007kn}. In combination with the impact factors the derivate acts on the impact factors and effectively leads to a contribution to the eigenvalue which depends on the impact factors \cite{Ivanov:2005gn,Vera:2006un,Vera:2007kn,Schwennsen:2007hs}:
\begin{multline}
   \label{eq:omegaNLO}
   \chihat(n,\nu) = \asbar \chi_0\left(|n|,\frac{1}{2}+i\nu\right)   + \asbar^2 \Bigg[\chi_1\left(|n|,\frac{1}{2}+i\nu\right)\\
-\frac{\pi b_0}{2N_c}\chi_0\left(|n|,\frac{1}{2}+i\nu\right) \left\{-2\ln\mu_R^2-i\frac{\partial}{\partial\nu}\ln\frac{C_{n,\nu}(|\veckjone|,x_{J,1})}{C_{n,\nu}(|\veckjtwo|,x_{J,2})}\right\}\Bigg],
\end{multline}
where
\begin{align}
  \chi_1(n,\gamma) =& \phantom{+}\mathcal{S}\chi_0(n,\gamma) + \frac{3}{2}\zeta(3)-\frac{\beta_0}{8N_c}\chi_0^2(n,\gamma)\non
& +\frac{1}{4}\left[\psi''\left(\gamma+\frac{n}{2}\right)+\psi''\left(1-\gamma+\frac{n}{2}\right)-2\phi(n,\gamma)-2\phi(n,1-\gamma)\right]\non
&- \frac{\pi^2\cos(\pi\gamma)}{4\sin^2(\pi\gamma)(1-2\gamma)}\Bigg\{\left[3+\left(1+\frac{N_f}{N_c^3}\right)\frac{2+3\gamma(1-\gamma)}{(3-2\gamma)(1+2\gamma)}\right]\delta_{n,0}\non
&\hspace{2cm}-\left(1+\frac{N_f}{N_c^3}\right)\frac{\gamma(1-\gamma)}{2(3-2\gamma)(1+2\gamma)}\delta_{n,2}\Bigg\},
\label{eq:nlokernel}
\end{align}
with the constant ${\mathcal S} = (4 - \pi^2 + 5 {\beta_0}/{N_c})/12$. $\zeta(n)=\sum_{k=1}^\infty k^{-n}$ is the Riemann zeta function while the function $\phi$ reads
\begin{multline}
  \phi(n,\gamma) = \sum_{k=0}^\infty \frac{(-1)^{k+1}}{k+\gamma+\frac{n}{2}}\Bigg(\psi'(k+n+1)-\psi'(k+1)\\
+ (-1)^{k+1}\left[\beta'(k+n+1)+\beta'(k+1)\right]
 +\frac{\psi(k+1)-\psi(k+n+1)}{k+\gamma+\frac{n}{2}}\Bigg),
\end{multline}
with
\begin{equation}
  \beta'(\gamma) = \frac{1}{4}\left[\psi'\left(\frac{1+\gamma}{2}\right)-\psi'\left(\frac{\gamma}{2}\right)\right].
\end{equation}

At NLL accuracy, only the leading order vertex coefficients \eqref{eq:cnnuLO} enter in the derivative term of \eqref{eq:omegaNLO}:
\begin{equation}
   \label{eq:dnuerivative}
-2\ln\mu_R^2-i\frac{\partial}{\partial\nu}\ln\frac{C^{\rm (LL)}_{n,\nu}(|\veckjone|,x_{J,1})}{\left(C^{\rm (LL)}_{n,\nu}(|\veckjtwo|,x_{J,2})\right)^*}
= 2\ln\frac{|\veckjone|\cdot|\veckjtwo|}{\mu_R^2} .
\end{equation}

\subsubsection{Collinear improved Green's function}

There are methods to improve the NLL BFKL kernel for $n=0$ by imposing compatibility with the DGLAP equation \cite{Gribov:1972ri,Lipatov:1974qm,Altarelli:1977zs,Dokshitzer:1977sg} in the collinear limit \cite{Salam:1998tj,Ciafaloni:1998iv,Ciafaloni:1999yw,Ciafaloni:2003rd}. They are known under the name $\omega$-shift because essentially poles in $\gamma=1/2+i\nu$ and $1-\gamma$ are shifted by $\omega/2$ with some compensation terms ensuring that the result is not changed at fixed order (having in mind that $\omega\sim\asbar\chi_0$). 
The different attempts are very similar, and here we use the most transparent method presented in \cite{Salam:1998tj}. In fact, based on previous experience \cite{Enberg:2005eq, Vera:2007kn,Schwennsen:2007hs} we use scheme 3 of \cite{Salam:1998tj}. The new kernel $\asbar\chi^{(1)}(\gamma,\omega)$ with shifted poles replaces $\asbar\chi_0(\gamma,0)+\asbar^2\chi_1(\gamma,0)$ and $\chihat(0,\nu)$ is obtained by solving the implicit equation 
\begin{equation}
  \label{eq:chisalam}
  \chihat(0,\nu) = \asbar\chi^{(1)}(\gamma,\chihat(0,\nu))
\end{equation}
for $\chihat(n,\nu)$ numerically.

In general the additional $\nu$-derivative term makes it necessary to recalculate the coefficients $d_{1,k}$ (defined in Ref.~\cite{Salam:1998tj}) but in our case the LL vertex does not contain any poles in $\gamma$ nor in $1-\gamma$ leaving the coefficients $d_{1,k}$ unchanged \footnote{The same is true for the NLL vertex (except for the $s_0$-correction term \eqref{eq:s0changeinphi}) as can be seen by using a closed contour in $\gamma$-plane around 0 or around 1, and numerically integrating integer powers of $\gamma$ or $1-\gamma$ times the vertex. Based on Cauchy's formula -- used here in reverse manner -- one can then obtain a numerical evaluation of the residue of arbitrary order, and show that they actually vanish.}. 
By introducing an $\omega$ dependence in the eigenvalue the pole in \eqref{eq:Gnnu} is no longer a simple one such that the residue in fact reads
\begin{equation}
  \label{eq:Gresummed}
  G_{0,0,\nu_1,\nu_2}(\shat) = \left(1-\left.\frac{\partial\chi^{(1)}(\frac{1}{2}+i\nu_1,\omega)}{\partial\omega}\right|_{\omega=\chihat(0,\nu_1)}\right)^{-1} \left(\frac{\shat}{s_0}\right)^{\chihat(0,\nu_1)}\delta(\nu_1-\nu_2) .
\end{equation}

\subsubsection{Approximate energy-momentum conservation in BFKL}
\label{Ap_emc}

We would like to finish this section with the following important observation.
Energy-momentum conservation is not fulfilled in any truncated BFKL treatment
(i.e.\ LL BFKL, or NLL BFKL, etc.) while it is preserved in any truncated fixed
order treatment \`a la DGLAP (LO, NLO, etc.).  Our approach uses a
semi-analytical resummed solution of the BFKL equation at NLL and does not
allow, in a direct way, for the implementation of a procedure based on the
iteration of the BFKL kernel (in the spirit of Ref.~\cite{Orr:1997im}) in which
energy-momentum conservation could be imposed step by step. Exact
energy-momentum conservation is beyond the scope of our pure BFKL treatment.
Therefore, our results are expected to be valid only in a limited range of
relative rapidity between the two Mueller Navelet jets, away from the
kinematical bounds.  Nevertheless, this violation within BFKL approach is less
dramatic when going further in the truncation (note that BFKL and DGLAP are
expected to converge to same result when going higher and higher in the order of
perturbation). We thus expect that in the region far from the kinematical limit,
such energy-momentum conservation effects would not introduce significant
corrections to our NLL BFKL results.

\section{Results}
\label{sec:Results}

We now present results for the LHC at the design center of mass energy
$\sqrt{s}=14\,{\rm TeV}$. Motivated by a recent CMS study \cite{Cerci:2008xv} we
restrict the rapidities of the Mueller Navelet jets to the region $3<|y_J|<5$.
We shall show the differential cross section with respect to the relative
rapidity variable $Y = y_{J,1} - y_{J,2}$ which therefore takes values between 6
and 10.  Note that, since the maximum possible rapidity of a jet with 50 GeV of
transverse energy (see below) is $y_{\mathrm{max}} = 5.6$, values of $Y \simeq
10$ are quite close to the kinematical boundary and the corresponding
predictions may suffer some uncertainties because of the fact that momentum
conservation is not exactly fulfilled in this BFKL approach.

We consider Mueller Navelet jets with $|\veckj|=35\,{\rm GeV}$, and $|\veckj|=50\,{\rm GeV}$ respectively. Due to our method of calculation and the factorization between the two Mueller Navelet jets  we can can combine the building blocks to two symmetric scenarios ($|\veckjone|=|\veckjtwo|=35\,{\rm GeV}$ or $|\veckjone|=|\veckjtwo|=50\,{\rm GeV}$) and one asymmetric scenario of $|\veckjone|=35\,{\rm GeV}$, $|\veckjtwo|=50\,{\rm GeV}$ (plus the `mirrored' process $|\veckjone|=50\,{\rm GeV}$, $|\veckjtwo|=35\,{\rm GeV}$) even though in doing so one mixes different choices for $\mu_R$. But since $\alpha_s$ only varies by $\sim 4\%$ between 35\,GeV and 50\,GeV we give the according result as well.

In Ref. \cite{Andersen:2001kta} it is argued based on Ref.\cite{Frixione:1997ks}
that imposing $|\veckjone|>E$, $|\veckjtwo|>E+D$ for $D\to0$ large logarithms of
non-BFKL origin make a fixed order calculation unstable.  Even though in a pure
BFKL framework these logarithms do not show up at all, they -- and their
resummation -- might be of significant impact on the result of a BFKL
calculation. Therefore, $D\ne 0$ is preferred also in the BFKL framework to be
safe from these unknown contributions.  However, we start the presentation of
our result with these symmetric scenarios but in order to be conservative and to
avoid the region where initial state radiations might require peculiar treatment
involving resummations \`a la Sudakov, which are beyond the scope of our work
(and not implemented in fixed NLO calculation neither), we will close in
Sec.~\ref{sec:asymmetriccase} with our results in the asymmetric case for which
such resummations effects are clearly not required.

In all cases we choose the number of active flavors to be five ($N_f=5$) with
$\Lambda_{\rm QCD}=221.2\,{\rm MeV}$ such that $\alpha_s(M_Z^2)=0.1176$.

The Monte Carlo integration \cite{Hahn:2004fe} itself is error-prone.
This error in Monte Carlo integration can be reduced to less than 1\% with a large number of sample points, so as to be practically negligible in comparison with other sources
of uncertainties. In practice, due to hardware/ time limitations 
we will display results for a Monte Carlo integration setup (for details see Sec.~\ref{sec:cubaparameters}) which aims
for an accuracy of the order of 1\%.

However, as we show in what follows, there are more serious uncertainties due to the renormalization scale $\mu_R$ which we choose as $\mu_R=\sqrt{|\veckjone|\cdot |\veckjtwo|}$. To study the dependence on it we vary $\mu_R$ by factors $2$ and $1/2$ respectively. The same we do for the energy scale $\sqrt{s_0}$.
We investigate the effect of the uncertainty of PDFs for asymmetric errors as defined in Eqs.~(51,52) of \cite{Martin:2009iq} with the eigenvector set ensuring all data sets being described within their $90\%$ confidence level limits.

Below we present our results. The logic of their presentation is the following. For each kinematical situation, we discuss the physical
observables and their uncertainties: cross-section encoded in ${\cal C}_0$, azimuthal decorrelation encoded in ${\cal C}_1/{\cal C}_0,$ ${\cal C}_2/{\cal C}_1$ and ${\cal C}_2/{\cal C}_1\,.$ We then give
further details on the additional underlying quantities ${\cal C}_1\,,$ ${\cal C}_2\,.$
We use the same color coding for all plots, namely blue shows the pure LL result, brown the pure NLL result, green the combination of LL vertices with the collinear improved NLL Green's function, red the full NLL vertices with the collinear improved NLL Green's function. Whenever we show curves for scales $\mu_R=\mu_F$ or $s_0$ changed by factors 2 or $1/2$, the thick curve corresponds to the scale changed by factor 2.

\subsection{$|\veckjone|=|\veckjtwo|=35\,{\rm GeV}$}
\label{sec:case3535}

\begin{figure}[h!]
  \centering
  \psfrag{varied}{}
  \psfrag{cubaerror}{}
  \psfrag{C1}{$\mathcal{C}_1 \left[\frac{\rm nb}{{\rm GeV}^2}\right] $}
  \psfrag{Y}{$Y$}
  \includegraphics[width=9cm]{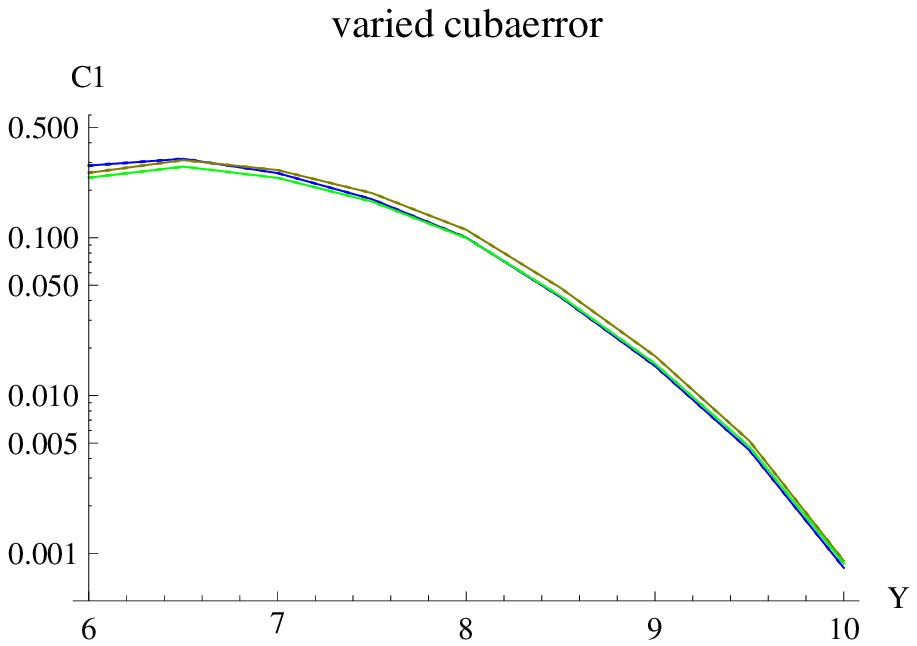}
  \caption{Coefficient $\mathcal{C}_1$ in dependence on $Y$ for $|\veckjone|=|\veckjtwo|=35\,{\rm GeV}$. The errors due to the Monte Carlo integration -- though hardly visible -- are given as error bands. The tabled values are shown in Tab.~\ref{tab:c13535}.}
  \label{fig:c13535}
\end{figure}

\begin{figure}[h!]
  \centering
  \psfrag{varied}{}
  \psfrag{cubaerror}{}
  \psfrag{C2}{$\mathcal{C}_2 \left[\frac{\rm nb}{{\rm GeV}^2}\right] $}
  \psfrag{Y}{$Y$}
  \includegraphics[width=9cm]{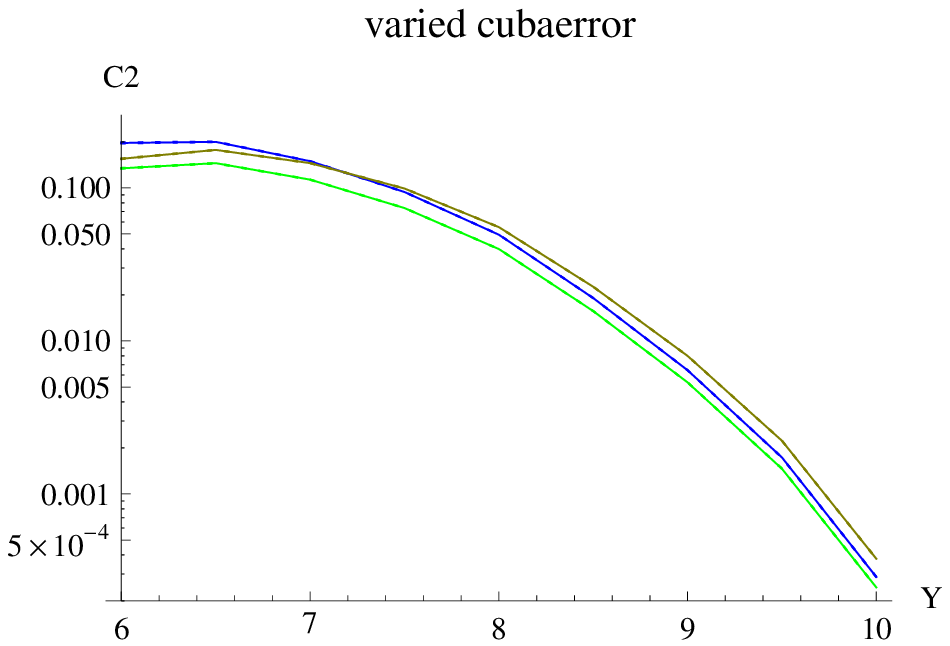}
  \caption{Coefficient $\mathcal{C}_2$ in dependence on $Y$ for $|\veckjone|=|\veckjtwo|=35\,{\rm GeV}$. The errors due to the Monte Carlo integration -- though hardly visible -- are given as error bands. The tabled values are shown in Tab.~\ref{tab:c23535}.}
  \label{fig:c23535}
\end{figure}

Thus the first thing to look at is the differential cross section as defined in Eq.~\eqref{def:dsigma}. The result of our calculation is shown in Fig.~\ref{fig:c03535} (the according tabled values are shown in Tab.~\ref{tab:c03535} in the Appendix). 
 The decrease of the cross section at large values of $Y \gtrsim 7$ is mostly
 an effect of the upper kinematical cut on the rapidity of the single jets.
 This is true also for the coefficients $\mathcal{C}_1$ and $\mathcal{C}_2$ in
 Figs.~\ref{fig:c13535} and~\ref{fig:c23535}.

\begin{figure}[h!]
  \centering
  \psfrag{varied}{}
  \psfrag{s0}{}\psfrag{cubaerror}{}\psfrag{pdfset}{}\psfrag{mu}{}
  \psfrag{deltaC0}{$\delta\mathcal{C}_0 \left[\frac{\rm nb}{{\rm GeV}^2}\right] $}
  \psfrag{Y}{$Y$}
  \includegraphics[width=15cm]{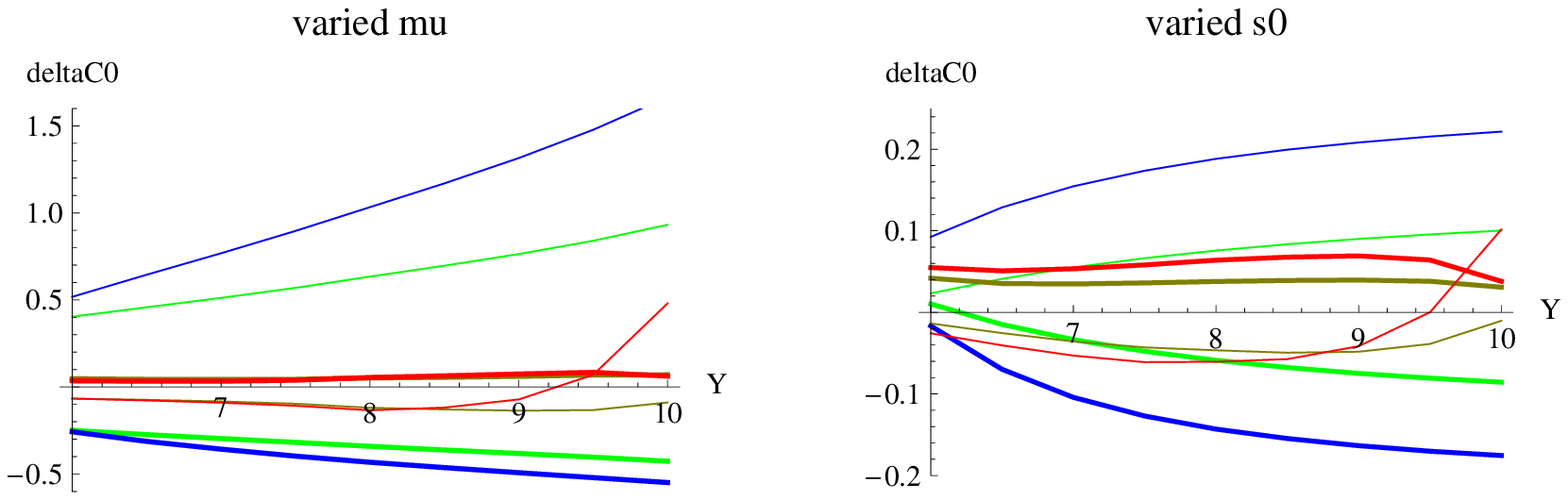}
  \caption{Relative effect of changing $\mu_R=\mu_F$ by factors 2 (thick) and $1/2$ (thin) respectively (left), and $\sqrt{s_0}$ (right) by factors 2 (thick) and $1/2$ (thin) resp. on the differential cross section in dependence on $Y$ for $|\veckjone|=|\veckjtwo|=35\,{\rm GeV}$. 
The tabled values are shown in Tabs.~\ref{tab:c03535_mu} and \ref{tab:c03535_s0}.
  }
  \label{fig:c03535rel_mu_s0}
\end{figure}

\begin{figure}[h!]
  \centering
  \psfrag{varied}{}
  \psfrag{s0}{}\psfrag{cubaerror}{}\psfrag{pdfset}{}\psfrag{mu}{}
  \psfrag{deltaC0}{$\delta\mathcal{C}_0 \left[\frac{\rm nb}{{\rm GeV}^2}\right] $}
  \psfrag{Y}{$Y$}
  \includegraphics[width=15cm]{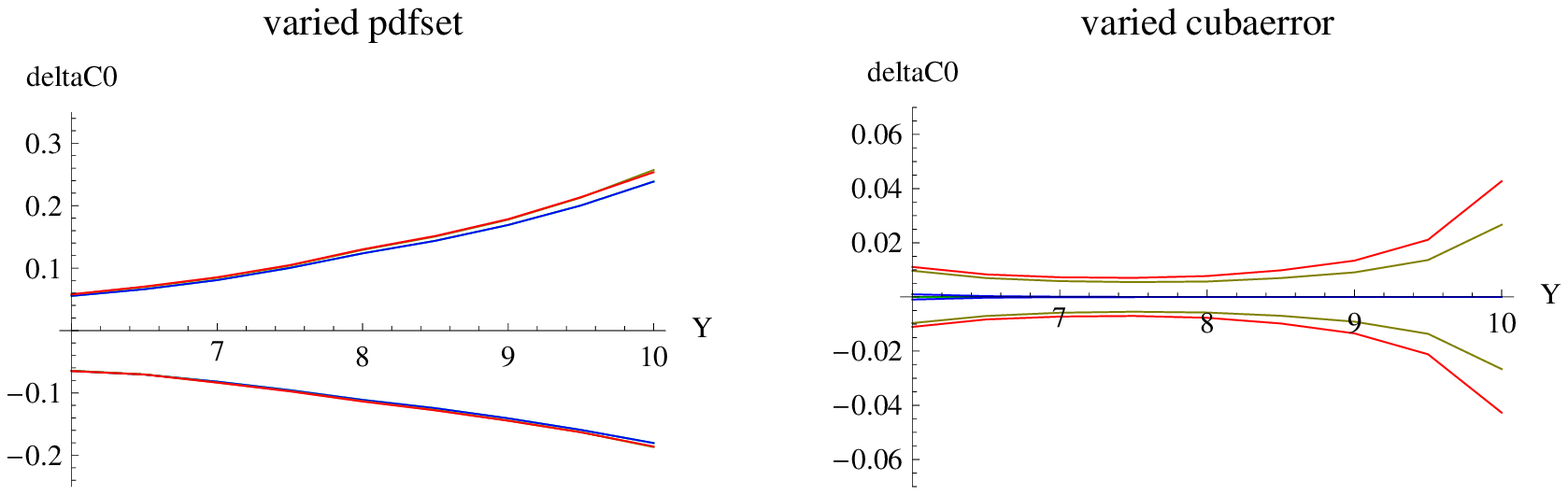}
  \caption{Relative effect of the PDF (left) and Monte Carlo (right) errors on the differential cross section in dependence on $Y$ for $|\veckjone|=|\veckjtwo|=35\,{\rm GeV}$. The tabled values are shown in Tabs.~\ref{tab:c03535_pdf} and \ref{tab:c03535}.}
  \label{fig:c03535rel_pdf_cuba}
\end{figure}

The purely numerical error due to the Monte Carlo integration of the NLL vertices is mainly below $2\%$ (see Fig.~\ref{fig:c03535rel_pdf_cuba}), and only for very large $Y$ of the order of $2-5\%$. We varied the renormalization and factorization scale by factors $2$ and $1/2$ to investigate the $\mu_R$ dependence. A full scan over this interval is not possible due to the CPU time consumption of the evaluation for a single choice of $\mu_R$. The results are displayed in Fig.~\ref{fig:c03535rel_mu_s0}. As one would expect, the full NLL result depends less on $\mu_R$ than the LL result or the combination of LL vertices with resummed NLL Green's function which was so far state-of-the-art \cite{Vera:2007kn,Marquet:2007xx}. 
However, the results obtained for the
 three choices of $\mu_R$ studied here seem to suggest that, as expected,
after inclusion of the NLL vertices the 
$\mu_R$ dependence is no longer monotone and flattens out, resulting in 
higher stability of NLL results. 
Another important scale is the energy $s_0$ introduced by the Mellin transformation from energy to $\omega$ space which is necessary to formulate the BFKL equation. Like $\mu_R$ it is an artificial scale which in an all order calculation would not affect the result. Indeed, the dependence is reduced when the NLL corrections to the vertices are taken into account (see Fig.~\ref{fig:c03535rel_mu_s0}). 

The dependence on PDF uncertainties is shown in Fig.~\ref{fig:c03535rel_pdf_cuba}. This significant sensitivity is almost identical for pure LL, pure NLL, combined LL vertices with collinear improved NLL Green's function and for full NLL vertices combined with the collinear improved NLL Green's function. 

\begin{figure}[h!]
   \centering
   \psfrag{varied}{}
   \psfrag{cubaerror}{}
   \psfrag{C1C0}{$\frac{\mathcal{C}_1}{\mathcal{C}_0}$}
   \psfrag{Y}{$Y$}
   \includegraphics[width=8.8cm]{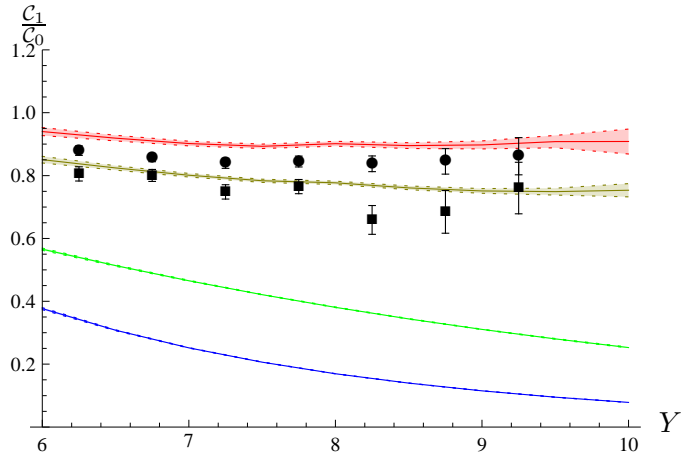}
   \caption{$\langle \cos \varphi\rangle$ in dependence on $Y$ for $|\veckjone|=|\veckjtwo|=35\,{\rm GeV}$. The errors due to the Monte Carlo integration are given as error bands. The tabled values are shown in Tab.~\ref{tab:c1c03535}. As dots are shown the results of Ref.~\cite{Cerci:2008xv} obtained with \textsc{Pythia} \cite{Sjostrand:2006za}. As squares are shown the results of Ref.~\cite{Cerci:2008xv} obtained with \textsc{Herwig} \cite{Marchesini:1991ch}.}
   \label{fig:c1c03535}
 \end{figure}

\begin{figure}[h!]
  \centering
  \psfrag{varied}{}
  \psfrag{cubaerror}{}
  \psfrag{C2C0}{$\frac{\mathcal{C}_2}{\mathcal{C}_0}$}
  \psfrag{Y}{$Y$}
  \includegraphics[width=9cm]{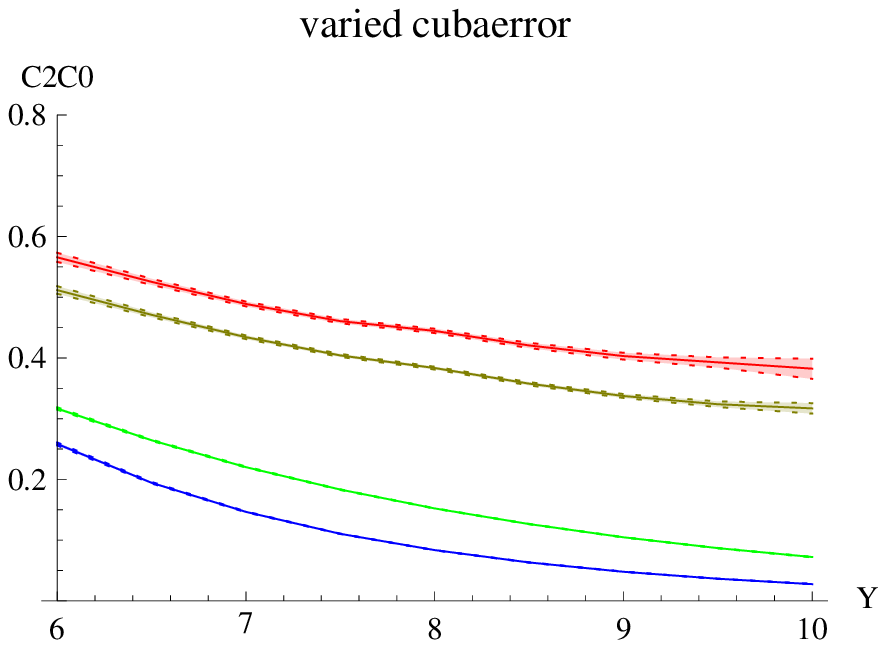}
  \caption{$\langle \cos 2\varphi\rangle$ in dependence on $Y$ for $|\veckjone|=|\veckjtwo|=35\,{\rm GeV}$. The errors due to the Monte Carlo integration are given as error bands. The tabled values are shown in Tab.~\ref{tab:c2c03535}.}
  \label{fig:c2c03535}
\end{figure}

The azimuthal decorrelation, described by coefficients defined in Eq.~(\ref{def:decor}),
 has often be predicted to be a striking feature of BFKL physics. Our results displayed in Fig.~\ref{fig:c1c03535} for $\langle \cos \varphi \rangle$ and in Fig.~\ref{fig:c2c03535} for $\langle \cos 2\varphi \rangle$ explicitly show that our inclusion of NLL vertices leads to an enormous correlation in the azimuthal angle (for completeness our results for ${\cal C}_1$ and ${\cal C}_2$ coefficients alone are displayed respectively in Fig.~\ref{fig:c13535} and Fig.~\ref{fig:c23535}).

\begin{figure}[h!]
   \centering
   \psfrag{varied}{}
   \psfrag{mu}{}
   \psfrag{C1C0}{$\frac{\mathcal{C}_1}{\mathcal{C}_0}$}
   \psfrag{Y}{$Y$}
   \includegraphics[width=15cm]{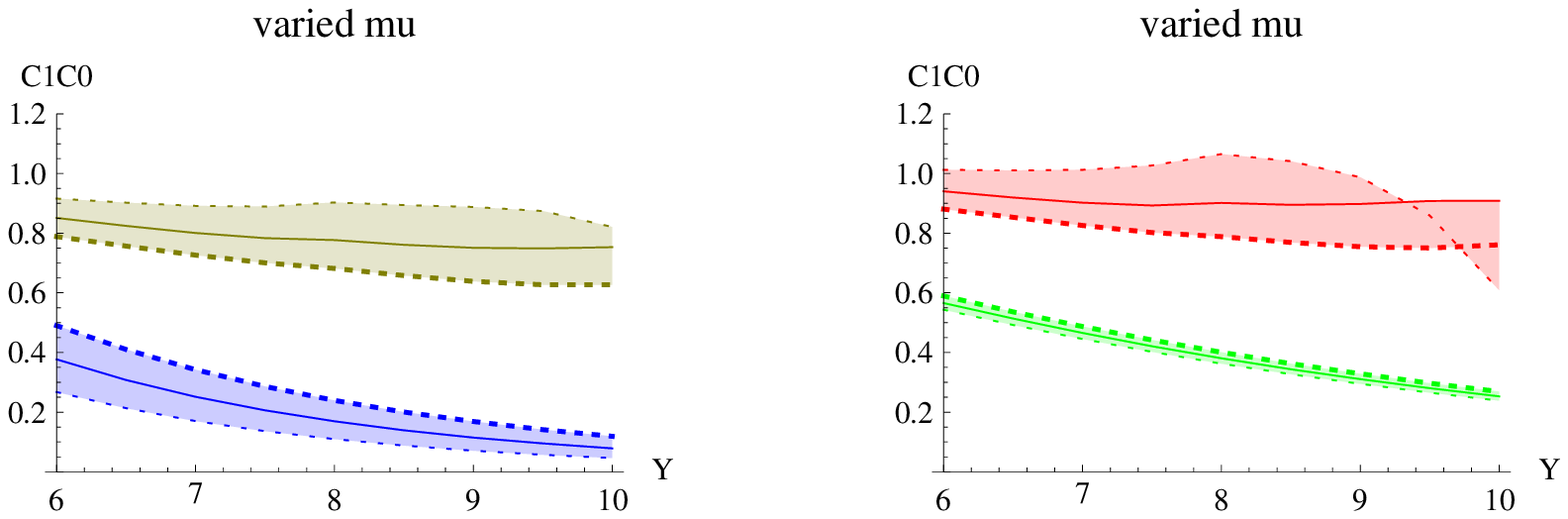}
   \caption{Effect of changing $\mu_R=\mu_F$ by factors 2 and $1/2$ respectively on $\langle \cos \varphi\rangle$ in dependence on $Y$ for $|\veckjone|=|\veckjtwo|=35\,{\rm GeV}$. The tabled values are shown in Tab.~\ref{tab:c1c03535_mu}}
   \label{fig:c1c03535_mu}
\end{figure}

\begin{figure}[h!]
  \centering
  \psfrag{varied}{}
  \psfrag{s0}{}
  \psfrag{C1C0}{$\frac{\mathcal{C}_1}{\mathcal{C}_0}$}
  \psfrag{Y}{$Y$}
  \includegraphics[width=15cm]{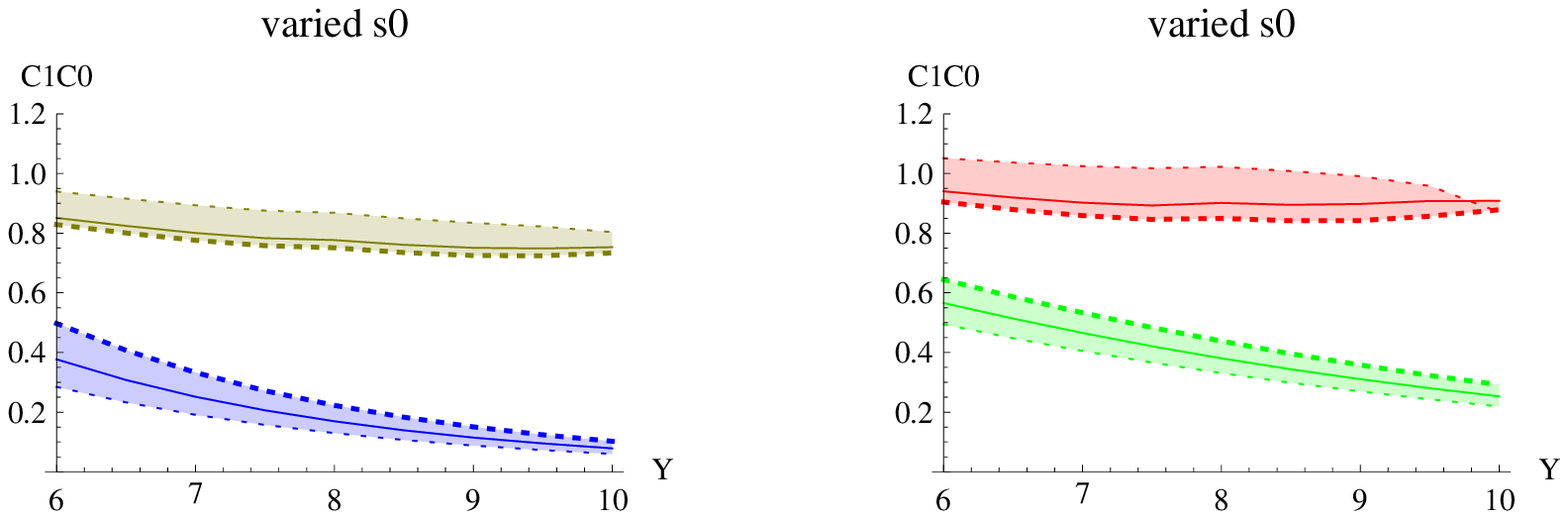}
  \caption{Effect of changing $\sqrt{s_0}$ by factors 2 and $1/2$ respectively on $\langle \cos \varphi\rangle$ in dependence on $Y$ for $|\veckjone|=|\veckjtwo|=35\,{\rm GeV}$. The tabled values are shown in Tab.~\ref{tab:c1c03535_s0}.}
  \label{fig:c1c03535_s0}
\end{figure}

\begin{figure}[h!]
  \centering
  \psfrag{varied}{}
  \psfrag{mu}{}
  \psfrag{C2C0}{$\frac{\mathcal{C}_2}{\mathcal{C}_0}$}
  \psfrag{Y}{$Y$}
  \includegraphics[width=15cm]{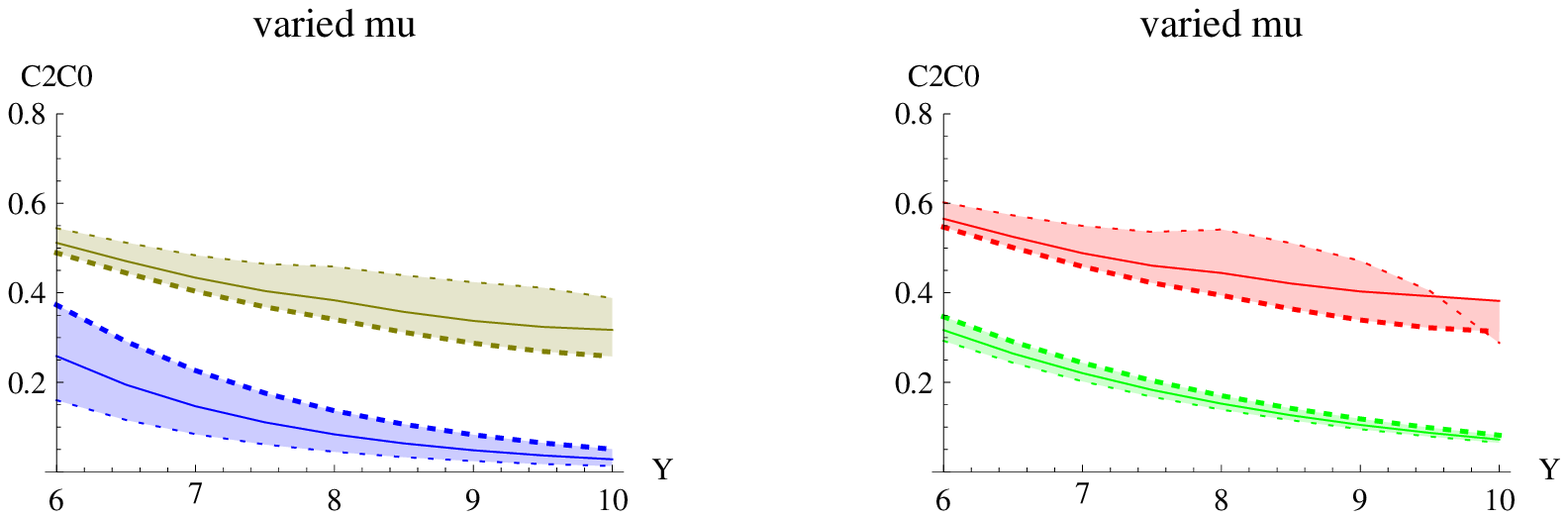}
  \caption{Effect of changing $\mu_R=\mu_F$ by factors 2 and $1/2$ respectively on $\langle \cos 2\varphi\rangle$ in dependence on $Y$ for $|\veckjone|=|\veckjtwo|=35\,{\rm GeV}$. The tabled values are shown in Tab.~\ref{tab:c2c03535_mu}.}
  \label{fig:c2c03535_mu}
\end{figure}

\begin{figure}[h!]
  \centering
  \psfrag{varied}{}
  \psfrag{s0}{}
  \psfrag{C2C0}{$\frac{\mathcal{C}_2}{\mathcal{C}_0}$}
  \psfrag{Y}{$Y$}
  \includegraphics[width=15cm]{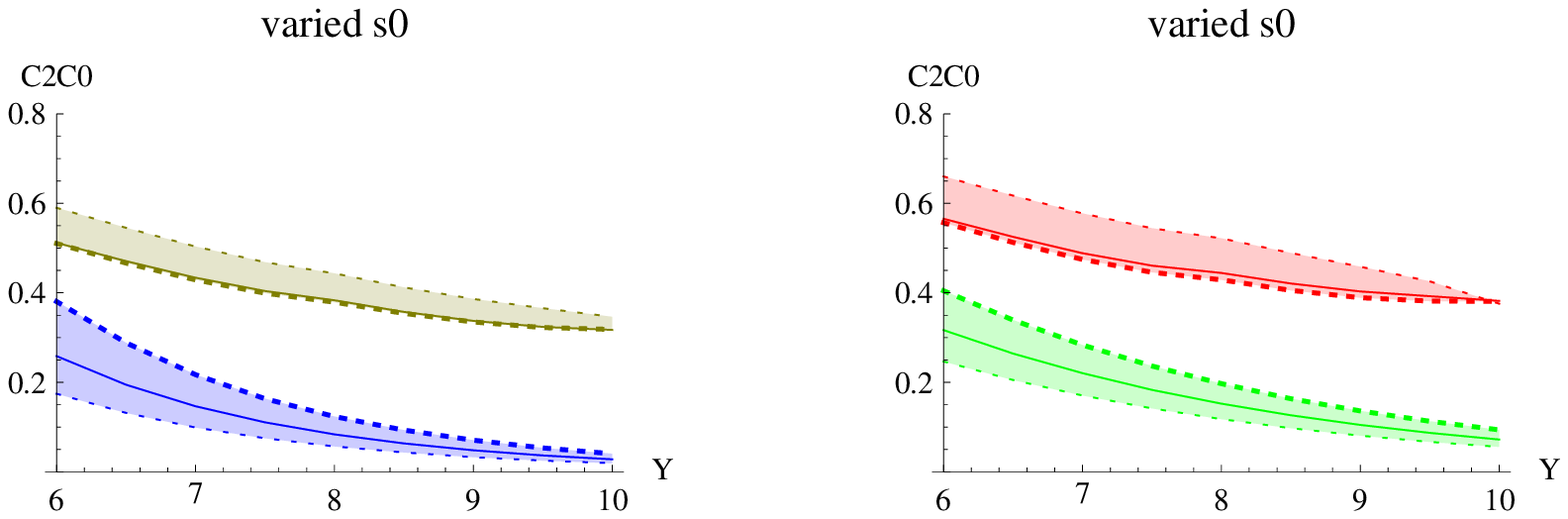}
  \caption{Effect of changing $\sqrt{s_0}$ by factors 2 and $1/2$ respectively on $\langle \cos 2\varphi\rangle$ in dependence on $Y$ for $|\veckjone|=|\veckjtwo|=35\,{\rm GeV}$. The tabled values are shown in Tab.~\ref{tab:c2c03535_s0}.}
  \label{fig:c2c03535_s0}
\end{figure}

 In particular, $\langle \cos \varphi \rangle$ shown in Fig.~\ref{fig:c1c03535} is rather close to the typical values predicted by LO-DGLAP Monte Carlos \textsc{Pythia} \cite{Sjostrand:2006za}
and \textsc{Herwig} \cite{Marchesini:1991ch}, used for CMS studies \cite{Cerci:2008xv}.
Note that \textsc{Herwig} has the tendency to predict more decorrelation, presumably because since it implements more radiations than \textsc{Pythia}, it has the phenomenological
effect
to involve some kind of NLO-DGLAP corrections, which enhance the decorrelation. The various sources of uncertainty of our results are shown for $\langle \cos \varphi \rangle$ in 
Fig.~\ref{fig:c1c03535_mu} (variation of $\mu_R=\mu_F$), 
Fig.~\ref{fig:c1c03535_s0} (variation of $s_0$), 
and for $\langle \cos 2 \varphi \rangle$ in Fig.~\ref{fig:c2c03535_mu},
and Fig.~\ref{fig:c2c03535_s0} 
accordingly.

Not only is the $Y$ dependence much flatter for the NLL vertices. But the mean value for $\cos\varphi$ itself is very close to 1, especially for the NLL calculation including the collinear improved Green's function. 
 Actually, in the collinear improved approach and for low values of $\mu_R$,
 $\langle \cos\varphi \rangle = \mathcal{C}_1/\mathcal{C}_0$ can exceed unity.
 This feature has to be ascribed to the fact that the collinear improvement is
 justified and applied only to the $n=0$ conformal spin of the kernel, with the
 effect of lowering $\mathcal{C}_0$ in a large region of the phase space, while
 keeping $\mathcal{C}_1$ unchanged. For this reason, the predictions of angular
 dependent quantities are more trustable in the pure NLL approach (without
 collinear resummation).

\begin{figure}[h!]
  \centering
  \psfrag{varied}{}
  \psfrag{mu}{}
  \psfrag{C2C1}{$\frac{\mathcal{C}_2}{\mathcal{C}_1}$}
  \psfrag{Y}{$Y$}
  \includegraphics[width=15cm]{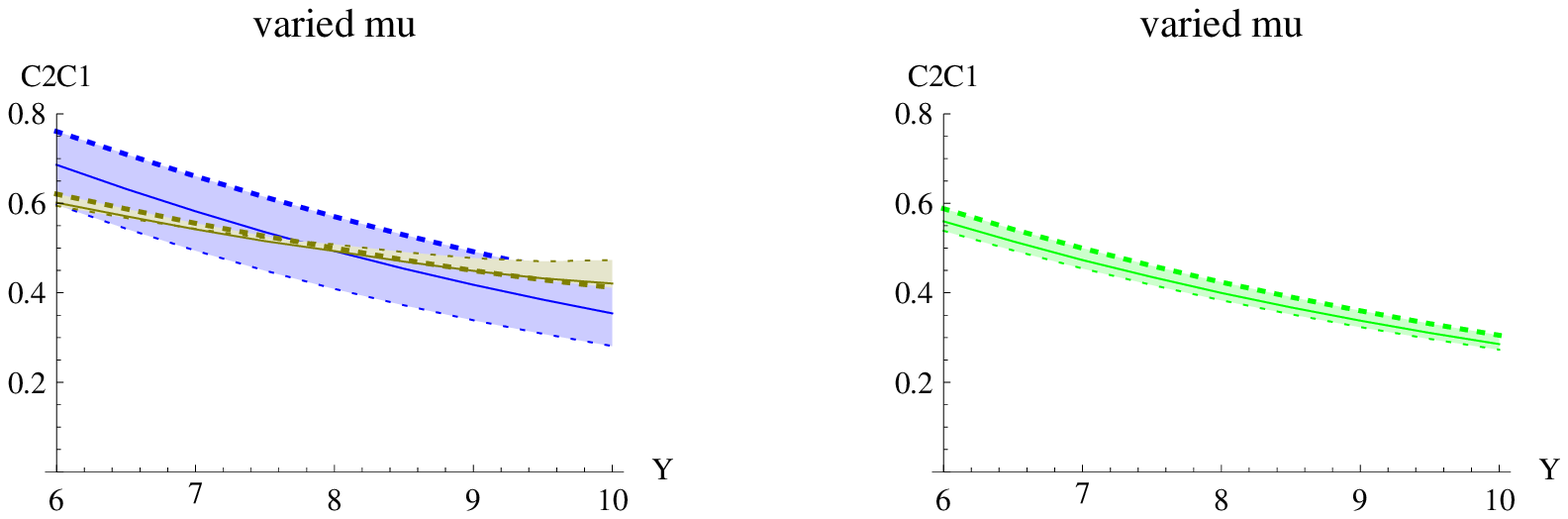}
  \caption{Effect of changing $\mu_R=\mu_F$ by factors 2 and $1/2$ respectively on $\langle \cos 2\varphi\rangle / \langle \cos \varphi\rangle$ in dependence on $Y$ for $|\veckjone|=|\veckjtwo|=35\,{\rm GeV}$. The tabled values are shown in Tab.~\ref{tab:c2c13535_mu}.}
  \label{fig:c2c13535_mu}
\end{figure}

\begin{figure}[h!]
  \centering
  \psfrag{varied}{}
  \psfrag{s0}{}
  \psfrag{C2C1}{$\frac{\mathcal{C}_2}{\mathcal{C}_1}$}
  \psfrag{Y}{$Y$}
  \includegraphics[width=15cm]{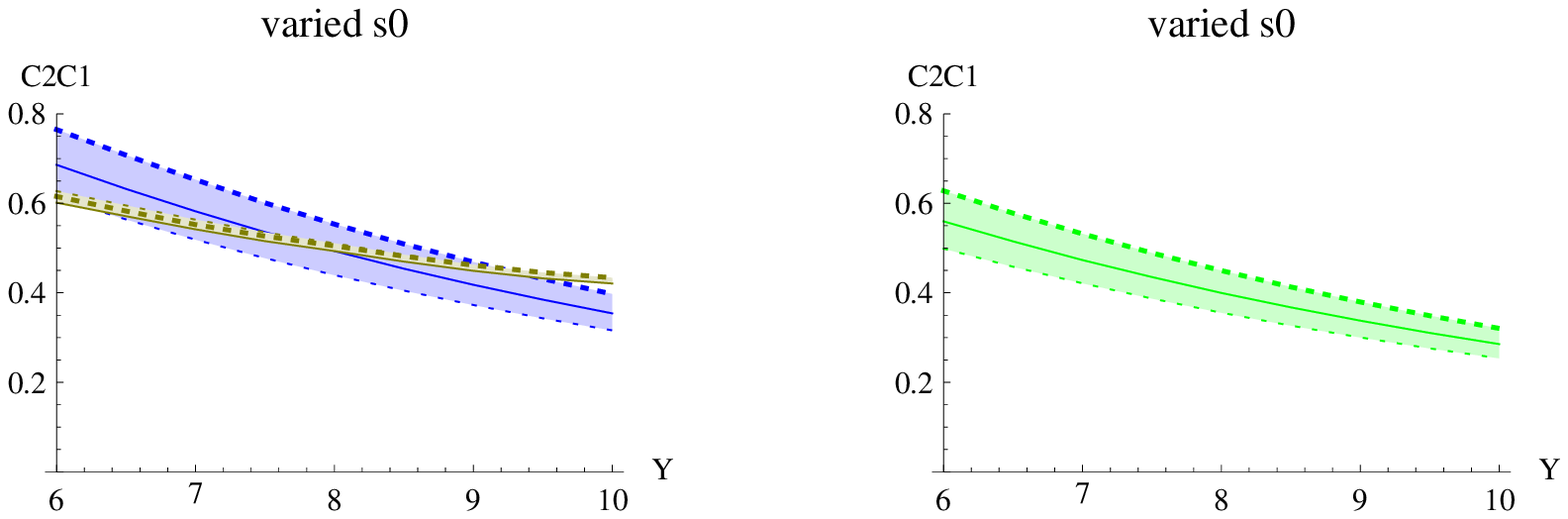}
  \caption{Effect of changing $\sqrt{s_0}$ by factors 2 and $1/2$ respectively on $\langle \cos 2\varphi\rangle / \langle \cos \varphi\rangle$ in dependence on $Y$ for $|\veckjone|=|\veckjtwo|=35\,{\rm GeV}$.  The tabled values are shown in Tab.~\ref{tab:c2c13535_s0}.}
  \label{fig:c2c13535_s0}
\end{figure}

\begin{figure}[h!]
  \centering
  \psfrag{varied}{}
  \psfrag{s0}{}\psfrag{cubaerror}{}\psfrag{pdfset}{}\psfrag{mu}{}
  \psfrag{deltaC1}{$\delta\mathcal{C}_1 \left[\frac{\rm nb}{{\rm GeV}^2}\right] $}
  \psfrag{Y}{$Y$}
  \includegraphics[width=15cm]{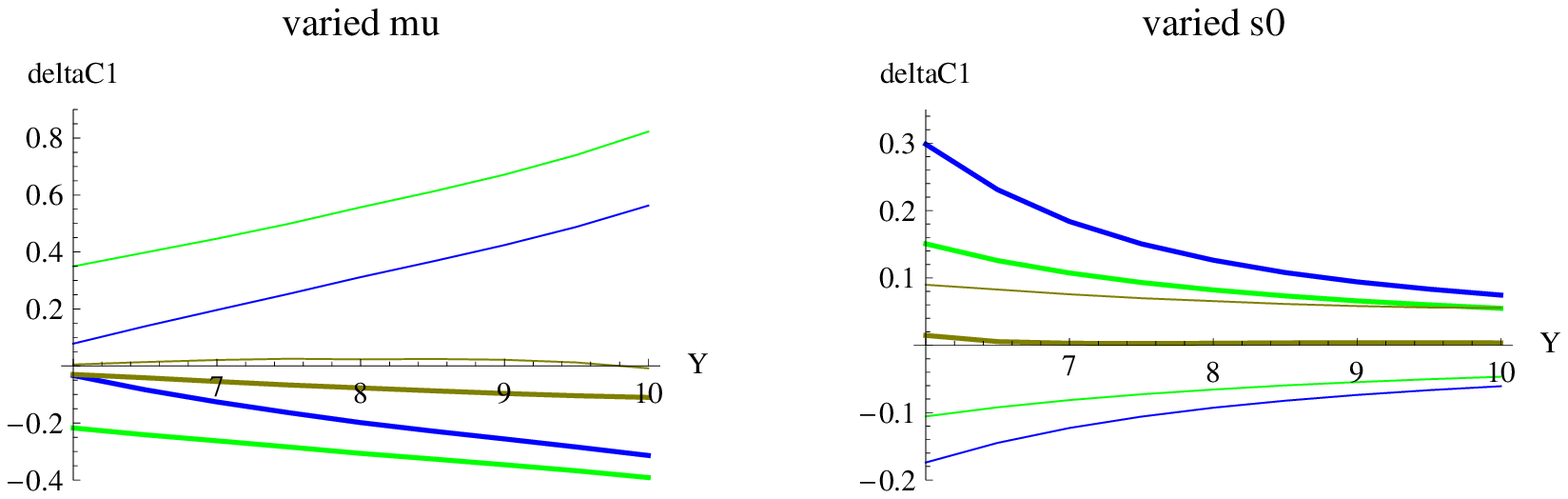}
  \caption{Relative effect of changing $\mu_R=\mu_F$ by factors 2 and $1/2$ respectively (left), and $\sqrt{s_0}$ (right) by factors 2 and $1/2$ respectively on the coefficient $\mathcal{C}_1$ in dependence on $Y$ for $|\veckjone|=|\veckjtwo|=35\,{\rm GeV}$. The tabled values are shown in Tabs.~\ref{tab:c13535_mu} and \ref{tab:c13535_s0}.}
  \label{fig:c13535rel_mu_s0}
\end{figure}

The $\mu_R$ dependence of $\mathcal{C}_1/\mathcal{C}_0$ (see Fig.~\ref{fig:c1c03535_mu}) is
puzzling, but has to be considered as a consequence of the $\mathcal{C}_0$ and $\mathcal{C}_1$ dependencies 
in Figs.~\ref{fig:c03535rel_mu_s0}
and \ref{fig:c13535rel_mu_s0} respectively.
A similar behavior can be observed for the $s_0$ dependence (see Fig.~\ref{fig:c1c03535_s0} for $\mathcal{C}_1/\mathcal{C}_0$ and Fig.~\ref{fig:c03535rel_mu_s0} and Fig.~\ref{fig:c13535rel_mu_s0} for ${\cal C}_0$ and ${\cal C}_1$ respectively).
While the $\mu_R$ and $s_0$ dependences for ${\cal C}_0$ and ${\cal C}_1$ are significantly reduced by the inclusion of the NLL vertices, in the ratio this is only true if compared to the pure LL calculation. The combination of LL vertices and NLL collinear improved NLL Green's function is less sensitive on changes of $s_0$ or $\mu_R$. The reason for this surprising `weakness' of the NLL result is that the changes of the LL vertices when changing $s_0$ or $\mu_R$ are not very sensitive on $n$ such that ratios of LL vertices are very stable against scale changes.
In contrast, the NLL correction -- especially the LL subtraction -- is very large (and negative) for the $n=0$ component while of minor significance for $n>0$ such that effects of a scale change for NLL vertices do not vanish by considering ratios $\mathcal{C}_n/\mathcal{C}_0$. The special role of $n=0$ becomes apparent if one compares the situation to $\mathcal{C}_{n\ne 0}/\mathcal{C}_{m\ne 0}$ (see Figs.~\ref{fig:c2c13535_s0}, \ref{fig:c2c13535_mu}) where the expected advantage of the full NLL calculation is clearly visible. 

\begin{figure}[h!]
  \centering
  \psfrag{varied}{}
  \psfrag{pdfset}{}
  \psfrag{C1C0}{$\frac{\mathcal{C}_1}{\mathcal{C}_0}$}
  \psfrag{Y}{$Y$}
  \includegraphics[width=15cm]{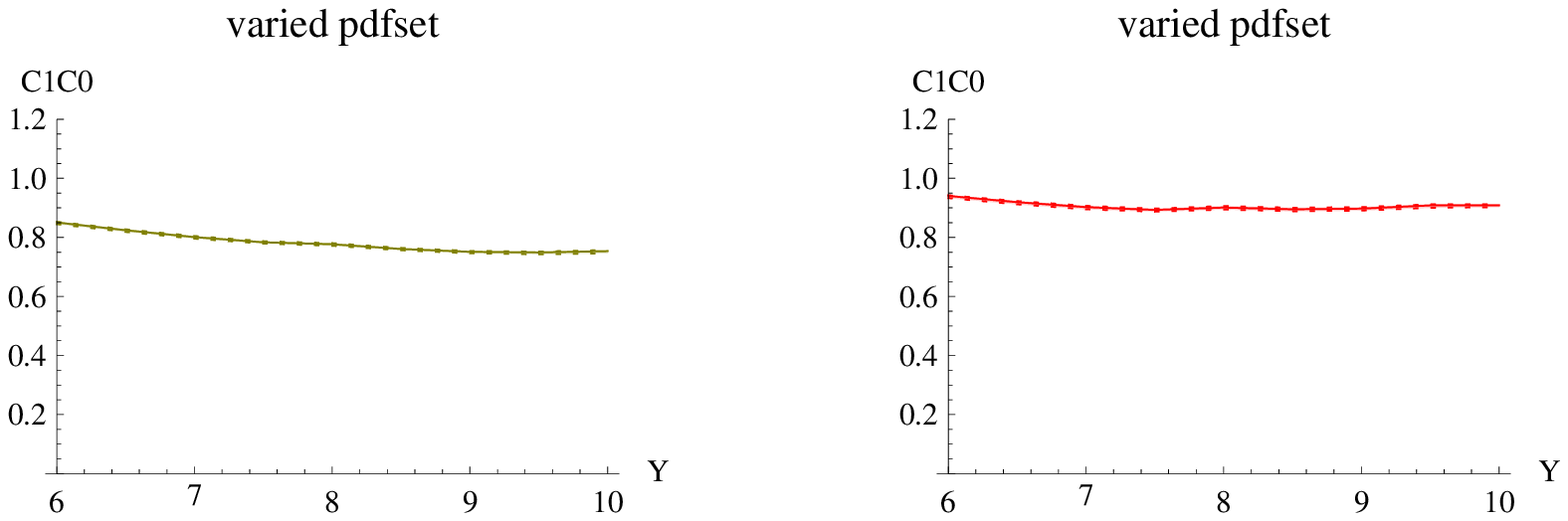}
  \caption{Effect of the PDF errors on $\langle \cos \varphi\rangle$ in dependence on $Y$ for $|\veckjone|=|\veckjtwo|=35\,{\rm GeV}$. The tabled values are shown in Tab.~\ref{tab:c1c03535_pdf}.}
  \label{fig:c1c03535_pdf}
\end{figure}

For both scales, the curves exceeding 1 belong to smaller scales (in the calculation with the collinear improved Green's function) which seem to be very disfavored in full NLL BFKL calculations as already discussed in \cite{Ivanov:2005gn,Ivanov:2006gt,Caporale:2007vs}.
The dependence on the PDFs completely drops out for LL vertices, and also for NLL vertices it is negligible as can be seen in Fig.~\ref{fig:c1c03535_pdf}. 

\begin{figure}[h!]
  \centering
  \psfrag{varied}{}
  \psfrag{s0}{}\psfrag{cubaerror}{}\psfrag{pdfset}{}\psfrag{mu}{}
  \psfrag{deltaC2}{$\delta\mathcal{C}_2 \left[\frac{\rm nb}{{\rm GeV}^2}\right] $}
  \psfrag{Y}{$Y$}
  \includegraphics[width=15cm]{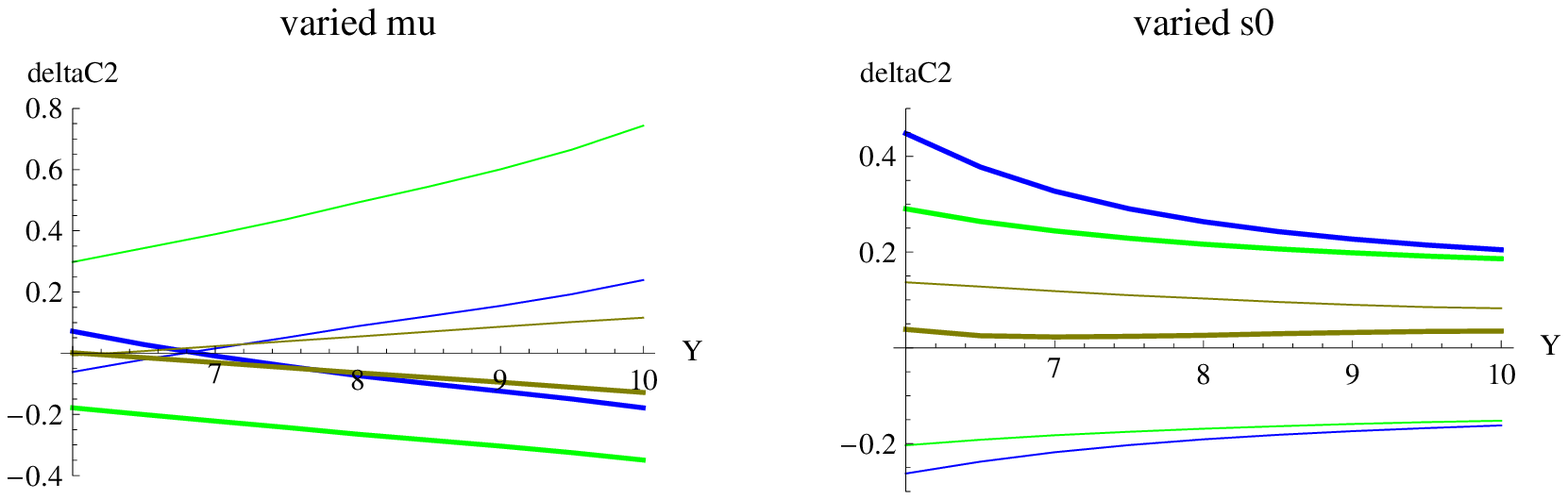}
  \caption{Relative effect of changing $\mu_R=\mu_F$ by factors 2 and $1/2$ respectively (left), and $\sqrt{s_0}$ (right) by factors 2 and $1/2$ respectively on the coefficient $\mathcal{C}_2$ in dependence on $Y$ for $|\veckjone|=|\veckjtwo|=35\,{\rm GeV}$. The tabled values are shown in Tabs.~\ref{tab:c23535_mu} and \ref{tab:c23535_s0}.}
  \label{fig:c23535rel_mu_s0}
\end{figure}

 A similar rather large dependency on $\mu_R=\mu_F$ and $s_0$ is obtained for ${\cal C}_2/{\cal C}_0\,,$
although it does not lead to values of $\langle \cos 2 \varphi \rangle$ close to 1, as can be seen from Figs.~\ref{fig:c2c03535_mu},
\ref{fig:c2c03535_s0}, based on detailed studies of coefficients
${\cal C}_0$ and
 ${\cal C}_2$ displayed respectively in Figs.~\ref{fig:c03535rel_mu_s0} and \ref{fig:c23535rel_mu_s0}.

\begin{figure}[h!]
  \centering
  \psfrag{varied}{}
  \psfrag{cubaerror}{}
  \psfrag{C2C1}{$\frac{\mathcal{C}_2}{\mathcal{C}_1}$}
  \psfrag{Y}{$Y$}
  \includegraphics[width=9cm]{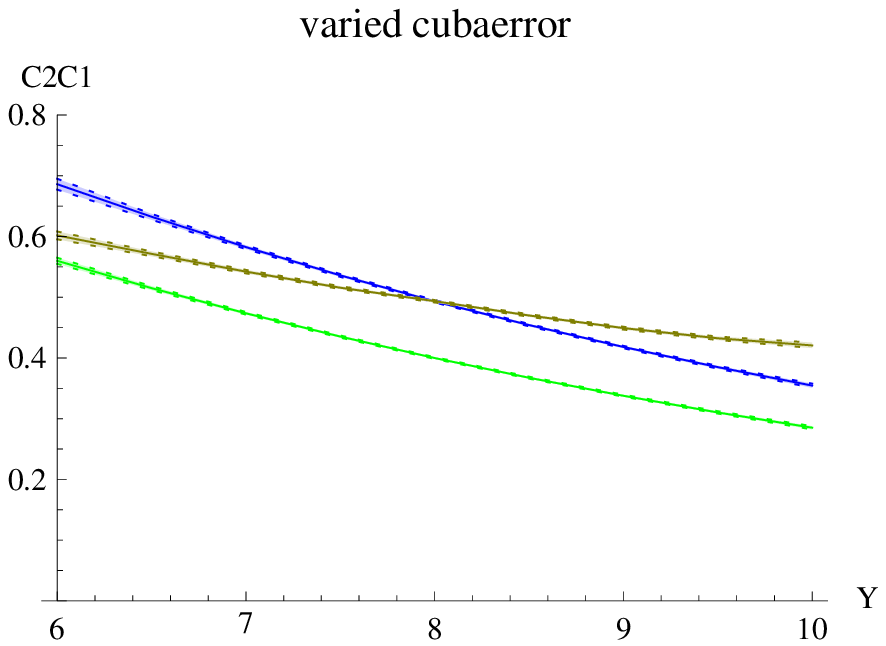}
  \caption{$\langle \cos 2\varphi\rangle / \langle \cos \varphi\rangle$ in dependence on $Y$ for $|\veckjone|=|\veckjtwo|=35\,{\rm GeV}$.  The errors due to the Monte Carlo integration -- though hardly visible -- are given as error bands. The tabled values are shown in Tab.~\ref{tab:c2c13535}.}
  \label{fig:c2c13535}
\end{figure}

In Ref.~\cite{Vera:2007kn} it has been proposed to also study other
observables
 $\mathcal{C}_m/\mathcal{C}_n$ with $m\ne 0\ne n$.
 This is motivated by the fact that the main source of uncertainty of the Green's function is associated with the $n=0$ component. This observation is not altered by the inclusion of NLL vertices, as we display in Figs.~\ref{fig:c2c13535}-\ref{fig:c2c13535_s0}. Moreover, Fig.~\ref{fig:c2c13535_mu} shows that the effect of changing $\mu_R=\mu_F$ leads to modifications of similar size for pure NLL and combined LL vertices with NLL Green's function predictions, while the changing of $s_0$ (see Fig.~\ref{fig:c2c13535_s0})
 leads to a reduced dependency with respect to $s_0$ of the pure NLL prediction. The PDF dependence for all ratios $\mathcal{C}_m/\mathcal{C}_n$ cancels in the same manner such that there is no use plotting it more than once (see Fig.~\ref{fig:c1c03535_pdf} for $\mathcal{C}_1/\mathcal{C}_0$).

\begin{figure}[h!]
  \centering
  \psfrag{varied}{}
  \psfrag{s0}{}\psfrag{cubaerror}{}\psfrag{pdfset}{}\psfrag{mu}{}
  \psfrag{deltaC1}{$\delta\mathcal{C}_1 \left[\frac{\rm nb}{{\rm GeV}^2}\right] $}
  \psfrag{Y}{$Y$}
  \includegraphics[width=15cm]{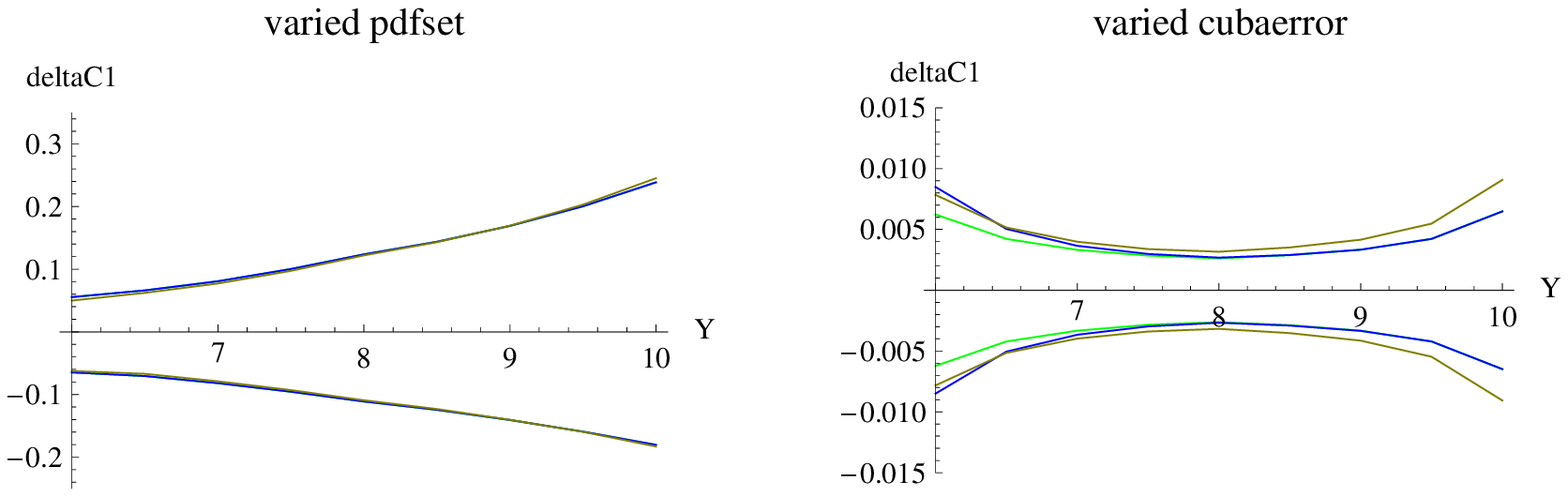}
  \caption{Relative effect of the PDF (left) and Monte Carlo (right) errors on the coefficient $\mathcal{C}_1$ in dependence on $Y$ for $|\veckjone|=|\veckjtwo|=35\,{\rm GeV}$. The tabled values are shown in Tabs.~\ref{tab:c13535_pdf} and \ref{tab:c13535}.}
  \label{fig:c13535rel_pdf_cuba}
\end{figure}

\begin{figure}[h!]
  \centering
  \psfrag{varied}{}
  \psfrag{s0}{}\psfrag{cubaerror}{}\psfrag{pdfset}{}\psfrag{mu}{}
  \psfrag{deltaC2}{$\delta\mathcal{C}_2 \left[\frac{\rm nb}{{\rm GeV}^2}\right] $}
  \psfrag{Y}{$Y$}
  \includegraphics[width=15cm]{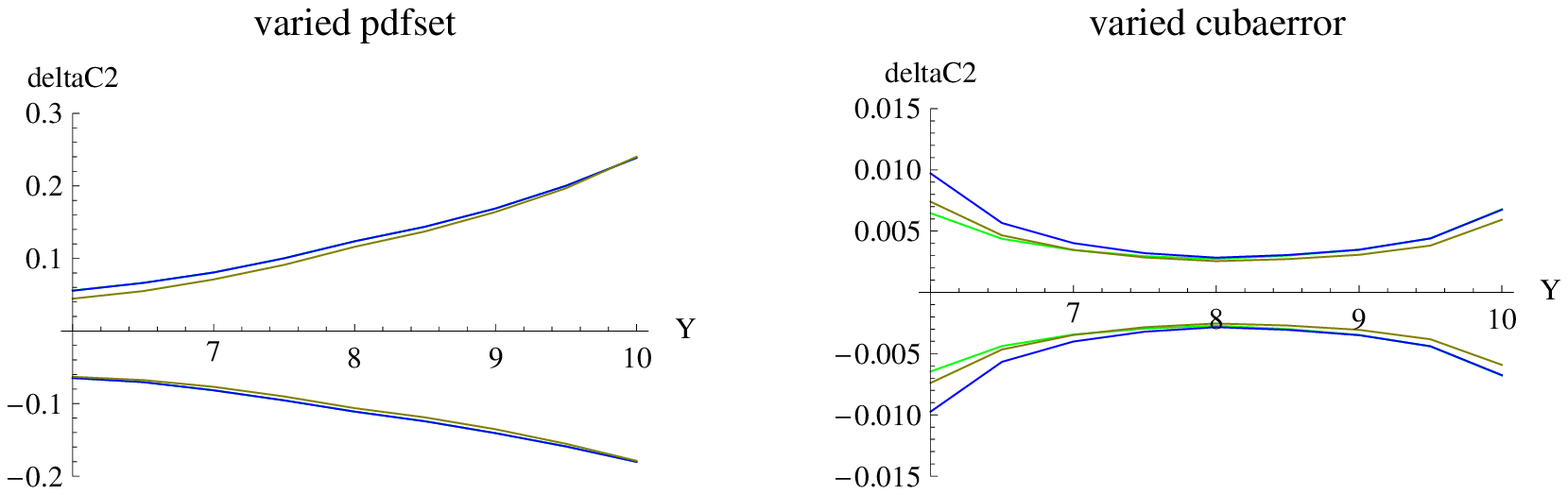}
  \caption{Relative effect of the PDF (left) and Monte Carlo (right) errors on the coefficient $\mathcal{C}_2$ in dependence on $Y$ for $|\veckjone|=|\veckjtwo|=35\,{\rm GeV}$. The tabled values are shown in Tabs.~\ref{tab:c23535_pdf} and \ref{tab:c23535}.}
  \label{fig:c23535rel_pdf_cuba}
\end{figure}

A priori, one would expect the numeric uncertainties for the ${\cal C}_{n>0}$ calculations including the NLL vertices to be larger because the coefficients $C_{n,\nu}$ contain an azimuthal integration which in the case of $n=0$ becomes trivial while for $n>0$ has to be carried out. This is indeed true for the NLL corrections alone, but since these corrections are more significant for $n=0$ than for $n>0$, in the sum together with the error-free LL vertices the opposite turns out to be true (the Monte Carlo errors are smaller than $1\%$ for ${\cal C}_{1,2}$ as shown in Figs.~\ref{fig:c13535rel_pdf_cuba}, and \ref{fig:c23535rel_pdf_cuba} respectively).

We would like to draw the reader's attention to the analogous effects for the kernel. For both the BFKL kernel and the Mueller Navelet vertices the NLL corrections to the $n=0$ component around $\nu=0$ are very large and negative while the relative corrections for $n>0$ are positive, much smaller than for $n=0$ and slowly increasing with $n$.

Finally, for other kinematical configurations to be discussed below, the same kind of
PDF uncertainties appear, and Monte Carlo errors are of similar order. Because of that, we will not display the corresponding curves.

\begin{figure}[h!]
  \centering
  \psfrag{varied}{}
  \psfrag{cubaerror}{}
  \psfrag{C0}{$\mathcal{C}_0 \left[\frac{\rm nb}{{\rm GeV}^2}\right] = \sigma$}
  \psfrag{Y}{$Y$}
  \includegraphics[width=9cm]{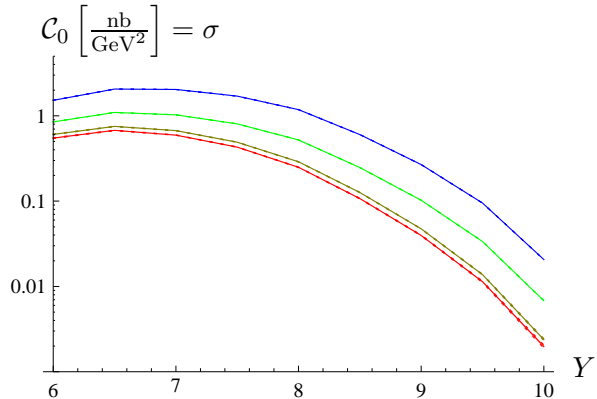}
  \caption{Differential cross section in dependence on $Y$ for $|\veckjone|=|\veckjtwo|=35\,{\rm GeV}$. The errors due to the Monte Carlo integration -- though hardly visible -- are given as error bands. The tabled values are shown in Tab.~\ref{tab:c03535}. 
Blue shows the pure LL result, brown the pure NLL result, green the combination of LL vertices with the collinear improved NLL Green's function, red the full NLL vertices with the collinear improved NLL Green's function. The same color coding is used in all subsequent plots.
}
  \label{fig:c03535}
\end{figure}

 \clearpage

\clearpage

\clearpage

\subsection{$|\veckjone|=|\veckjtwo|=50\,{\rm GeV}$}

\begin{figure}[h!]
  \centering
  \psfrag{varied}{}
  \psfrag{s0}{}\psfrag{cubaerror}{}\psfrag{pdfset}{}\psfrag{mu}{}
  \psfrag{deltaC0}{$\delta\mathcal{C}_0 \left[\frac{\rm nb}{{\rm GeV}^2}\right] $}
  \psfrag{Y}{$Y$}
  \includegraphics[width=15cm]{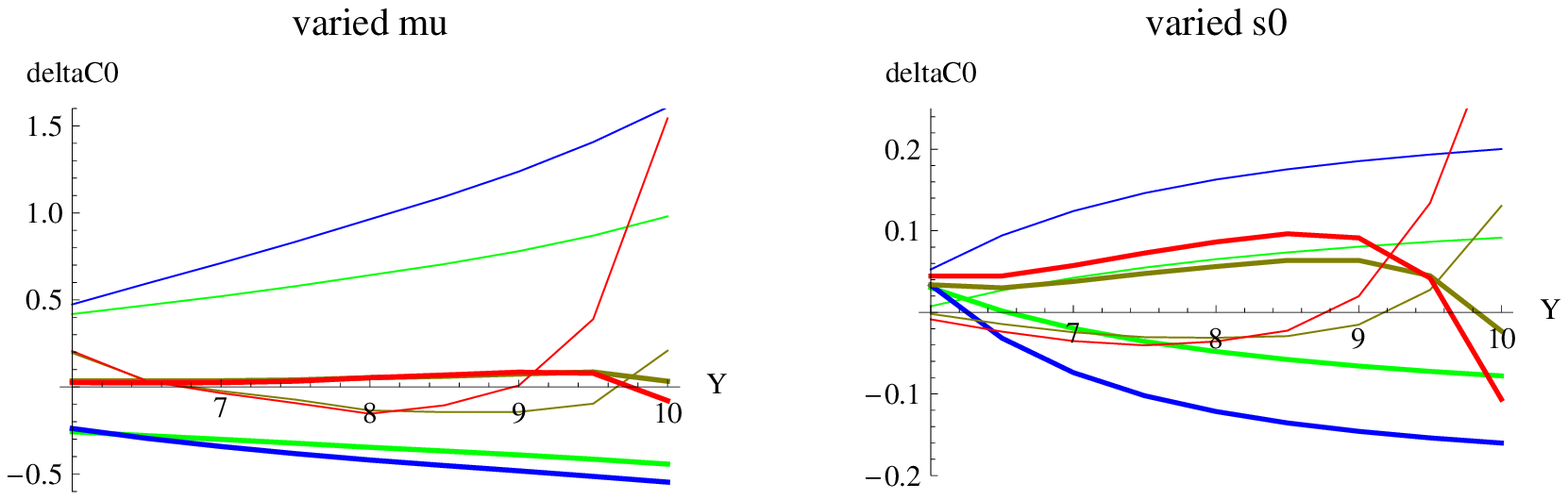}
  \caption{Relative effect of changing $\mu_R=\mu_F$ by factors 2 and $1/2$ respectively (left), and $\sqrt{s_0}$ (right) by factors 2 and $1/2$ respectively on the differential cross section in dependence on $Y$ for $|\veckjone|=|\veckjtwo|=50\,{\rm GeV}$. The tabled values are shown in Tabs.~\ref{tab:c05050_mu} and \ref{tab:c05050_s0}.}
  \label{fig:c05050rel_mu_s0}
\end{figure}

Going to larger jet scales, we meet more or less the same advantages and problems as for {35\,GeV}.
Again we start with
the differential cross section \eqref{def:dsigma}. The result  is shown in Fig.~\ref{fig:c05050} (the according tabled values are shown in Tab.~\ref{tab:c05050} in the Appendix). The dependences with respect to $\mu_R$, and $s_0$ are displayed in Fig.~\ref{fig:c05050rel_mu_s0}.

\begin{figure}[h!]
  \centering
  \psfrag{varied}{}
  \psfrag{cubaerror}{}
  \psfrag{C1C0}{$\frac{\mathcal{C}_1}{\mathcal{C}_0}$}
  \psfrag{Y}{$Y$}
  \includegraphics[width=9cm]{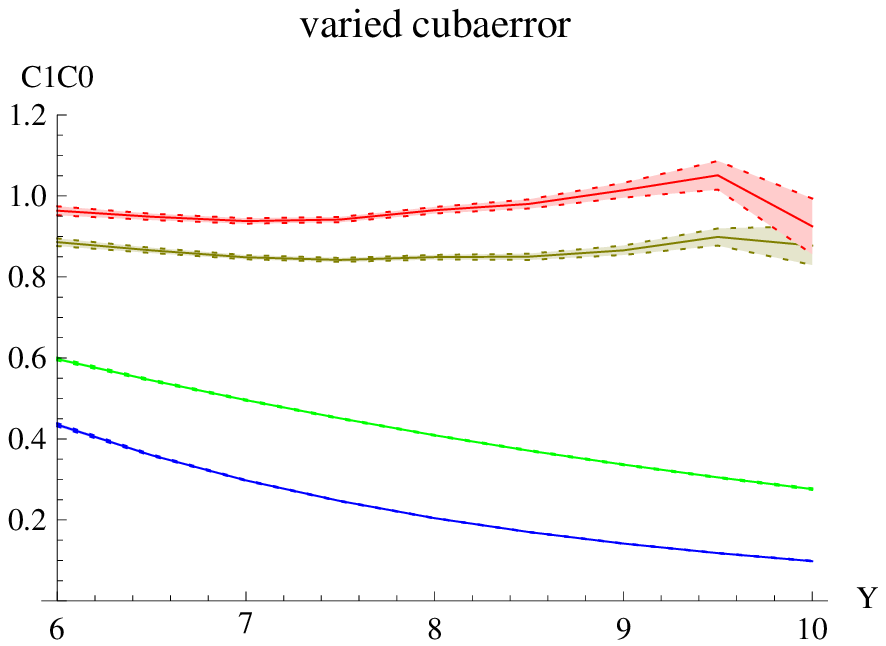}
  \caption{$\langle \cos \varphi\rangle$ in dependence on $Y$ for $|\veckjone|=|\veckjtwo|=50\,{\rm GeV}$. The errors due to the Monte Carlo integration are given as error bands. The tabled values are shown in Tab.~\ref{tab:c1c05050}.}
  \label{fig:c1c05050}
\end{figure}

\begin{figure}[h!]
  \centering
  \psfrag{varied}{}
  \psfrag{cubaerror}{}
  \psfrag{C2C0}{$\frac{\mathcal{C}_2}{\mathcal{C}_0}$}
  \psfrag{Y}{$Y$}
  \includegraphics[width=9cm]{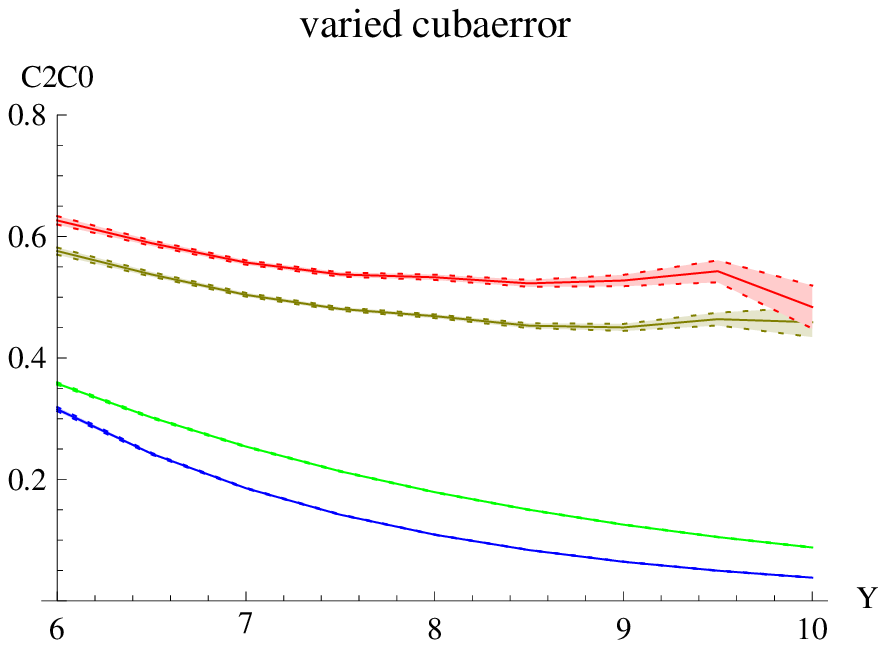}
  \caption{$\langle \cos 2\varphi\rangle$ in dependence on $Y$ for $|\veckjone|=|\veckjtwo|=50\,{\rm GeV}$. The errors due to the Monte Carlo integration are given as error bands. The tabled values are shown in Tab.~\ref{tab:c2c05050}.}
  \label{fig:c2c05050}
\end{figure}

\begin{figure}[h!]
  \centering
  \psfrag{varied}{}
  \psfrag{cubaerror}{}
  \psfrag{C1}{$\mathcal{C}_1 \left[\frac{\rm nb}{{\rm GeV}^2}\right] $}
  \psfrag{Y}{$Y$}
  \includegraphics[width=9cm]{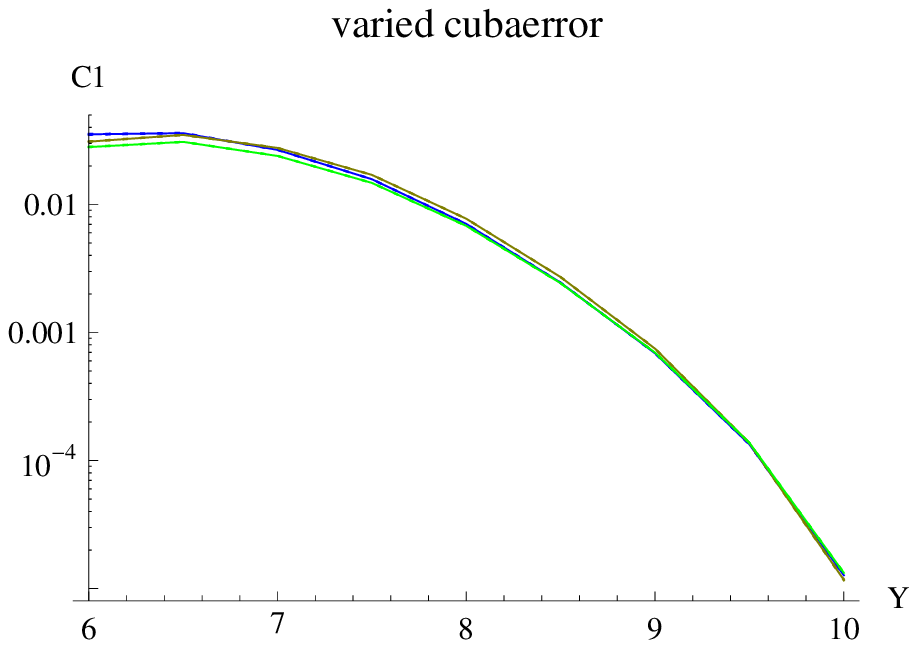}
  \caption{Coefficient $\mathcal{C}_1$ in dependence on $Y$ for $|\veckjone|=|\veckjtwo|=50\,{\rm GeV}$. The errors due to the Monte Carlo integration -- though hardly visible -- are given as error bands. The tabled values are shown in Tab.~\ref{tab:c15050}.}
  \label{fig:c15050}
\end{figure}

\begin{figure}[h!]
  \centering
  \psfrag{varied}{}
  \psfrag{cubaerror}{}
  \psfrag{C2}{$\mathcal{C}_2 \left[\frac{\rm nb}{{\rm GeV}^2}\right] $}
  \psfrag{Y}{$Y$}
  \includegraphics[width=9cm]{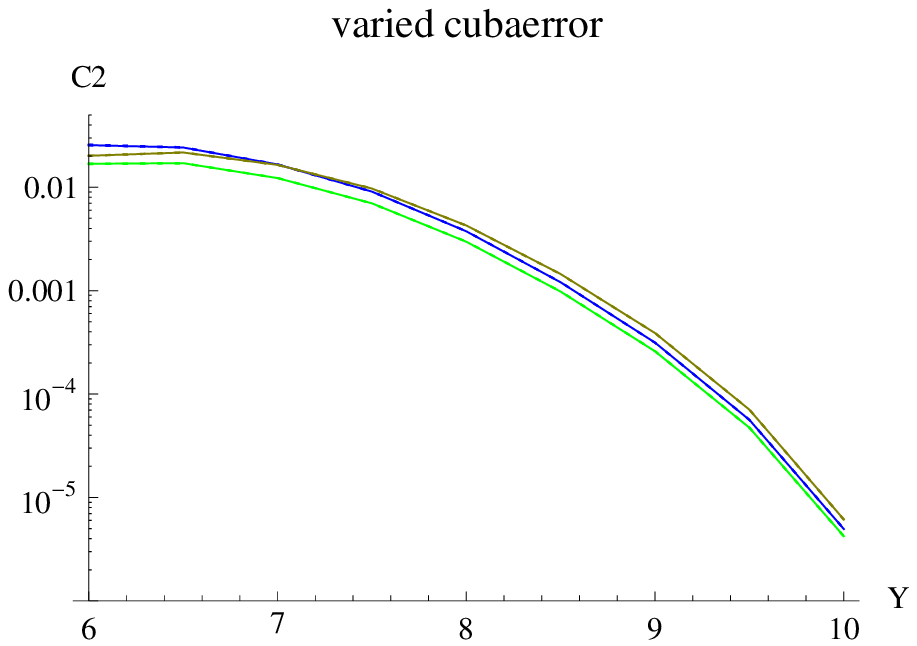}
  \caption{Coefficient $\mathcal{C}_2$ in dependence on $Y$ for $|\veckjone|=|\veckjtwo|=50\,{\rm GeV}$. The errors due to the Monte Carlo integration -- though hardly visible -- are given as error bands. The tabled values are shown in Tab.~\ref{tab:c25050}.}
  \label{fig:c25050}
\end{figure}

The azimuthal decorrelation is displayed in Fig.~\ref{fig:c1c05050} for $\langle \cos \varphi \rangle$ and in Fig.~\ref{fig:c2c05050} for $\langle \cos 2\varphi \rangle\,,$
again explicitly showing that inclusion of NLL vertices leads to an enormous correlation in the azimuthal angle
(for completeness our results for ${\cal C}_1$ and ${\cal C}_2$ coefficients alone are displayed respectively in Fig.~\ref{fig:c15050} and Fig.~\ref{fig:c25050}). Here, the angular correlation even has the tendency to increase with growing rapidity $Y$. This might be interpreted as the effect of stronger limited phase space for additional emissions at large energies and large transverse momenta of the produced jets (Note, that the cross section is a factor $\sim 10$ smaller at $Y=6$ compared to the previous configuration, and a factor $\sim 100$ smaller at $Y=10$).

\begin{figure}[h!]
  \centering
  \psfrag{varied}{}
  \psfrag{mu}{}
  \psfrag{C1C0}{$\frac{\mathcal{C}_1}{\mathcal{C}_0}$}
  \psfrag{Y}{$Y$}
  \includegraphics[width=15cm]{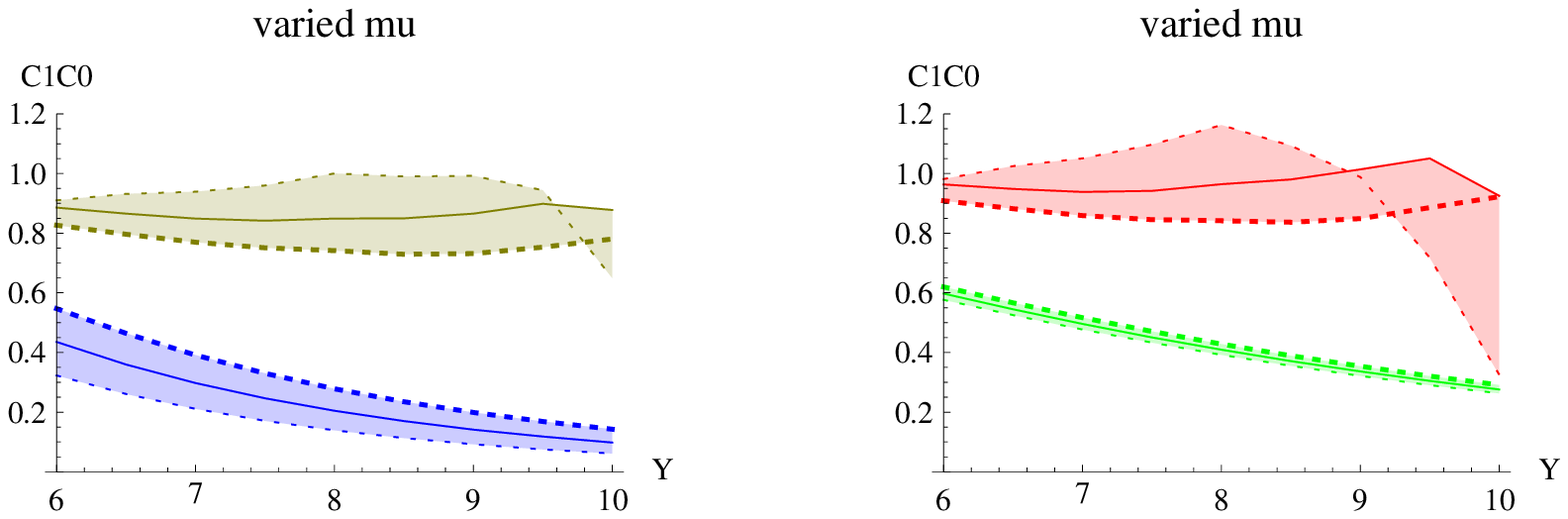}
  \caption{Effect of changing $\mu_R=\mu_F$ by factors 2 and $1/2$ respectively on $\langle \cos \varphi\rangle$ in dependence on $Y$ for $|\veckjone|=|\veckjtwo|=50\,{\rm GeV}$. The tabled values are shown in Tab.~\ref{tab:c1c05050_mu}.}
  \label{fig:c1c05050_mu}
\end{figure}

\begin{figure}[h!]
  \centering
  \psfrag{varied}{}
  \psfrag{s0}{}
  \psfrag{C1C0}{$\frac{\mathcal{C}_1}{\mathcal{C}_0}$}
  \psfrag{Y}{$Y$}
  \includegraphics[width=15cm]{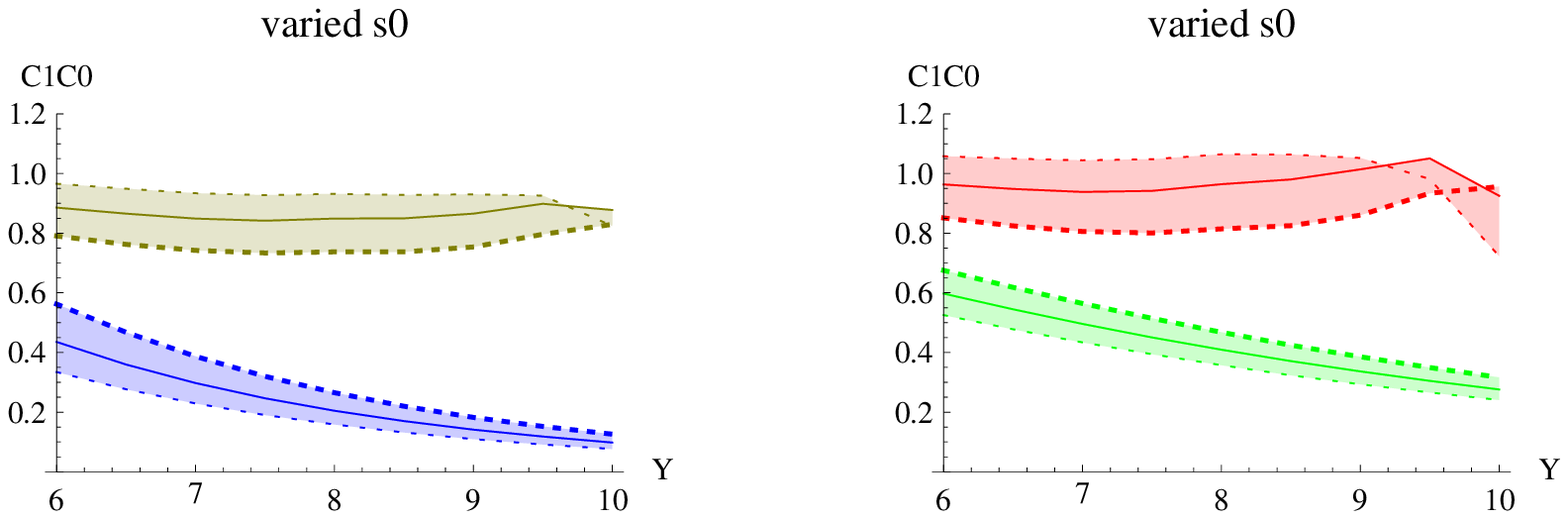}
  \caption{Effect of changing $\sqrt{s_0}$ by factors 2 and $1/2$ respectively on $\langle \cos \varphi\rangle$ in dependence on $Y$ for $|\veckjone|=|\veckjtwo|=50\,{\rm GeV}$. The tabled values are shown in Tab.~\ref{tab:c1c05050_s0}.}
  \label{fig:c1c05050_s0}
\end{figure}

\begin{figure}[h!]
  \centering
  \psfrag{varied}{}
  \psfrag{mu}{}
  \psfrag{C2C0}{$\frac{\mathcal{C}_2}{\mathcal{C}_0}$}
  \psfrag{Y}{$Y$}
  \includegraphics[width=15cm]{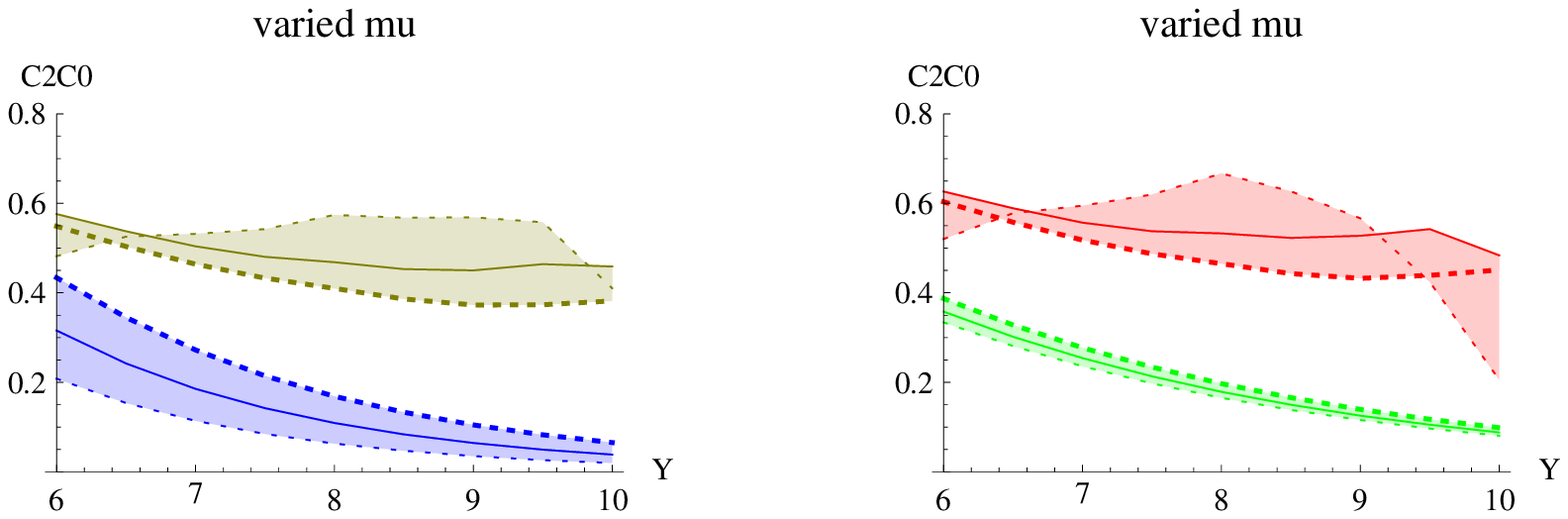}
  \caption{Effect of changing $\mu_R=\mu_F$ by factors 2 and $1/2$ respectively on $\langle \cos 2\varphi\rangle$ in dependence on $Y$ for $|\veckjone|=|\veckjtwo|=50\,{\rm GeV}$. The tabled values are shown in Tab.~\ref{tab:c2c05050_mu}.}
  \label{fig:c2c05050_mu}
\end{figure}

\begin{figure}[h!]
  \centering
  \psfrag{varied}{}
  \psfrag{s0}{}
  \psfrag{C2C0}{$\frac{\mathcal{C}_2}{\mathcal{C}_0}$}
  \psfrag{Y}{$Y$}
  \includegraphics[width=15cm]{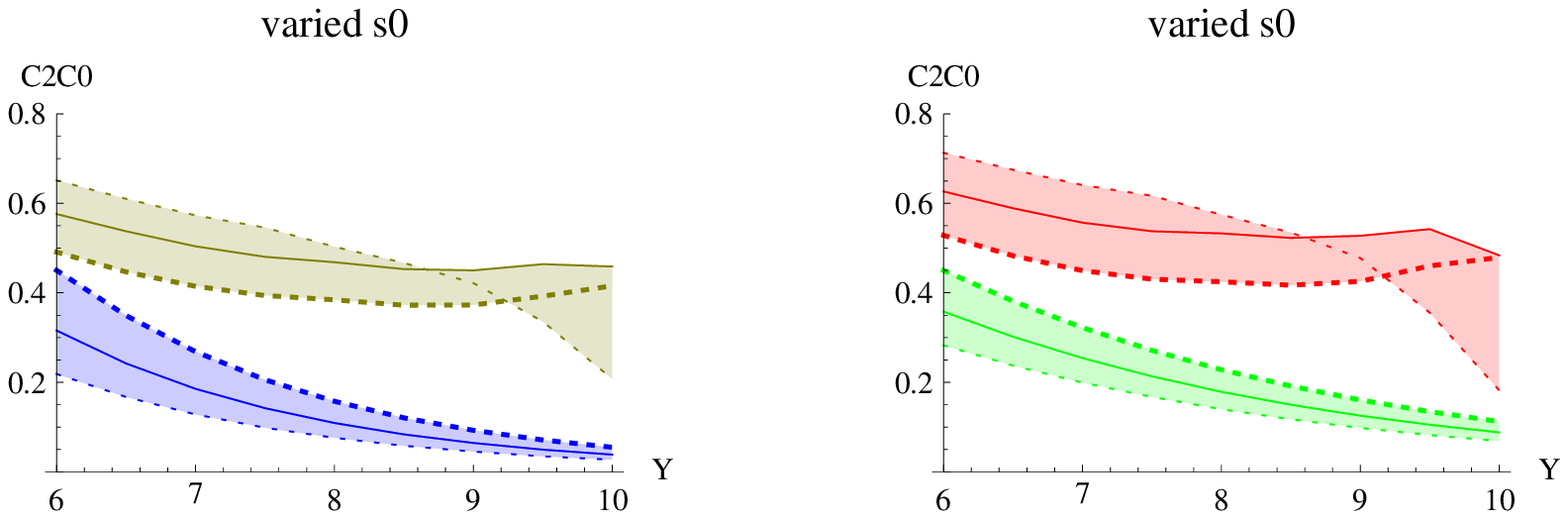}
  \caption{Effect of changing $\sqrt{s_0}$ by factors 2 and $1/2$ respectively on $\langle \cos 2\varphi\rangle$ in dependence on $Y$ for $|\veckjone|=|\veckjtwo|=50\,{\rm GeV}$. The tabled values are shown in Tab.~\ref{tab:c2c05050_s0}.}
  \label{fig:c2c05050_s0}
\end{figure}

 The various sources of uncertainty of our results are shown for $\langle \cos \varphi \rangle$ in Fig.~\ref{fig:c1c05050_mu} (variation of $\mu_R=\mu_F$), Fig.~\ref{fig:c1c05050_s0} (variation of $s_0$), and for $\langle \cos 2 \varphi \rangle$ in Fig.~\ref{fig:c2c05050_mu} (variation of $\mu_R=\mu_F$), Fig.~\ref{fig:c2c05050_s0} (variation of $s_0$).

\begin{figure}[h!]
  \centering
  \psfrag{varied}{}
  \psfrag{s0}{}\psfrag{cubaerror}{}\psfrag{pdfset}{}\psfrag{mu}{}
  \psfrag{deltaC1}{$\delta\mathcal{C}_1 \left[\frac{\rm nb}{{\rm GeV}^2}\right] $}
  \psfrag{Y}{$Y$}
  \includegraphics[width=15cm]{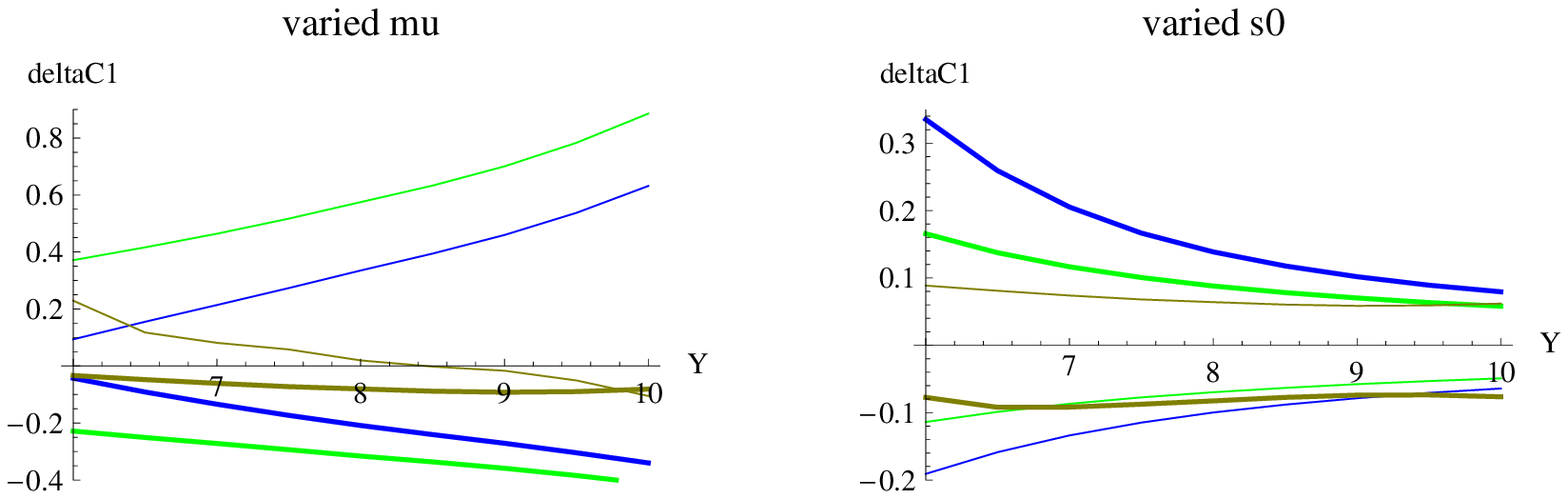}
  \caption{Relative effect of changing $\mu_R=\mu_F$ by factors 2 and $1/2$ respectively (left), and $\sqrt{s_0}$ (right) by factors 2 and $1/2$ respectively on the coefficient $\mathcal{C}_1$ in dependence on $Y$ for $|\veckjone|=|\veckjtwo|=50\,{\rm GeV}$. The tabled values are shown in Tabs.~\ref{tab:c15050_mu} and \ref{tab:c15050_s0}.}
  \label{fig:c15050rel_mu_s0}
\end{figure}

 The scale dependences of $\mathcal{C}_1/\mathcal{C}_0$ (see Figs.~\ref{fig:c1c05050_mu}, and \ref{fig:c1c05050_s0}) as well as of $\mathcal{C}_0$ (see Fig.~\ref{fig:c05050rel_mu_s0}) and $\mathcal{C}_1$ (see Fig.~\ref{fig:c15050rel_mu_s0}) alone reveal the same basic features as before, namely a non-monotone scale dependence of the NLL corrections and (more serious) unphysical results for $\langle \cos \varphi \rangle$ in case of the resummed NLL prediction for small $s_0$ and/ or $\mu_R=\mu_F$ scales.

\begin{figure}[h!]
  \centering
  \psfrag{varied}{}
  \psfrag{s0}{}\psfrag{cubaerror}{}\psfrag{pdfset}{}\psfrag{mu}{}
  \psfrag{deltaC2}{$\delta\mathcal{C}_2 \left[\frac{\rm nb}{{\rm GeV}^2}\right] $}
  \psfrag{Y}{$Y$}
  \includegraphics[width=15cm]{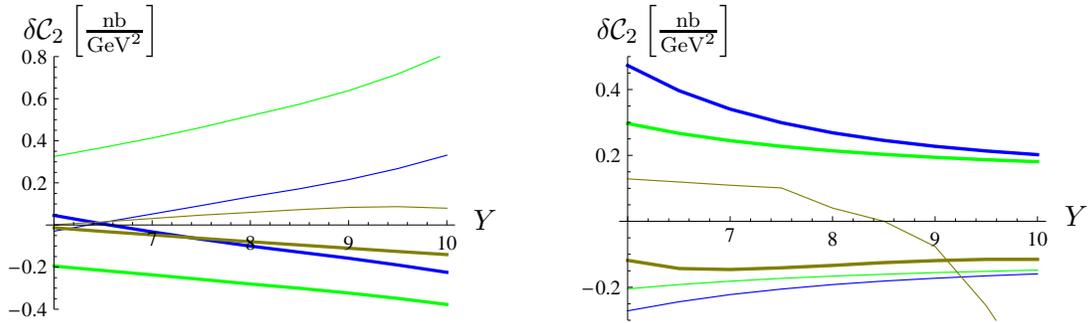}
  \caption{Relative effect of changing $\mu_R=\mu_F$ by factors 2 and $1/2$ respectively (left), and $\sqrt{s_0}$ (right) by factors 2 and $1/2$ respectively on the coefficient $\mathcal{C}_2$ in dependence on $Y$ for $|\veckjone|=|\veckjtwo|=50\,{\rm GeV}$. The tabled values are shown in Tabs.~\ref{tab:c25050_mu} and \ref{tab:c25050_s0}.}
  \label{fig:c25050rel_mu_s0}
\end{figure}

 A similar rather large dependency on $\mu_R=\mu_F$ and $s_0$ is obtained for ${\cal C}_2/{\cal C}_0$,
 as can be seen from Figs.~\ref{fig:c2c05050_mu},
\ref{fig:c2c05050_s0}, based on detailed studies of coefficients
${\cal C}_0$ and
 ${\cal C}_2$ displayed respectively in Fig.~\ref{fig:c05050rel_mu_s0}
 and Fig.~\ref{fig:c25050rel_mu_s0}.

The problematic behavior for smaller scales of $s_0$ and/ or $\mu_R$ is more dramatic for $|\veckjone|=|\veckjtwo|=50\,{\rm GeV}$ (see {\it e.g} Figs.~\ref{fig:c1c05050_mu}, \ref{fig:c1c05050_s0}, \ref{fig:c2c05050_mu}, \ref{fig:c2c05050_s0}). Especially the $\mu_R$ dependence (see Fig.~\ref{fig:c1c05050_mu} seems to indicate that already the a priori natural scale $\mu_R=|\veckj|$ is too small.

\clearpage

\begin{figure}[h!]
  \centering
  \psfrag{varied}{}
  \psfrag{cubaerror}{}
  \psfrag{C0}{$\mathcal{C}_0 \left[\frac{\rm nb}{{\rm GeV}^2}\right] = \sigma$}
  \psfrag{Y}{$Y$}
  \includegraphics[width=9cm]{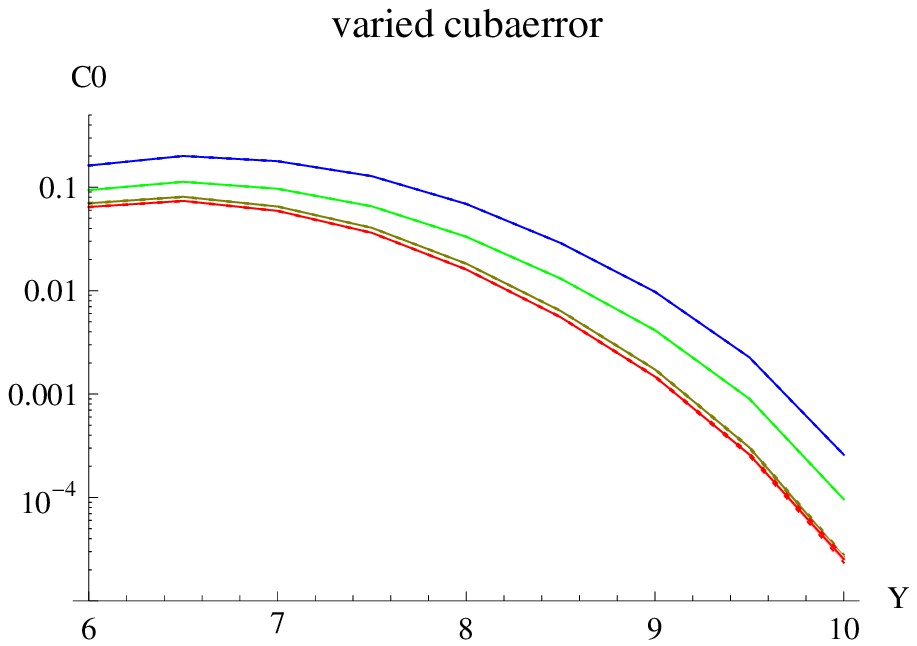}
  \caption{Differential cross section in dependence on $Y$ for $|\veckjone|=|\veckjtwo|=50\,{\rm GeV}$. The errors due to the Monte Carlo integration -- though hardly visible -- are given as error bands. The tabled values are shown in Tab.~\ref{tab:c05050}.}
  \label{fig:c05050}
\end{figure}

\begin{figure}[h!]
  \centering
  \psfrag{varied}{}
  \psfrag{cubaerror}{}
  \psfrag{C2C1}{$\frac{\mathcal{C}_2}{\mathcal{C}_1}$}
  \psfrag{Y}{$Y$}
  \includegraphics[width=9cm]{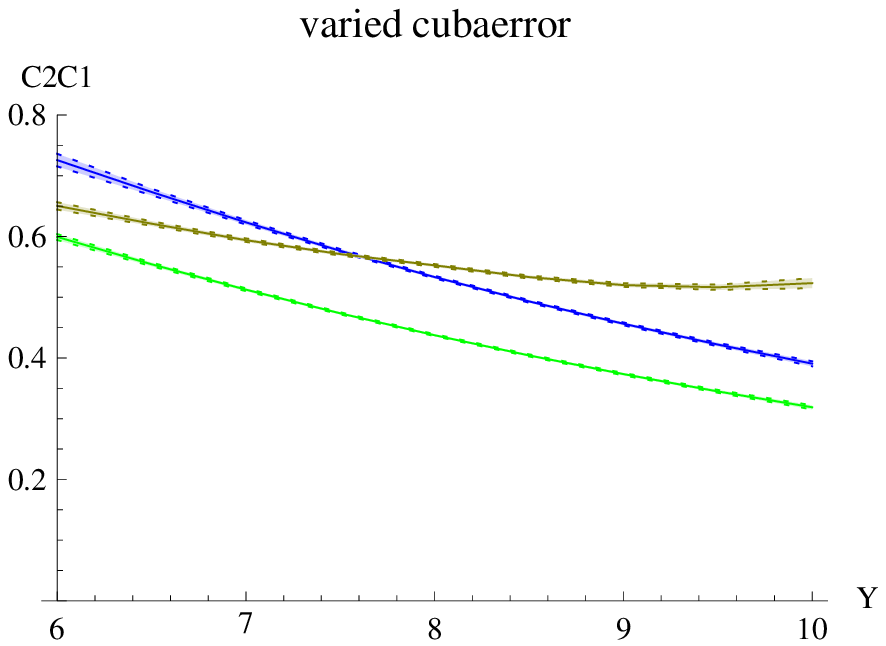}
  \caption{$\langle \cos 2\varphi\rangle / \langle \cos \varphi\rangle$ in dependence on $Y$ for $|\veckjone|=|\veckjtwo|=50\,{\rm GeV}$.  The errors due to the Monte Carlo integration -- though hardly visible -- are given as error bands. The tabled values are shown in Tab.~\ref{tab:c2c15050}.}
  \label{fig:c2c15050}
\end{figure}

\begin{figure}[h!]
  \centering
  \psfrag{varied}{}
  \psfrag{mu}{}
  \psfrag{C2C1}{$\frac{\mathcal{C}_2}{\mathcal{C}_1}$}
  \psfrag{Y}{$Y$}
  \includegraphics[width=15cm]{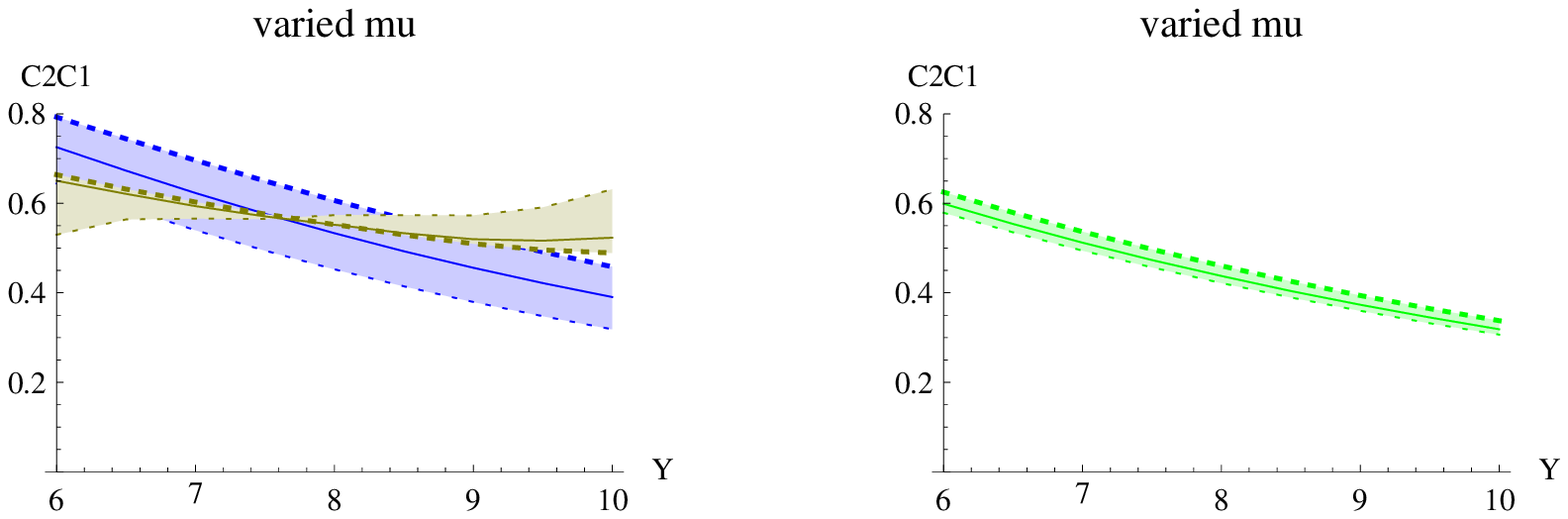}
  \caption{Effect of changing $\mu_R=\mu_F$ by factors 2 and $1/2$ respectively on $\langle \cos 2\varphi\rangle / \langle \cos \varphi\rangle$ in dependence on $Y$ for $|\veckjone|=|\veckjtwo|=50\,{\rm GeV}$. The tabled values are shown in Tab.~\ref{tab:c2c15050_mu}.}
  \label{fig:c2c15050_mu}
\end{figure}

\begin{figure}[h!]
  \centering
  \psfrag{varied}{}
  \psfrag{s0}{}
  \psfrag{C2C1}{$\frac{\mathcal{C}_2}{\mathcal{C}_1}$}
  \psfrag{Y}{$Y$}
  \includegraphics[width=15cm]{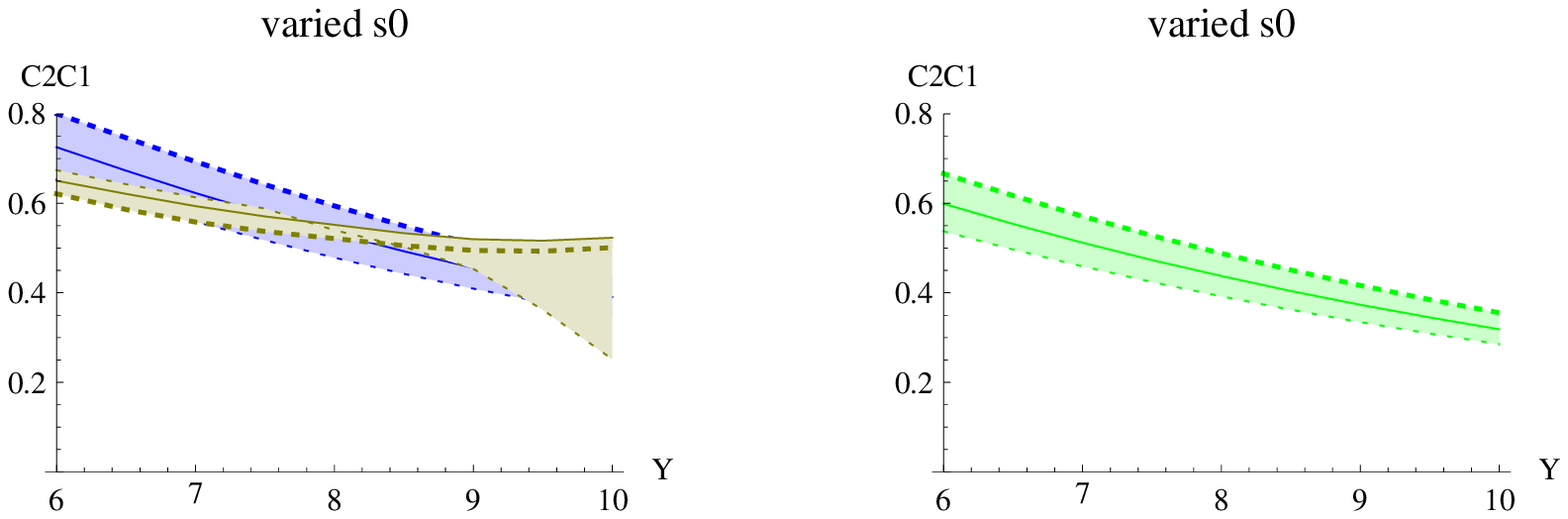}
  \caption{Effect of changing $\sqrt{s_0}$ by factors 2 and $1/2$ respectively on $\langle \cos 2\varphi\rangle / \langle \cos \varphi\rangle$ in dependence on $Y$ for $|\veckjone|=|\veckjtwo|=50\,{\rm GeV}$.  The tabled values are shown in Tab.~\ref{tab:c2c15050_s0}.}
  \label{fig:c2c15050_s0}
\end{figure}

\clearpage

\clearpage 

\clearpage

\subsection{$|\veckjone|=35\,{\rm GeV}$, $|\veckjtwo|=50\,{\rm GeV}$}
\label{sec:asymmetriccase}

\begin{figure}[h!]
  \centering
  \psfrag{varied}{}
  \psfrag{cubaerror}{}
  \psfrag{C0}{$\mathcal{C}_0 \left[\frac{\rm nb}{{\rm GeV}^2}\right] = \sigma$}
  \psfrag{Y}{$Y$}
  \includegraphics[width=9cm]{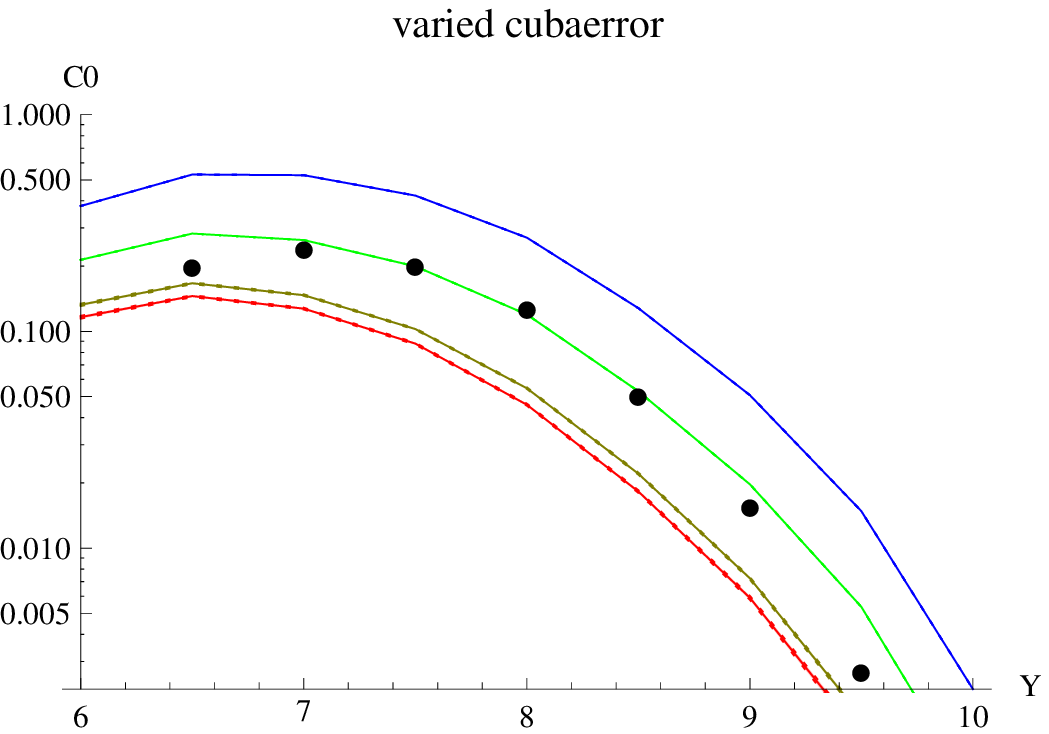}
  \caption{Differential cross section in dependence on $Y$ for $|\veckjone|=35\,{\rm GeV}$, $|\veckjtwo|=50\,{\rm GeV}$. The errors due to the Monte Carlo integration -- though hardly visible -- are given as error bands. The tabled values are shown in Tab.~\ref{tab:c03550}.
  As dots are shown the results of Ref.~\cite{Fontannaz} obtained with \textsc{Dijet} \cite{Aurenche:2008dn}.}
  \label{fig:c03550}
\end{figure}

\begin{figure}[h!]
   \centering
   \psfrag{varied}{}
   \psfrag{cubaerror}{}
   \psfrag{C1C0}{$\frac{\mathcal{C}_1}{\mathcal{C}_0}=\langle \cos \varphi\rangle$}
   \psfrag{Y}{$Y$}
   \includegraphics[width=9cm]{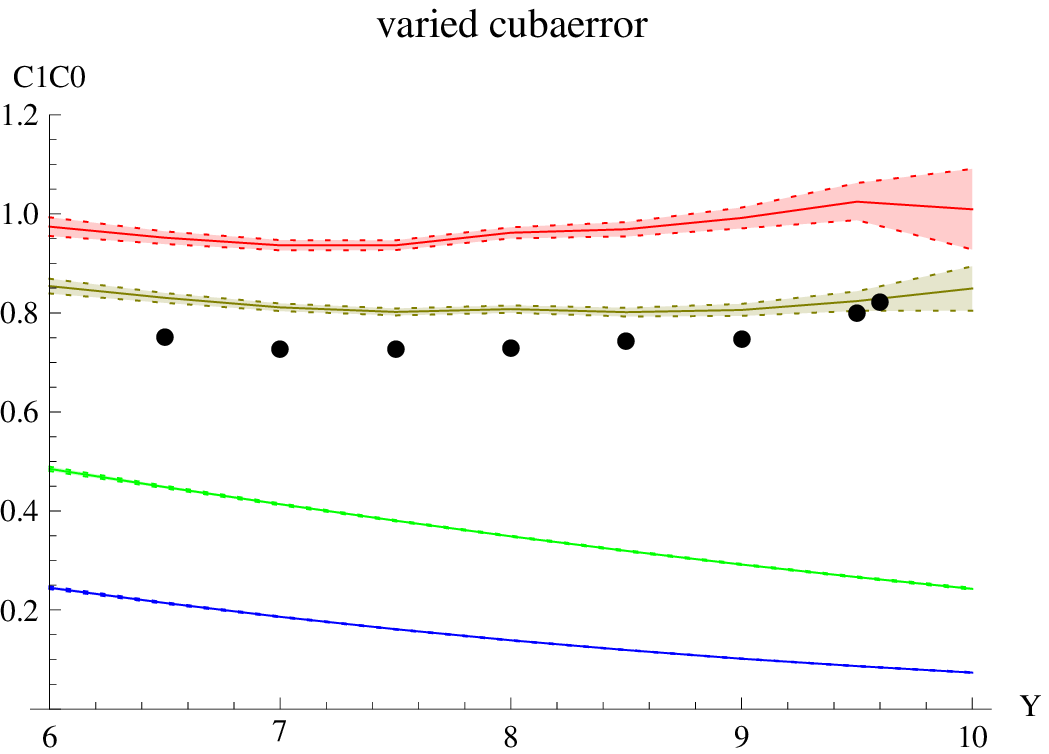}
   \caption{$\langle \cos \varphi\rangle$ in dependence on $Y$ for $|\veckjone|=35\,{\rm GeV}$, $|\veckjtwo|=50\,{\rm GeV}$. The errors due to the Monte Carlo integration are given as error bands. The tabled values are shown in Tab.~\ref{tab:c1c03550}. As dots are shown the results of Ref.~\cite{Fontannaz} obtained with \textsc{Dijet} \cite{Aurenche:2008dn}.}
   \label{fig:c1c03550}
\end{figure}

\begin{figure}[h!]
   \centering
   \psfrag{varied}{}
   \psfrag{cubaerror}{}
   \psfrag{C2C0}{$\frac{\mathcal{C}_2}{\mathcal{C}_0}=\langle \cos 2\varphi\rangle$}
   \psfrag{Y}{$Y$}
  \includegraphics[width=9cm]{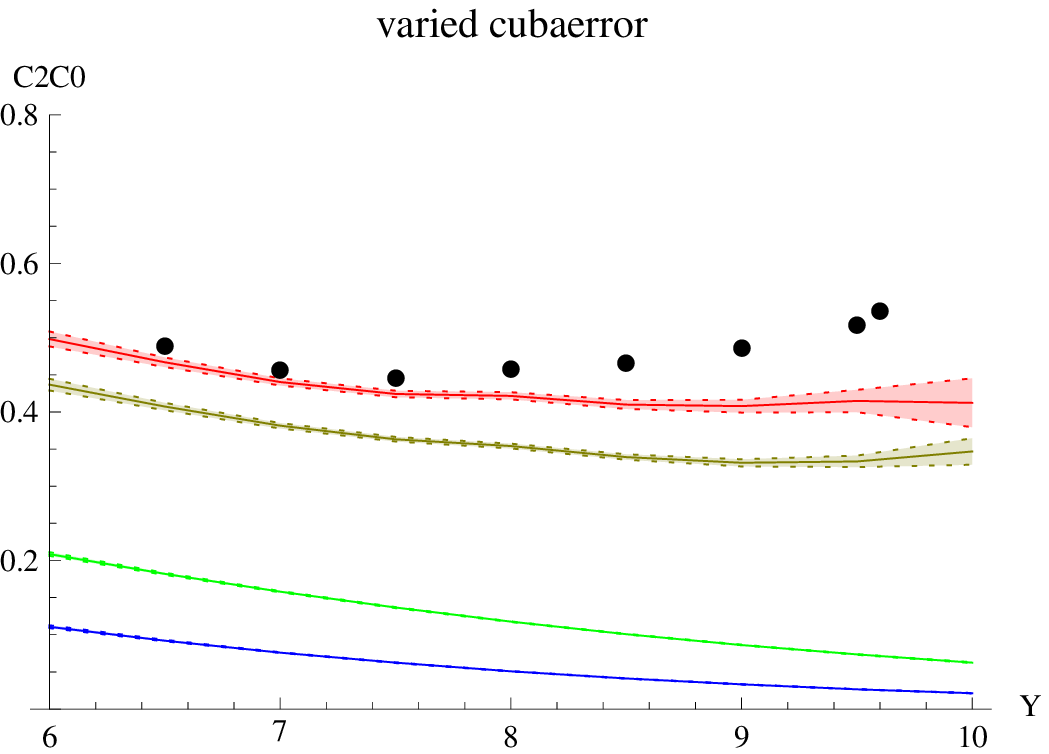}
   \caption{$\langle \cos 2\varphi\rangle$ in dependence on $Y$ for $|\veckjone|=35\,{\rm GeV}$, $|\veckjtwo|=50\,{\rm GeV}$. The errors due to the Monte Carlo integration are given as error bands. The tabled values are shown in Tab.~\ref{tab:c2c03550}. As dots are shown the results of Ref.~\cite{Fontannaz} obtained with \textsc{Dijet} \cite{Aurenche:2008dn}.}
   \label{fig:c2c03550}
 \end{figure}

\begin{figure}[h!]
   \centering
   \psfrag{varied}{}
   \psfrag{cubaerror}{}
   \psfrag{C2C1}{$\frac{\mathcal{C}_2}{\mathcal{C}_1}$}
   \psfrag{Y}{$Y$}
  \includegraphics[width=9cm]{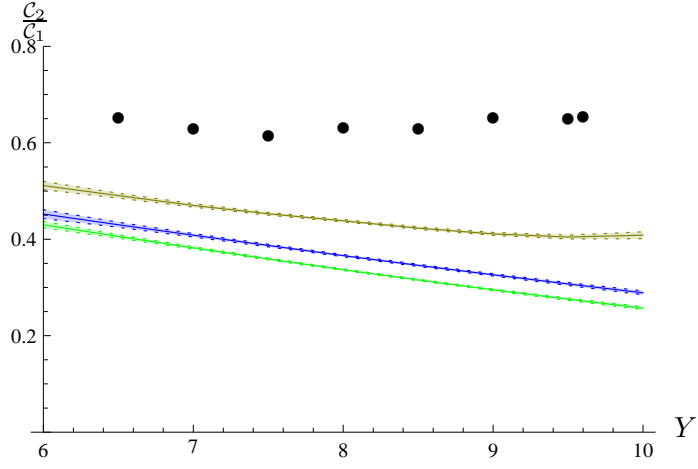}
   \caption{$\langle \cos 2\varphi\rangle / \langle \cos \varphi\rangle$ in dependence on $Y$ for $|\veckjone|=35\,{\rm GeV}$, $|\veckjtwo|=50\,{\rm GeV}$.  The errors due to the Monte Carlo integration -- though hardly visible -- are given as error bands. The tabled values are shown in Tab.~\ref{tab:c2c13550}. As dots are shown the results of Ref.~\cite{Fontannaz} obtained with \textsc{Dijet} \cite{Aurenche:2008dn}.}
   \label{fig:c2c13550}
\end{figure}

We end up with the consideration of the asymmetric case, 
 which we investigate in order to provide a comparison with NLO-DGLAP predictions \cite{Fontannaz} obtained through the NLO-DGLAP partonic generator \textsc{Dijet} \cite{Aurenche:2008dn}. These prediction are very sensitive to the precise compensation between the real and the virtual contribution, and a symmetric cut leads to some kind of
Sudakov resummation effects which are not completely under control at the moment \cite{Fontannaz:2001nq}, even leading to a negative cross-section for $|\veckjone|=|\veckjtwo|=35\,{\rm GeV}$. These prediction are much more stable in the asymmetric configuration. Our own predictions  for the cross-section, for $\langle \cos \varphi \rangle \,,$
$\langle \cos 2 \varphi \rangle \,,$ and $\langle \cos 2 \varphi \rangle / \langle \cos \varphi \rangle$ are given respectively in
 Figs.~\ref{fig:c03550}, \ref{fig:c1c03550}, \ref{fig:c2c03550} and
\ref{fig:c2c13550}.

\begin{figure}[h!]
  \centering
  \psfrag{varied}{}
  \psfrag{s0}{}\psfrag{cubaerror}{}\psfrag{pdfset}{}\psfrag{mu}{}
  \psfrag{deltaC0}{$\delta\mathcal{C}_0 \left[\frac{\rm nb}{{\rm GeV}^2}\right] $}
  \psfrag{Y}{$Y$}
  \includegraphics[width=15cm]{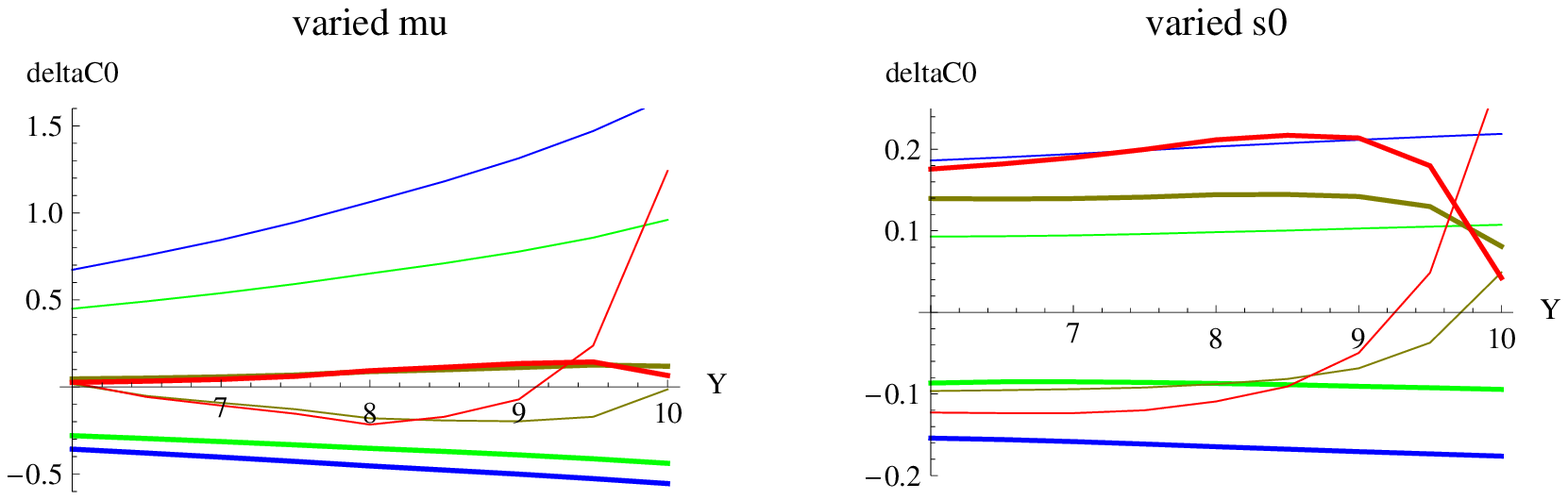}
  \caption{Relative effect of changing $\mu_R=\mu_F$ by factors 2 and $1/2$ respectively (left), and $\sqrt{s_0}$ (right) by factors 2 and $1/2$ respectively on the differential cross section in dependence on $Y$ for $|\veckjone|=35\,{\rm GeV},\;\;|\veckjtwo|=50\,{\rm GeV}$. The tabled values are shown in Tabs.~\ref{tab:c03550_mu} and \ref{tab:c03550_s0}.}
  \label{fig:c03550rel_mu_s0}
\end{figure}

\begin{figure}[h!]
   \centering
   \psfrag{varied}{}
   \psfrag{mu}{}
   \psfrag{C1C0}{$\frac{\mathcal{C}_1}{\mathcal{C}_0}=\langle \cos \varphi\rangle$}
   \psfrag{Y}{$Y$}
   \includegraphics[width=15cm]{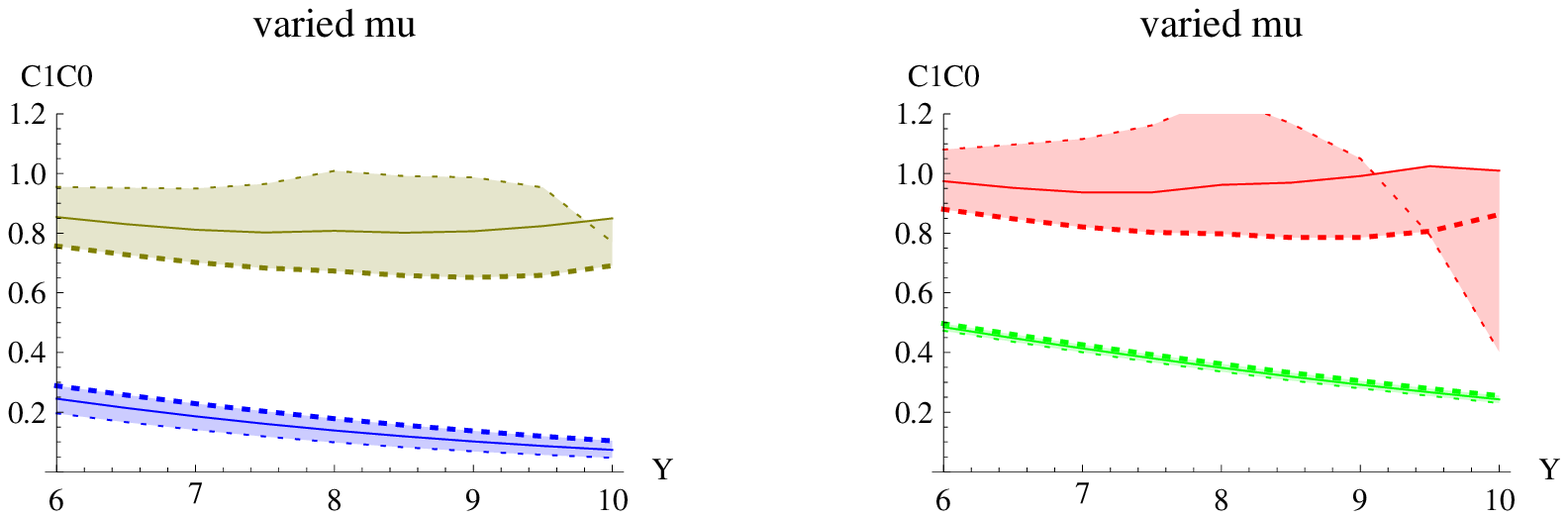}
   \caption{Effect of changing $\mu_R=\mu_F$ by factors 2 and $1/2$ respectively on $\langle \cos \varphi\rangle$ in dependence on $Y$ for $|\veckjone|=35\,{\rm GeV}$, $|\veckjtwo|=50\,{\rm GeV}$. The tabled values are shown in Tab.~\ref{tab:c1c03550_mu}}
   \label{fig:c1c03550_mu}
\end{figure}

\begin{figure}[h!]
  \centering
  \psfrag{varied}{}
  \psfrag{s0}{}
  \psfrag{C1C0}{$\frac{\mathcal{C}_1}{\mathcal{C}_0}=\langle \cos \varphi\rangle$}
  \psfrag{Y}{$Y$}
  \includegraphics[width=15cm]{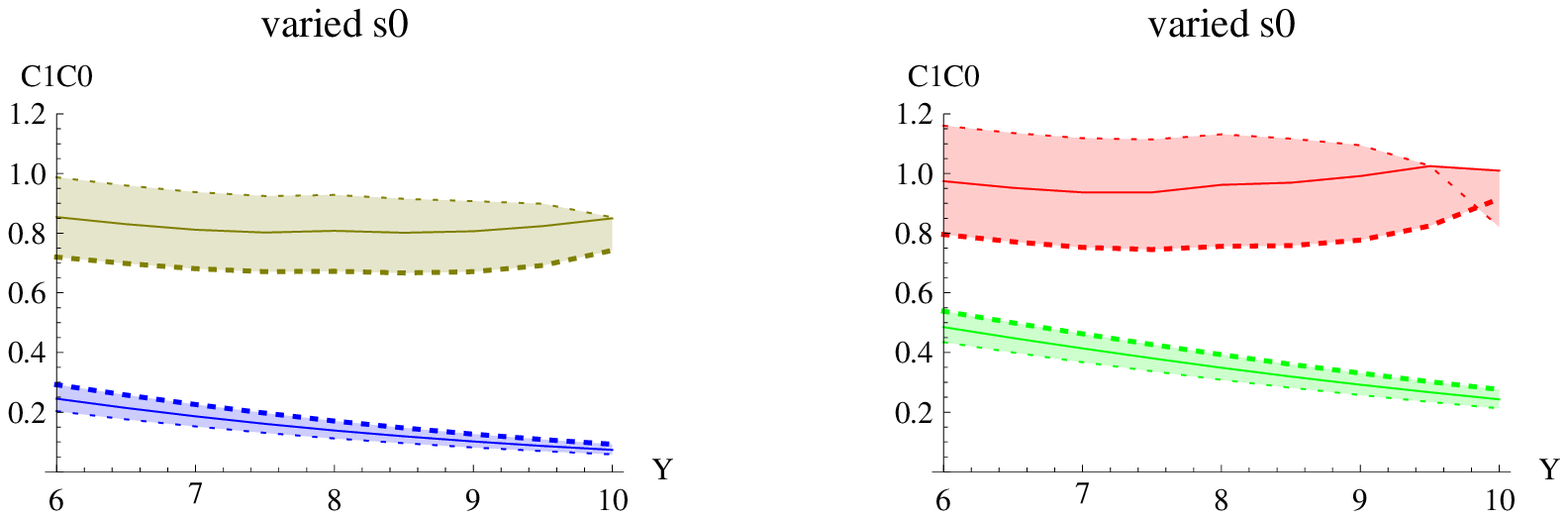}
 \caption{Effect of changing $\sqrt{s_0}$ by factors 2 and $1/2$ respectively on $\langle \cos \varphi\rangle$ in dependence on $Y$ for $|\veckjone|=35\,{\rm GeV}$, $|\veckjtwo|=50\,{\rm GeV}$. The tabled values are shown in Tab.~\ref{tab:c1c03550_s0}.}
  \label{fig:c1c03550_s0}
\end{figure}

\begin{figure}[h!]
   \centering
   \psfrag{varied}{}
   \psfrag{mu}{}
   \psfrag{C2C0}{$\frac{\mathcal{C}_2}{\mathcal{C}_0}=\langle \cos 2\varphi\rangle$}
   \psfrag{Y}{$Y$}
   \includegraphics[width=15cm]{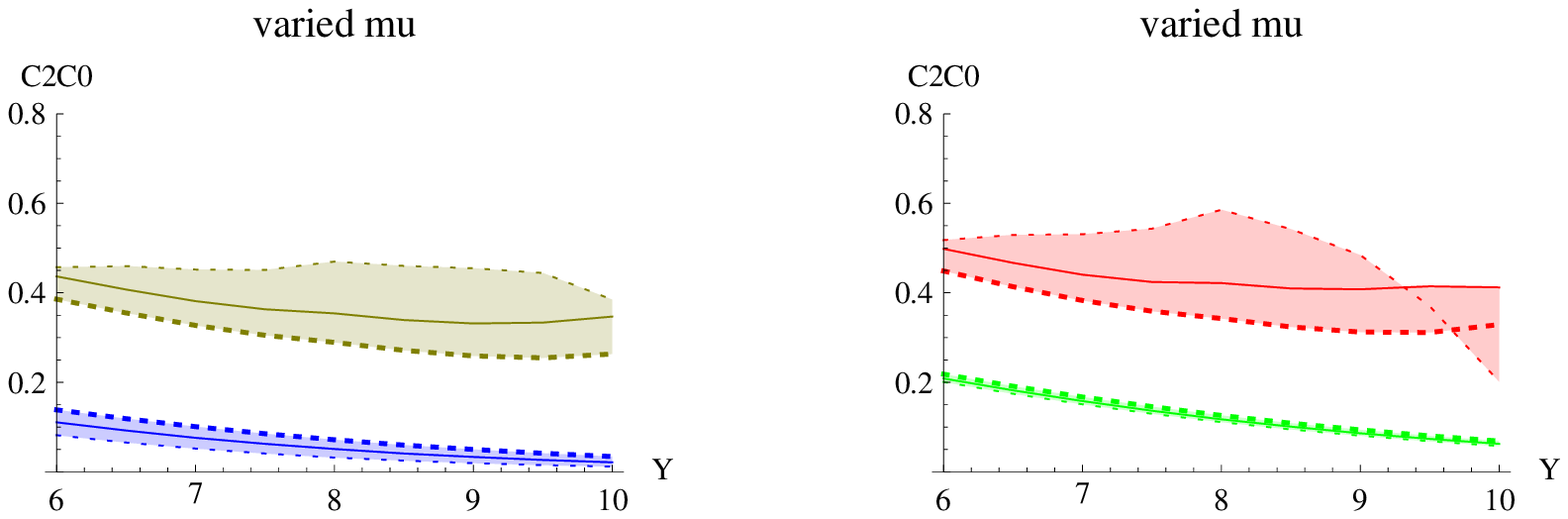}
   \caption{Effect of changing $\mu_R=\mu_F$ by factors 2 and $1/2$ respectively on $\langle \cos 2\varphi\rangle$ in dependence on $Y$ for $|\veckjone|=35\,{\rm GeV}$, $|\veckjtwo|=50\,{\rm GeV}$. The tabled values are shown in Tab.~\ref{tab:c2c03550_mu}}
   \label{fig:c2c03550_mu}
\end{figure}

\begin{figure}[h!]
  \centering
  \psfrag{varied}{}
  \psfrag{s0}{}
  \psfrag{C2C0}{$\frac{\mathcal{C}_2}{\mathcal{C}_0}=\langle \cos 2\varphi\rangle$}
  \psfrag{Y}{$Y$}
  \includegraphics[width=15cm]{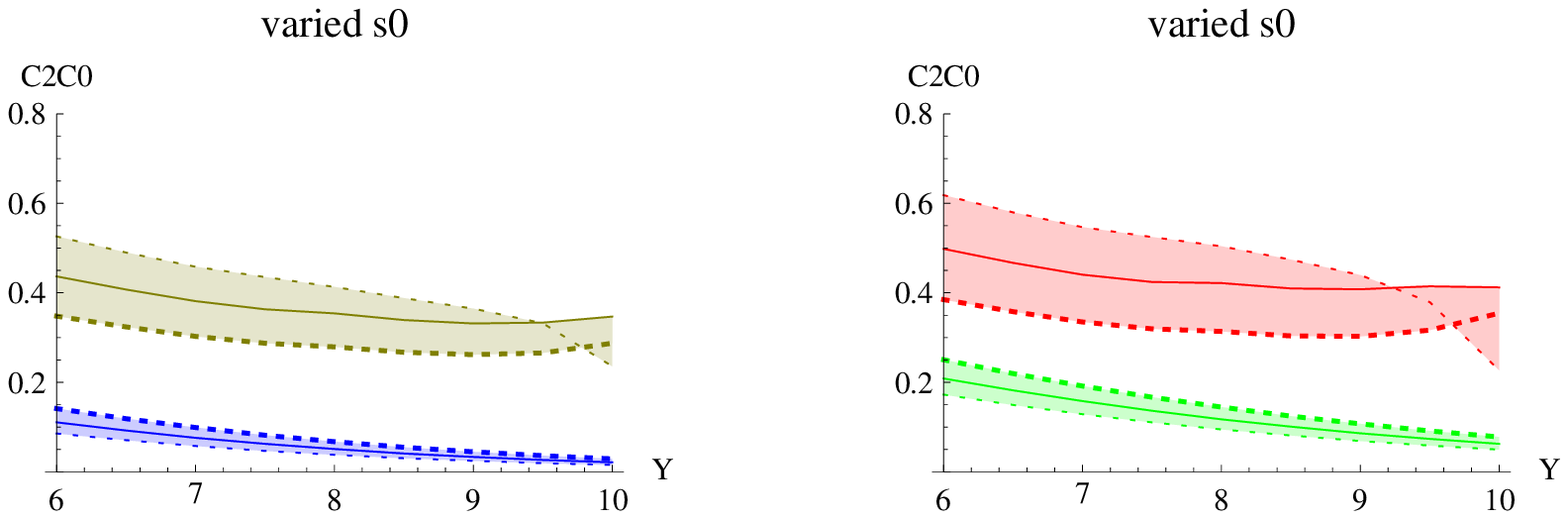}
  \caption{Effect of changing $\sqrt{s_0}$ by factors 2 and $1/2$ respectively on $\langle \cos 2\varphi\rangle$ in dependence on $Y$ for $|\veckjone|=35\,{\rm GeV}$, $|\veckjtwo|=50\,{\rm GeV}$. The tabled values are shown in Tab.~\ref{tab:c2c03550_s0}.}
  \label{fig:c2c03550_s0}
\end{figure}

\begin{figure}[h!]
   \centering
   \psfrag{varied}{}
   \psfrag{mu}{}
   \psfrag{C2C1}{$\frac{\mathcal{C}_2}{\mathcal{C}_1}$}
   \psfrag{Y}{$Y$}
   \includegraphics[width=15cm]{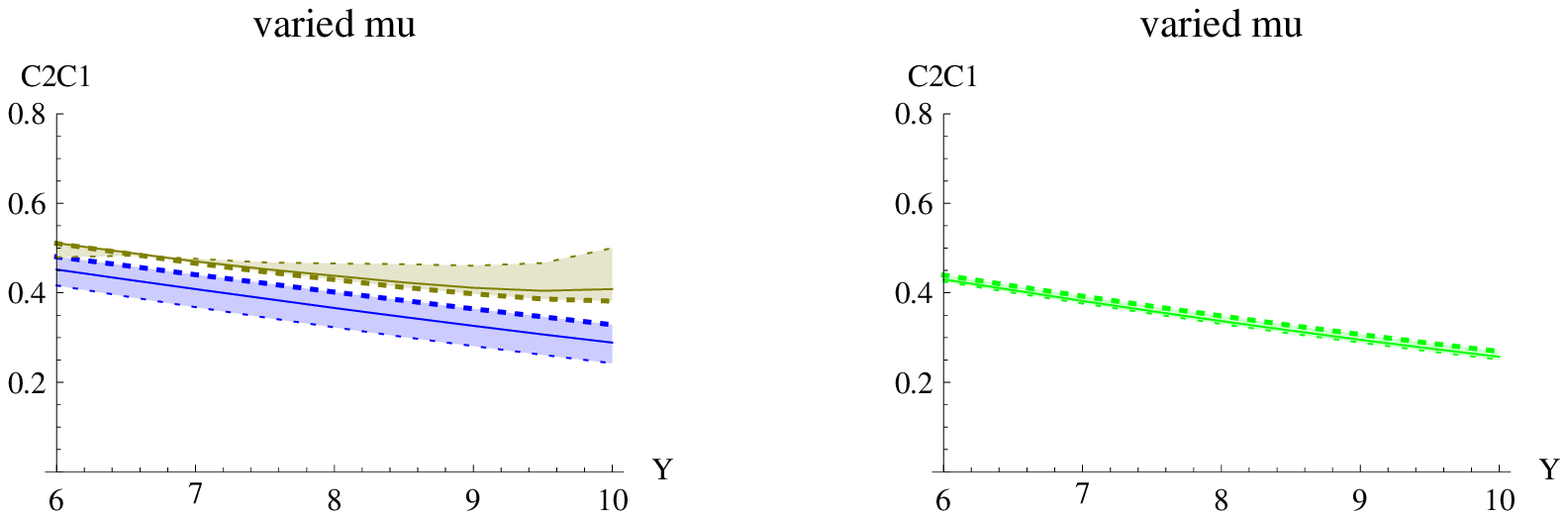}
   \caption{Effect of changing $\mu_R=\mu_F$ by factors 2 and $1/2$ respectively on $\langle \cos 2\varphi\rangle / \langle \cos \varphi\rangle$ in dependence on $Y$ for $|\veckjone|=35\,{\rm GeV}$, $|\veckjtwo|=50\,{\rm GeV}$. The tabled values are shown in Tab.~\ref{tab:c2c13550_mu}}
   \label{fig:c2c13550_mu}
 \end{figure}

\begin{figure}[h!]
  \centering
  \psfrag{varied}{}
  \psfrag{s0}{}
  \psfrag{C2C1}{$\frac{\mathcal{C}_2}{\mathcal{C}_1}$}
  \psfrag{Y}{$Y$}
  \includegraphics[width=15cm]{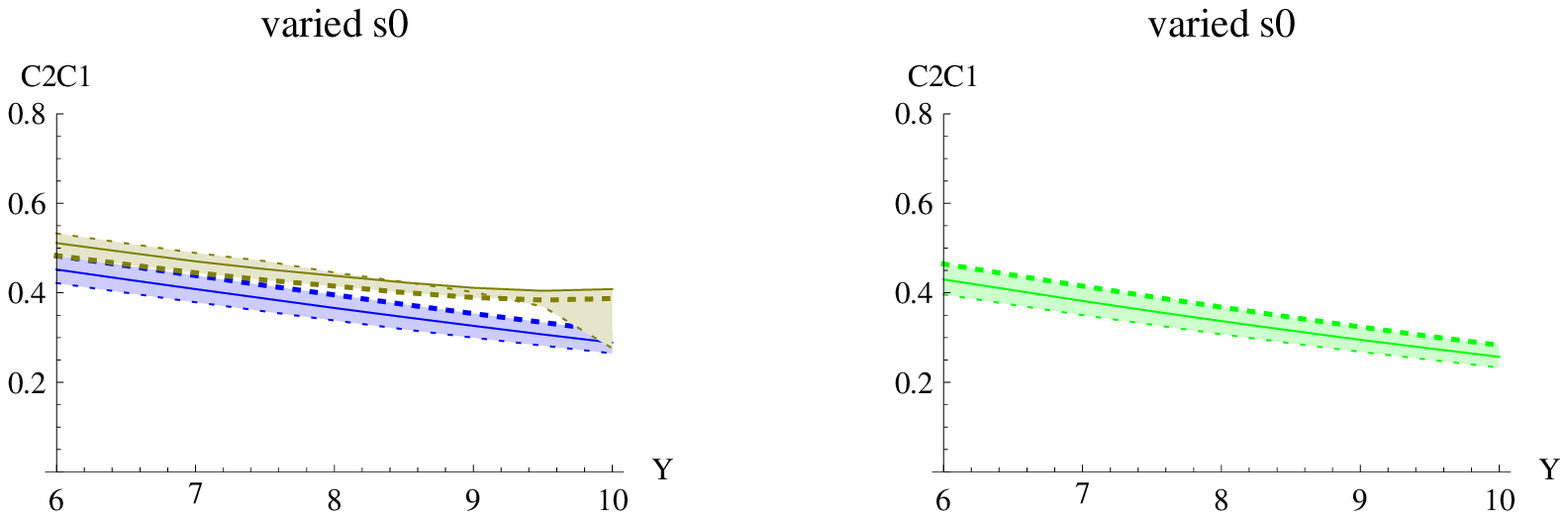}
  \caption{Effect of changing $\sqrt{s_0}$ by factors 2 and $1/2$ respectively on $\langle \cos 2\varphi\rangle / \langle \cos \varphi\rangle$ in dependence on $Y$ for $|\veckjone|=35\,{\rm GeV}$, $|\veckjtwo|=50\,{\rm GeV}$.  The tabled values are shown in Tab.~\ref{tab:c2c13550_s0}.}
  \label{fig:c2c13550_s0}
\end{figure}

\begin{figure}[h!]
  \centering
  \psfrag{varied}{}
  \psfrag{cubaerror}{}
  \psfrag{C1}{$\mathcal{C}_1 \left[\frac{\rm nb}{{\rm GeV}^2}\right] $}
  \psfrag{Y}{$Y$}
  \includegraphics[width=9cm]{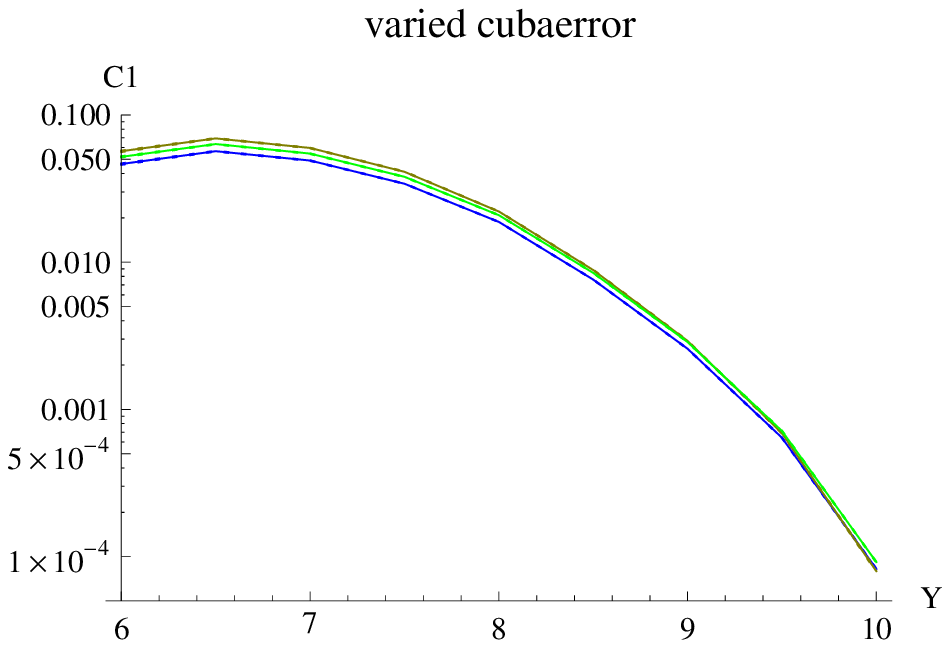}
  \caption{Coefficient $\mathcal{C}_1$ in dependence on $Y$ for $|\veckjone|=35\,{\rm GeV}$, $|\veckjtwo|=50\,{\rm GeV}$. The errors due to the Monte Carlo integration -- though hardly visible -- are given as error bands. The tabled values are shown in Tab.~\ref{tab:c13550}.}
  \label{fig:c13550}
\end{figure}

\begin{figure}[h!]
  \centering
  \psfrag{varied}{}
  \psfrag{s0}{}\psfrag{cubaerror}{}\psfrag{pdfset}{}\psfrag{mu}{}
  \psfrag{deltaC1}{$\delta\mathcal{C}_1 \left[\frac{\rm nb}{{\rm GeV}^2}\right] $}
  \psfrag{Y}{$Y$}
  \includegraphics[width=15cm]{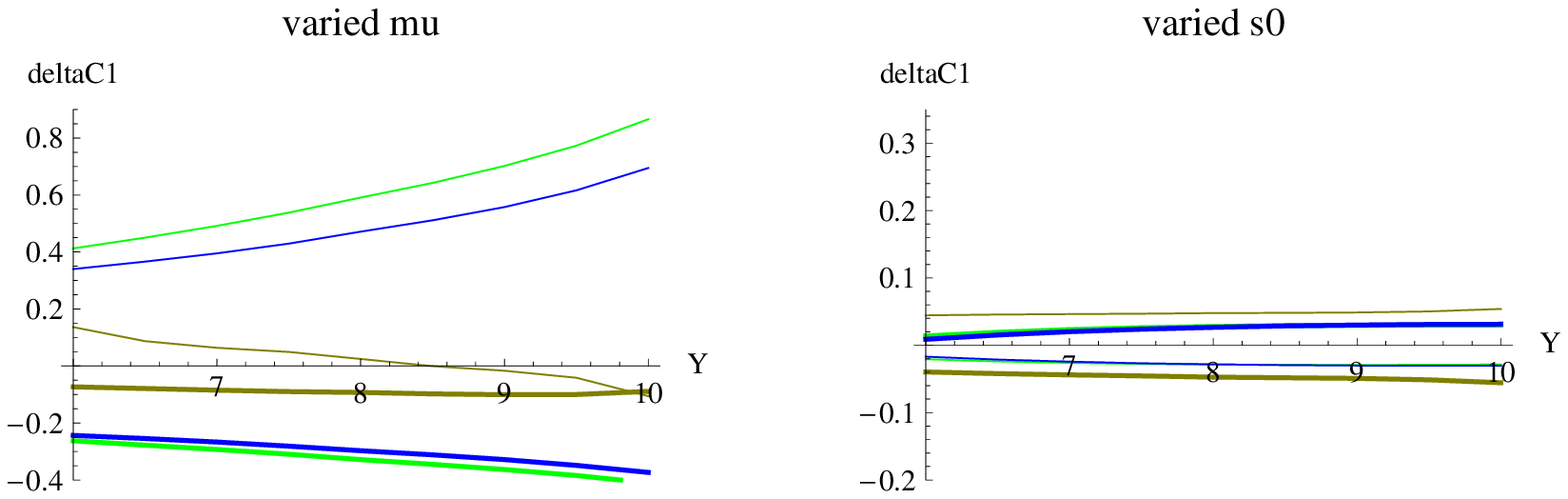}
  \caption{Relative effect of changing $\mu_R=\mu_F$ by factors 2 and $1/2$ respectively (left), and $\sqrt{s_0}$ (right) by factors 2 and $1/2$ respectively on the coefficient $\mathcal{C}_1$ in dependence on $Y$ for $|\veckjone|=35\,{\rm GeV},\;\;|\veckjtwo|=50\,{\rm GeV}$. The tabled values are shown in Tabs.~\ref{tab:c13550_mu} and \ref{tab:c13550_s0}.}
  \label{fig:c13550rel_mu_s0}
\end{figure}

\begin{figure}[h!]
  \centering
  \psfrag{varied}{}
  \psfrag{cubaerror}{}
  \psfrag{C2}{$\mathcal{C}_2 \left[\frac{\rm nb}{{\rm GeV}^2}\right] $}
  \psfrag{Y}{$Y$}
  \includegraphics[width=9cm]{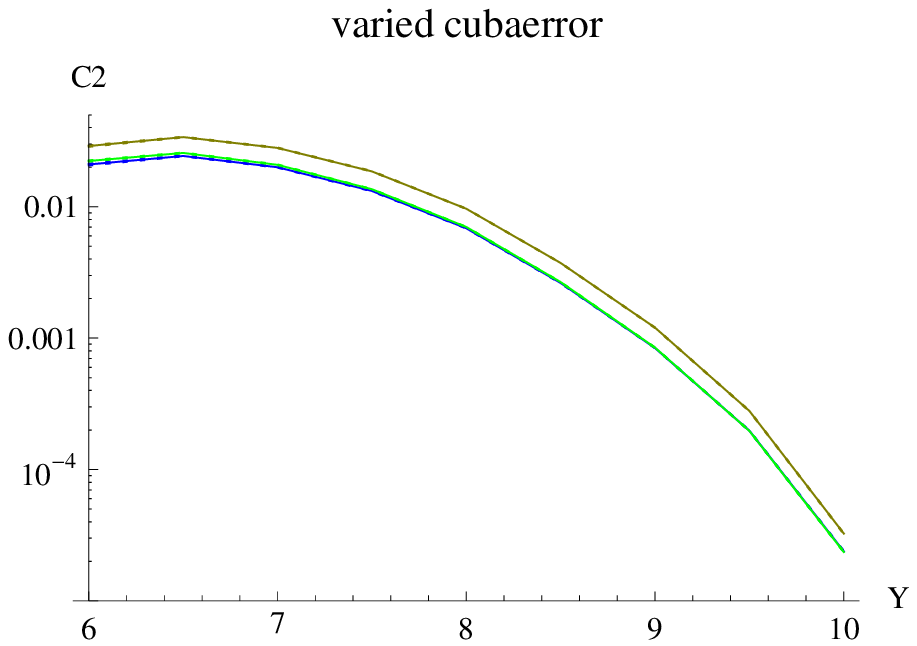}
  \caption{Coefficient $\mathcal{C}_2$ in dependence on $Y$ for $|\veckjone|=35\,{\rm GeV}$, $|\veckjtwo|=50\,{\rm GeV}$. The errors due to the Monte Carlo integration -- though hardly visible -- are given as error bands. The tabled values are shown in Tab.~\ref{tab:c23550}.}
  \label{fig:c23550}
\end{figure}

\begin{figure}[h!]
  \centering
  \psfrag{varied}{}
  \psfrag{s0}{}\psfrag{cubaerror}{}\psfrag{pdfset}{}\psfrag{mu}{}
  \psfrag{deltaC2}{$\delta\mathcal{C}_2 \left[\frac{\rm nb}{{\rm GeV}^2}\right] $}
  \psfrag{Y}{$Y$}
  \includegraphics[width=15cm]{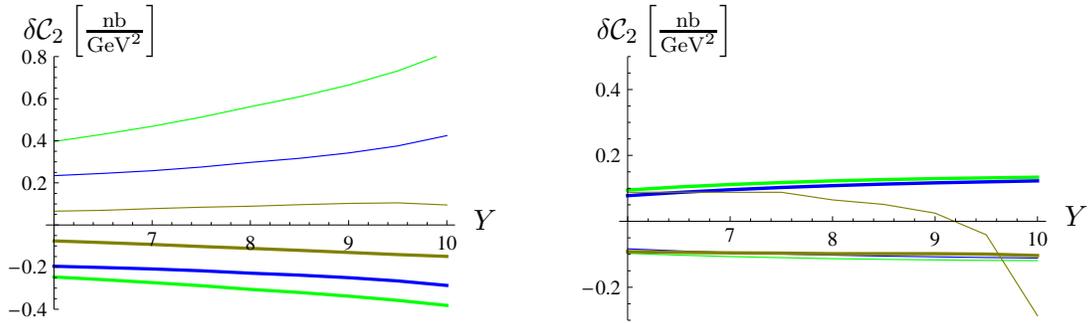}
  \caption{Relative effect of changing $\mu_R=\mu_F$ by factors 2 and $1/2$ respectively (left), and $\sqrt{s_0}$ (right) by factors 2 and $1/2$ respectively on the coefficient $\mathcal{C}_2$ in dependence on $Y$ for $|\veckjone|=35\,{\rm GeV},\;\;|\veckjtwo|=50\,{\rm GeV}$. The tabled values are shown in Tabs.~\ref{tab:c23550_mu} and \ref{tab:c23550_s0}.}
  \label{fig:c23550rel_mu_s0}
\end{figure}

Due to the factorization, the sensitivity of our prediction with respect to $s_0$, $\mu_R$ is similar to the two previous symmetrical configurations, as shown in Figs.~\ref{fig:c03550rel_mu_s0} for ${\cal C}_0$, in Figs.~\ref{fig:c1c03550_mu}, \ref{fig:c1c03550_s0} for ${\cal C}_1/{\cal C}_0\,,$ in Figs.~\ref{fig:c2c03550_mu}, \ref{fig:c2c03550_s0} for ${\cal C}_2/{\cal C}_0$
and in Figs.~\ref{fig:c2c13550_mu}, \ref{fig:c2c13550_s0}
for ${\cal C}_2/{\cal C}_1$.
In Figs.~\ref{fig:c13550}, \ref{fig:c13550rel_mu_s0} and Figs.~\ref{fig:c23550}, \ref{fig:c23550rel_mu_s0}, detailed studies for separate coefficients ${\cal C}_1$ and ${\cal C}_2$ are displayed.

One sees from Fig.~\ref{fig:c03550} that our pure NLL prediction, as well
as our resummed NLL prediction, are a bit below the NLO-DGLAP prediction,
while the LL prediction is much higher than the NLO-DGLAP prediction.
The combined LL vertices plus resummed NLL Green's function is rather close to the NLO-DGLAP prediction. One may expect that including higher order corrections in both DGLAP and BFKL approaches would make them converging. We note however that comparing both kinds of treatment should be done with some cautious. Indeed, the NLO-DGLAP
involves scales which are smaller than the scale which we consider:
we take $\mu_R=\sqrt{|\veckjone|\cdot |\veckjtwo|}$ which is similar
to $(|\veckjone|+ |\veckjtwo|)/2$, while the NLO-DGLAP calculation uses the scale
$(|\veckjone|+ |\veckjtwo|)/4\,.$
Changing this scale from $(|\veckjone|+ |\veckjtwo|)/4\,$ to
$(|\veckjone|+ |\veckjtwo|)/8$ leads to a variation of the order of 5\% in the NLO-DGLAP prediction.
Our treatment, especially when considering the azimuthal decorrelation, favors higher scales, like $\sqrt{|\veckjone|\cdot |\veckjtwo|}\sim (|\veckjone|+ |\veckjtwo|)/2$ or even $2 \sqrt{|\veckjone|\cdot |\veckjtwo|}\sim |\veckjone|+ |\veckjtwo|\,.$

The azimuthal decorrelation, which is expected to be the best signal,
is predicted to be similar in magnitude and shape both from our pure NLL prediction and our resummed NLL prediction and from the NLO-DGLAP approach, as can be seen from Figs.~\ref{fig:c1c03550} and \ref{fig:c2c03550}.
Note however that the uncertainties of our predictions are rather high.
Anyway, the general trend is clear: the azimuthal decorrelation is much lower than
expected from a LL BFKL treatment or from a mixed treatment with LL vertices combined with NLL Green's function. It is also rather flat with $Y\,.$ The only observable which still remain different when comparing pure NLL approaches (the resummed NLL approach makes no difference here since it only affects ${\cal C}_0$) with NLO-DGLAP
is the ratio $\langle \cos 2 \varphi \rangle / \langle \cos \varphi \rangle$ for which the NLO-DGLAP is still significantly higher than the NLL prediction, as can be seen from Fig.~\ref{fig:c2c13550}.

\clearpage

\clearpage 

\clearpage

\section{Conclusions}
\label{sec:Summary}

We have implemented at full NLL order the Mueller Navelet jets cross-section as well as their
relative  
azimuthal angle dependency.
In contrast to the general belief, the effect of NLL corrections to the vertex function is very important, of the same order as the one obtained when passing from LL to NLL Green's function.
The importance of NLL corrections to the impact factor observed in the present paper is analogous to recent results obtained at NLL in diffractive double $\rho$-electroproduction
\cite{Ivanov:2005gn, Ivanov:2006gt}.
Interestingly, the full NLL calculations for $\langle \cos \varphi\rangle$ and $\langle \cos 2\varphi\rangle$ are quite close to a calculation \cite{Fontannaz} using \textsc{Dijet} \cite{Aurenche:2008dn} which is based on DGLAP dynamics and to a dedicated study \cite{Cerci:2008xv} using \textsc{Pythia} \cite{Sjostrand:2006za} and \textsc{Herwig} \cite{Marchesini:1991ch}.
The uncertainty due to changes in $\mu_R$ (and $s_0$) is drastically reduced for all $\mathcal{C}_n$ when one takes into account the NLL Mueller Navelet vertices. The uncertainty
due to PDFs are also moderate. As a consequence, our results for the cross-section are very stable.

However, for azimuthal decorrelation the dependence on $\mu_R$ (and $s_0$) is still sizeable.
In the case of the NLL Green's function with collinear improvement one observes that $\langle \cos \varphi\rangle$ can exceed 1 for certain choices of the parameters, in particular for low values of $\mu_R=\mu_F\,,$ taken to be smaller than the ``natural'' value $\sqrt{|k_{J,1}|\, \cdot|k_{J,2}|}\,.$
One might also think of a collinear improvement of the vertices \cite{Ciafaloni:2003rd} but the Mueller Navelet vertex for fixed $|\veckj|$ does not have poles in $\gamma$ nor $1-\gamma$,
so there is no room for such a treatment. The resummation of soft initial radiation might be of relevance for the azimuthal correlation as well.
This is left for further investigations, and in this work we rather consider the full NLL calculation without additional collinear resummation to be our solid prediction, while the `collinear improvement' as it stands is not appropriate to study azimuthal dependences.

At present, there is little experience with the effect of NLL impact factors.
To the best of our knowledge, up to now, the only full NLL BFKL calculation existing in the literature is the vector meson production in virtual photon collisions \cite{Ivanov:2005gn,Ivanov:2006gt,Caporale:2007vs}, which is very sensitive to NLL corrections to the impact factor and for which very large values for $s_0$ and $\mu_R$ are preferred.
In \cite{Caporale:2007vs} it has been shown that a collinear improved treatment combined with 
the application of the principle of minimal sensitivity \cite{Stevenson:1980du, Stevenson:1981vj}
reduces this large values to more ``natural'' values. Still, $\mu_R$ larger than the ``natural'' values
are favored \cite{Caporale:2007vs}. In the present case, with the scales $\mu_R$ and $s_0$ set by the jet scale, 
we get azimuthal correlations which are rather
similar to DGLAP dynamics predictions (although, as we already mentioned, the DGLAP prediction are based on smaller scales). To conclude, contrarily to the expectation, it thus seems that the azimuthal decorrelation is almost not enhanced by an increasing rapidity. This suggests that the study of Mueller Navelet jets is probably not the best place to exhibit differences between BFKL and DGLAP dynamics.

\acknowledgments

We acknowledge discussions with Jochen Bartels, Salim Cerci, David d'Enterria,
Bernard Pire, Agust{\'i}n Sabio Vera, Kirsten Sachs, Gian Paolo Vacca. 
We especially thank Michel Fontannaz for many explanations of the 
DGLAP-based approach and for providing us his own predictions.
This work is supported in part by the Polish Grant N202 249235, the French-Polish
scientific agreement Polonium, by the grant ANR-06-JCJC-0084 and by the ECO-NET
program, contract 12584QK, and by a PRIN grant (MIUR, Italy).

\appendix

\section{Details on the numerical implementation}

\subsection{Programs used}

We implemented all numerical calculations in \textsc{Mathematica}. To this purpose we used the according interfaces for the MSTW 2008 PDFs \cite{Martin:2009iq} and for version 1.5 of \textsc{Cuba} \cite{Hahn:2004fe} which we used for numerical integration.

\subsection{Choice of Parameters}
\label{sec:cubaparameters}

\textsc{Cuba} provides different integration routines which we also used to cross-check the results of the Monte Carlo integration. However, for the final results we used the Vegas routine of \textsc{Cuba} with an aimed precision of $10^{-2}$ and a maximal number of $500\,000$ points per integration. To use a Monte Carlo integrator, all integration intervals have to be mapped on finite intervals. For the transverse momentum integrations we used the mapping $|\veck|=|\veckj|\tan(\xi\pi/2)$.

The cancellations which analytically have been shown in Refs.~\cite{Bartels:2001ge,Bartels:2002yj} numerically can corrupt the integration due to the limited precision of a computer. In all these cases, were the integration interval have been mapped to the compact interval $[0,1]$, we used a cut off of $10^{-5}$ where the cut off dependence becomes negligible.

\subsubsection{The $\nu$-grid}

Due to the complicate matrix element, the PDF evaluation, and the implementation in \textsc{Mathematica} instead of a dedicated stand-alone code the Monte Carlo integration is very time consuming. Therefor, the choice of the $\nu$-values at which the coefficients $C_{n,\nu}$ \eqref{eq:mastercnnu} are evaluated is crucial.

We are guided by the shape of the BFKL Green's function which is peaked around $\nu = 0$ and then monotonically falls. The smaller $Y$ the slower the decrease. Even though the minimal $Y$ in this study is 6, we want our coefficients to be prepared also for smaller $Y$'s. We choose a maximal $\nu_{\rm max}$ such that an integration up to $\nu_{\rm max}$ of just the NLL BFKL Green's function at $Y=4$ for $n=0$ reproduces $96\%$ of the integration over the full $\nu$-range. For the case of $Y=6$ it reproduces $99.97\%$ of the full integral.

The coefficients $C_{n,\nu}$ are oscillating like $\exp(i\nu\ln\veckji^2)$ but more important is the product of the two which has an oscillating part with a frequency $\nu_{\rm oscillation}=\pi/\ln\frac{|\veckjone|}{|\veckjtwo|}$. This frequency is zero for $|\veckjone|=|\veckjtwo|$ but for the example of $|\veckjone|=35\,{\rm GeV}$ and $|\veckjtwo|=50\,{\rm GeV}$ we chose a step width for $\nu$ of $\nu_{\rm oscillation}/4$. 

For large $Y$ it is really the small region close to $\nu=0$ which matters. Therefore we sample this region in more detail according to the shape of the NLL BFKL Green's function for $n=0$ and $Y=7$. The final $\nu$-grid reads
\begin{multline}
  \{0, 0.0334439, 0.0671152, 0.101257, 0.136128, 0.172017, 0.209264, 0.248284, \\
0.289594, 0.333866, 0.382007, 0.435281, 0.495535, 0.565607, 0.65013, 0.75725, \\
0.902736, 1.12137, 1.50735, 2.44882, \frac{2\pi}{4\ln\frac{10}{7}}, \frac{3\pi}{4\ln\frac{10}{7}}, \ldots, \frac{49\pi}{4\ln\frac{10}{7}}\}
\end{multline}

For the final integration over $\nu$ the product $C_{n,\nu}(|\veckjone|,x_{J,1})C_{n,\nu}^*(|\veckjtwo|,x_{J,2})$ is interpolated by cubic splines.

\subsection{Grouping the integrand}

In this section we describe how the NLL contribution to the coefficients $C_{n,\nu}$, as defined in Eq.~\eqref{eq:mastercnnu}, is arranged.

The jet defining function $\mathcal{S}_J^{(3)}$ given in Eq.~\eqref{eq:jetdefcone} consists of three parts which we label $\mathcal{S}_{J,a}^{(3)}$, $\mathcal{S}_{J,b}^{(3)}$ and $\mathcal{S}_{J,c}^{(3)}$. With this separation $V_{\rm q}^{(1)}$, given in Eq.~\eqref{eq:Vqright}, consists of 17 elementary blocks which we denote by $V_{\rm q}[i]$ (suppressing for the time being all further arguments and indices) , where $i=1,\ldots,17$. In the same spirit we decompose $V_{\rm g}^{(1)}$ given in Eq.~\eqref{eq:Vgright} in its 25 elementary blocks $V_{\rm g}[i]$. 

It is useful to replace the integration variable $\veck$ by $\veck\to \veckj-\veck$ in the integrands $V_{\rm q}[5]$ and $V_{\rm q}[7]$ ($V_{\rm g}[23]$ and $V_{\rm g}[25]$). Moreover we split up $V_{\rm q}[3]$ ($V_{\rm g}[21]$) and create $V_{\rm q}[18]$ ($V_{\rm g}[26]$) where in the new elementary blocks the integrand $\veck'$ is replaced by $\veckj-\veck'$. Then we make the following replacements
\begin{align}
  V_{\rm q}[3] \quad\rightarrow\quad & V_{\rm q}[3]\Theta\left(z-\frac{1}{2}\right)+V_{\rm q}[18]\Theta\left(\frac{1}{2}-z\right) \\
  V_{\rm g}[21] \quad\rightarrow\quad & V_{\rm g}[21]\Theta\left(z-\frac{1}{2}\right)+V_{\rm g}[26]\Theta\left(\frac{1}{2}-z\right).
\end{align}

\clearpage
The elementary blocks are now grouped to 14 minimal basic blocks $B[i]$ which also contain the integrations from Eq.~\eqref{eq:mastercnnu}
\begin{align}
  B [1]=&\iiiintop V_{\rm q}[1]\\
  B [2]=&\iiiintop V_{\rm q}[2]\\
  B [3]=&\iiiintop \left(V_{\rm q}[3]\Theta\left(z-\frac{1}{2}\right)+V_{\rm q}[4]+V_{\rm q}[6]\right)\\
  B [4]=&\iiiintop \left(V_{\rm q}[18]\Theta\left(\frac{1}{2}-z\right)+V_{\rm q}[5]+V_{\rm q}[7]\right)\\
  B [5]=&\iiiintop \left(V_{\rm q}[11]+V_{\rm q}[12]+V_{\rm q}[13]+V_{\rm q}[15]+\frac{1}{2}\left(V_{\rm q}[16]+V_{\rm q}[17]\right)\right)\\
  B [6]=&\iiiintop \left(V_{\rm q}[8]+V_{\rm q}[9]+V_{\rm q}[10]+V_{\rm q}[14]+\frac{1}{2}\left(V_{\rm q}[16]+V_{\rm q}[17]\right)\right)\\
  B [7]=&\iiiintop V_{\rm g}[1]\\
  B [8]=&\iiiintop V_{\rm g}[2]\\
  B [9]=&\iiiintop \sum_{i=3}^6 V_{\rm g}[i]\\
  B[10]=&\iiiintop \sum_{i=7}^{10} V_{\rm g}[i]\\
  B[11]=&\iiiintop \left(V_{\rm g}[11]+V_{\rm g}[12]+V_{\rm g}[13]+V_{\rm g}[17]+\frac{1}{2}\left(V_{\rm g}[19]+V_{\rm g}[20]\right)\right)\\
  B[12]=&\iiiintop \left(V_{\rm g}[14]+V_{\rm g}[15]+V_{\rm g}[16]+V_{\rm g}[18]+\frac{1}{2}\left(V_{\rm g}[19]+V_{\rm g}[20]\right)\right)\\
  B[13]=&\iiiintop \left(V_{\rm g}[21]\Theta\left(z-\frac{1}{2}\right)+V_{\rm g}[22]+V_{\rm g}[24]\right)\\
  B[14]=&\iiiintop \left(V_{\rm g}[26]\Theta\left(\frac{1}{2}-z\right)+V_{\rm g}[23]+V_{\rm g}[25]\right),
\end{align}
where we made use of the short hand notation $\iiiint \equiv \fourint$.

We would like to point out that the inclusion of $V_{\rm q}[18]$ ($V_{\rm g}[26]$) in $B [4]$ ($B[14]$) is essential. Even though it is correctly stated after Eq.~(88) in Ref.~\cite{Bartels:2001ge} (and repeated in Ref.~\cite{Bartels:2002yj} after Eq.~(53)), that in the composite jet configuration the domain of integration shrinks like $z^2$ for $z\to 0$, the conclusion that this prevents a divergence is wrong. In fact, in the limit $z\to 0$ the integrand scales like $z^{-3}$ and only the sum of $V_{\rm q}[18]$ ($V_{\rm g}[26]$) and $V_{\rm q}[5]$ ($V_{\rm g}[23]$) cancels properly against $V_{\rm q}[7]$ ($V_{\rm g}[25]$) in the dangerous region.

Note that in case of fix $x_J$ for $B[1]$ and $B[7]$ no numerical integration is needed, while for $B[2]$ and $B[8]$ only one integration over $z$ has to be done. All four are proportional to the LL Mueller Navelet vertex regarding the transverse momentum dependences. The Dirac-$\delta$ in transverse momenta we always use for the $\veck$ integration. Only for contributions with $\delta^{(2)}(\veck'-\veckj)$ it is used for the $\veck'$ integration. 
The $x$-integration is trivially performed by evaluating the according Dirac $\delta$-distribution.

\section{Tabled values of diagrams}

To allow for later accurate comparisons, we give the values for all plots in this work. We mark the pure LL calculation by `LL', and the pure NLL one by `NLL'. The combination of LL vertices with the NLL collinear improved Green's function is denoted as `LL+', and the combination of NLL vertices with the NLL collinear improved Green's function as `NLL+'. Whenever in a figure the effect of the variation of one parameter is presented, in the according table the first column shows the central value, while the second and third show the change of this central value due to the varied parameter. For brevity we suppress the energy unit GeV in the headings of the tables.

\begin{table}[h]
  \centering
  \tiny
\begin{tabular}{l|l|l|l|l}
& LL &\multicolumn{3}{|c}{ NLL}\\
$Y$& 
& MC0  &  MC+ &  MC- \\ 
\hline 
  6   & 1.52 & 0.606 & 0.006 & -0.006 \\
  6.5 & 2.06 & 0.752 & 0.005 & -0.005 \\
  7   & 2.04 & 0.670 & 0.004 & -0.004 \\
  7.5 & 1.70 & 0.490 & 0.003 & -0.003 \\
  8   & 1.18 & 0.289 & 0.002 & -0.002 \\
  8.5 & 0.602 &  0.126 & 0.0009 & -0.0009 \\
  9   & 0.268 &  0.0474 & 0.0004 & -0.0004 \\
  9.5 & 0.0949 & 0.0138 & 0.0002 & -0.0002 \\
 10   & 0.0207 & 0.00238 & 0.00006 & -0.00006\\
\hline
& LL+ &\multicolumn{3}{|c}{NLL+}\\
$Y$& 
& MC0  &  MC+ &  MC- \\ 
\hline
  6   & 0.849 &  0.549 & 0.006 & -0.006 \\
  6.5 & 1.10 & 0.675 & 0.006 & -0.006 \\
  7   & 1.03 & 0.595 & 0.004 & -0.004 \\
  7.5 & 0.804 & 0.430 & 0.003 & -0.003 \\
  8   & 0.523 &  0.249 & 0.002 & -0.002 \\
  8.5 & 0.248 & 0.107 & 0.001 & -0.001 \\
  9   & 0.102 & 0.0396 & 0.0005 & -0.0005 \\
  9.5 & 0.0337 & 0.0114 & 0.0002 & -0.0002 \\
 10   & 0.00681 & 0.00198 & 0.00008 & -0.00008 
  \end{tabular}
  \caption{tabled values for Fig.~\ref{fig:c03535}, \ref{fig:c03535rel_pdf_cuba}.}
  \label{tab:c03535}
\end{table}

\begin{table}[h]
  \centering
  \tiny\begin{tabular}{l|l|l|l|l|l|l}
&\multicolumn{3}{|c}{ LL}&\multicolumn{3}{|c}{ NLL}\\
$Y$ & $\mu=35$ & $\mu=70$ & $\mu=17.5$ 
    & $\mu=35$ & $\mu=70$ & $\mu=17.5$ \\
\hline
  6   & 1.52 & -0.4 & 0.8 & 0.606 & 0.03 & -0.04 \\
  6.5 & 2.06 & -0.6 & 1. & 0.752 & 0.03 & -0.06 \\
  7   & 2.04 & -0.7 & 2. & 0.670 & 0.03 & -0.06 \\
  7.5 & 1.70 & -0.7 & 2. & 0.490 & 0.02 & -0.05 \\
  8   & 1.18 & -0.5 & 1. & 0.289 & 0.02 & -0.03 \\
  8.5 & 0.602 & -0.3 & 0.7 & 0.126 & 0.007 & -0.02 \\
  9   & 0.268 & -0.1 & 0.4 & 0.0474 & 0.003 & -0.006 \\
  9.5 & 0.0949 & -0.05 & 0.1 & 0.0138 & 0.001 & -0.002 \\
 10   & 0.0207 & -0.01 & 0.03 & 0.00238 & 0.0002 & -0.0002 \\
\hline
&\multicolumn{3}{|c}{ LL+}&\multicolumn{3}{|c}{ NLL+}\\
$Y$ & $\mu=35$ & $\mu=70$ & $\mu=17.5$ 
    & $\mu=35$ & $\mu=70$ & $\mu=17.5$ \\
\hline
  6   & 0.849 & -0.2 & 0.3 & 0.549 & 0.02 & -0.04 \\
  6.5 & 1.10 & -0.3 & 0.5 & 0.675 & 0.02 & -0.05 \\
  7   & 1.03 & -0.3 & 0.5 & 0.595 & 0.02 & -0.05 \\
  7.5 & 0.805 & -0.3 & 0.5 & 0.430 & 0.02 & -0.05 \\
  8   & 0.523 & -0.2 & 0.3 & 0.249 & 0.01 & -0.03 \\
  8.5 & 0.248 & -0.09 & 0.2 & 0.107 & 0.007 & -0.01 \\
  9   & 0.102 & -0.04 & 0.08 & 0.0396 & 0.003 & -0.003 \\
  9.5 & 0.0337 & -0.01 & 0.03 & 0.0114 & 0.0009 & 0.0008 \\
 10   & 0.00681 & -0.003 & 0.006 & 0.00197 & 0.0001 & 0.0009
\end{tabular}
\caption{tabled values for left figure of Fig.~\ref{fig:c03535rel_mu_s0}.}
  \label{tab:c03535_mu}
\end{table}

\begin{table}[h]
  \centering
  \tiny\begin{tabular}{l|l|l|l|l|l|l}
&\multicolumn{3}{|c}{ LL}&\multicolumn{3}{|c}{ NLL}\\
$Y$ & $\sqrt{s_0}=35$ & $\sqrt{s_0}=70$ & $\sqrt{s_0}=17.5$ 
    & $\sqrt{s_0}=35$ & $\sqrt{s_0}=70$ & $\sqrt{s_0}=17.5$ \\
\hline
  6   & 1.52 & -0.03 & 0.1 & 0.606 & 0.03 & -0.008 \\
  6.5 & 2.06 & -0.1 & 0.3 & 0.752 & 0.03 & -0.02 \\
  7   & 2.04 & -0.2 & 0.3 & 0.670 & 0.02 & -0.02 \\
  7.5 & 1.70 & -0.2 & 0.3 & 0.490 & 0.02 & -0.02 \\
  8   & 1.18 & -0.2 & 0.2 & 0.289 & 0.01 & -0.01 \\
  8.5 & 0.602 & -0.09 & 0.1 & 0.126 & 0.005 & -0.006 \\
  9   & 0.268 & -0.04 & 0.06 & 0.0474 & 0.002 & -0.002 \\
  9.5 & 0.0949 & -0.02 & 0.02 & 0.0138 & 0.0005 & -0.0005 \\
 10   & 0.0207 & -0.004 & 0.005 & 0.00238 & 0.00007 & -0.00002\\
\hline
&\multicolumn{3}{|c}{ LL+}&\multicolumn{3}{|c}{ NLL+}\\
$Y$ & $\sqrt{s_0}=35$ & $\sqrt{s_0}=70$ & $\sqrt{s_0}=17.5$ 
    & $\sqrt{s_0}=35$ & $\sqrt{s_0}=70$ & $\sqrt{s_0}=17.5$ \\
\hline
  6   & 0.849 & 0.009 & 0.02 & 0.549 & 0.03 & -0.01 \\
  6.5 & 1.10 & -0.02 & 0.05 & 0.675 & 0.03 & -0.03 \\
  7   & 1.03 & -0.03 & 0.06 & 0.595 & 0.03 & -0.03 \\
  7.5 & 0.805 & -0.04 & 0.05 & 0.430 & 0.03 & -0.03 \\
  8   & 0.523 & -0.03 & 0.04 & 0.249 & 0.02 & -0.02 \\
  8.5 & 0.248 & -0.02 & 0.02 & 0.107 & 0.007 & -0.006 \\
  9   & 0.102 & -0.008 & 0.009 & 0.0396 & 0.003 & -0.002 \\
  9.5 & 0.0337 & -0.003 & 0.003 & 0.0114 & 0.0007 & 0.000003 \\
 10   & 0.00681 & -0.0006 & 0.0007 & 0.00197 & 0.00008 & 0.0002
\end{tabular}
\caption{tabled values for right figure of Fig.~\ref{fig:c03535rel_mu_s0}.}
  \label{tab:c03535_s0}
\end{table}

\begin{table}[h]
  \centering
  \tiny\begin{tabular}{l|l|l|l|l|l|l}
&\multicolumn{3}{|c}{ LL}&\multicolumn{3}{|c}{ NLL}\\
$Y$& { PDF0} & { PDF+} & { PDF-}
& { PDF0} & { PDF+} & { PDF-}\\
\hline
  6   & 1.52 & 0.08 & -0.1 & 0.606 & 0.03 & -0.04 \\
  6.5 & 2.06 & 0.1 & -0.1 & 0.752 & 0.05 & -0.05 \\
  7   & 2.04 & 0.2 & -0.2 & 0.670 & 0.06 & -0.06 \\
  7.5 & 1.70 & 0.2 & -0.2 & 0.490 & 0.05 & -0.05 \\
  8   & 1.18 & 0.1 & -0.1 & 0.289 & 0.04 & -0.03 \\
  8.5 & 0.602 & 0.09 & -0.07 & 0.126 & 0.02 & -0.02 \\
  9   & 0.268 & 0.05 & -0.04 & 0.0474 & 0.008 & -0.007 \\
  9.5 & 0.0949 & 0.02 & -0.02 & 0.0138 & 0.003 & -0.002 \\
 10   & 0.0207 & 0.005 & -0.004 & 0.00238 & 0.0006 & -0.0004\\
\hline
&\multicolumn{3}{|c}{ LL+}&\multicolumn{3}{|c}{ NLL+}\\
$Y$& { PDF0} & { PDF+} & { PDF-}
& { PDF0} & { PDF+} & { PDF-}\\
\hline
  6   & 0.849 & 0.05 & -0.06 & 0.549 & 0.03 & -0.04 \\
  6.5 & 1.10 & 0.07 & -0.08 & 0.675 & 0.05 & -0.05 \\
  7   & 1.03 & 0.08 & -0.08 & 0.595 & 0.05 & -0.05 \\
  7.5 & 0.805 & 0.08 & -0.08 & 0.430 & 0.05 & -0.04 \\
  8   & 0.523 & 0.06 & -0.06 & 0.249 & 0.03 & -0.03 \\
  8.5 & 0.248 & 0.04 & -0.03 & 0.107 & 0.02 & -0.01 \\
  9   & 0.102 & 0.02 & -0.01 & 0.0396 & 0.007 & -0.006 \\
  9.5 & 0.0337 & 0.007 & -0.005 & 0.0114 & 0.002 & -0.002 \\
 10   & 0.00681 & 0.002 & -0.001 & 0.00197 & 0.0005 & -0.0004
\end{tabular}
\caption{tabled values for left figure of Fig.~\ref{fig:c03535rel_pdf_cuba}.}
  \label{tab:c03535_pdf}
\end{table}

\begin{table}[h]
  \centering
  \tiny\begin{tabular}{l|l|l|l|l|l|l}
&\multicolumn{3}{|c}{ LL}&\multicolumn{3}{|c}{ NLL}\\
$Y$& { MC0} & { MC+} & { MC-}
& { MC0} & { MC+} & { MC-}\\
\hline
  6   & 0.377 & 0.003 & -0.003 & 0.851 & 0.01 & -0.01 \\
  6.5 & 0.308 & 0.002 & -0.002 & 0.824 & 0.007 & -0.007 \\
  7   & 0.252 & 0.0009 & -0.0009 & 0.801 & 0.006 & -0.006 \\
  7.5 & 0.206 & 0.0006 & -0.0006 & 0.783 & 0.005 & -0.005 \\
  8   & 0.170 & 0.0005 & -0.0005 & 0.777 & 0.005 & -0.005 \\
  8.5 & 0.140 & 0.0004 & -0.0004 & 0.761 & 0.006 & -0.006 \\
  9   & 0.115 & 0.0004 & -0.0004 & 0.751 & 0.008 & -0.008 \\
  9.5 & 0.0949 & 0.0004 & -0.0004 & 0.749 & 0.01 & -0.01 \\
 10   & 0.0784 & 0.0005 & -0.0005 & 0.753 & 0.02 & -0.02\\
\hline
&\multicolumn{3}{|c}{ LL+}&\multicolumn{3}{|c}{ NLL+}\\
$Y$& { MC0} & { MC+} & { MC-}
& { MC0} & { MC+} & { MC-}\\
\hline
  6   & 0.566 & 0.004 & -0.004 & 0.940 & 0.01 & -0.01 \\
  6.5 & 0.514 & 0.002 & -0.002 & 0.919 & 0.009 & -0.009 \\
  7   & 0.465 & 0.002 & -0.002 & 0.902 & 0.007 & -0.007 \\
  7.5 & 0.421 & 0.001 & -0.001 & 0.893 & 0.007 & -0.007 \\
  8   & 0.381 & 0.001 & -0.001 & 0.901 & 0.008 & -0.008 \\
  8.5 & 0.344 & 0.001 & -0.001 & 0.895 & 0.009 & -0.009 \\
  9   & 0.311 & 0.001 & -0.001 & 0.898 & 0.01 & -0.01 \\
  9.5 & 0.280 & 0.001 & -0.001 & 0.908 & 0.02 & -0.02 \\
 10   & 0.253 & 0.002 & -0.002 & 0.908 & 0.04 & -0.04
\end{tabular}
\caption{tabled values for Fig.~\ref{fig:c1c03535}.}
  \label{tab:c1c03535}
\end{table}

\begin{table}[h]
  \centering
  \tiny\begin{tabular}{l|l|l|l|l|l|l}
&\multicolumn{3}{|c}{ LL}&\multicolumn{3}{|c}{ NLL}\\
$Y$ & $\mu=35$ & $\mu=70$ & $\mu=17.5$ 
    & $\mu=35$ & $\mu=70$ & $\mu=17.5$ \\
\hline
  6   & 0.377 & 0.1 & -0.1 & 0.851 & -0.06 & 0.07 \\
  6.5 & 0.308 & 0.1 & -0.09 & 0.824 & -0.07 & 0.08 \\
  7   & 0.252 & 0.09 & -0.08 & 0.801 & -0.07 & 0.09 \\
  7.5 & 0.206 & 0.08 & -0.07 & 0.783 & -0.08 & 0.1 \\
  8   & 0.170 & 0.07 & -0.06 & 0.777 & -0.1 & 0.1 \\
  8.5 & 0.140 & 0.06 & -0.05 & 0.761 & -0.1 & 0.1 \\
  9   & 0.115 & 0.05 & -0.04 & 0.751 & -0.1 & 0.1 \\
  9.5 & 0.0949 & 0.05 & -0.04 & 0.749 & -0.1 & 0.1 \\
 10   & 0.0784 & 0.04 & -0.03 & 0.753 & -0.1 & 0.07 \\
\hline
&\multicolumn{3}{|c}{ LL+}&\multicolumn{3}{|c}{ NLL+}\\
$Y$ & $\mu=35$ & $\mu=70$ & $\mu=17.5$ 
    & $\mu=35$ & $\mu=70$ & $\mu=17.5$ \\
\hline
  6   & 0.566 & 0.02 & -0.02 & 0.940 & -0.06 & 0.07 \\
  6.5 & 0.514 & 0.02 & -0.02 & 0.919 & -0.07 & 0.09 \\
  7   & 0.465 & 0.02 & -0.02 & 0.902 & -0.08 & 0.1 \\
  7.5 & 0.421 & 0.02 & -0.02 & 0.893 & -0.09 & 0.1 \\
  8   & 0.381 & 0.02 & -0.02 & 0.901 & -0.1 & 0.2 \\
  8.5 & 0.344 & 0.02 & -0.02 & 0.895 & -0.1 & 0.1 \\
  9   & 0.311 & 0.02 & -0.02 & 0.898 & -0.1 & 0.09 \\
  9.5 & 0.280 & 0.02 & -0.02 & 0.908 & -0.2 & -0.05 \\
 10   & 0.253 & 0.02 & -0.01 & 0.908 & -0.1 & -0.3
\end{tabular}
\caption{tabled values for Figs.~\ref{fig:c1c03535_mu}}
  \label{tab:c1c03535_mu}
\end{table}

\begin{table}[h]
  \centering
  \tiny\begin{tabular}{l|l|l|l|l|l|l}
&\multicolumn{3}{|c}{ LL}&\multicolumn{3}{|c}{ NLL}\\
$Y$ & $\sqrt{s_0}=35$ & $\sqrt{s_0}=70$ & $\sqrt{s_0}=17.5 $
    & $\sqrt{s_0}=35$ & $\sqrt{s_0}=70$ & $\sqrt{s_0}=17.5$ \\
\hline
  6   & 0.377 & 0.1 & -0.09 & 0.851 & -0.02 & 0.09 \\
  6.5 & 0.308 & 0.1 & -0.07 & 0.824 & -0.02 & 0.09 \\
  7   & 0.252 & 0.08 & -0.06 & 0.801 & -0.02 & 0.09 \\
  7.5 & 0.206 & 0.07 & -0.05 & 0.783 & -0.03 & 0.09 \\
  8   & 0.170 & 0.05 & -0.04 & 0.777 & -0.03 & 0.09 \\
  8.5 & 0.140 & 0.04 & -0.03 & 0.761 & -0.03 & 0.09 \\
  9   & 0.115 & 0.04 & -0.03 & 0.751 & -0.03 & 0.08 \\
  9.5 & 0.0949 & 0.03 & -0.02 & 0.749 & -0.02 & 0.07 \\
 10   & 0.0784 & 0.02 & -0.02 & 0.753 & -0.02 & 0.05\\
\hline
&\multicolumn{3}{|c}{ LL+}&\multicolumn{3}{|c}{ NLL+}\\
$Y$& $\sqrt{s_0}=35$ & $\sqrt{s_0}=70$ & $\sqrt{s_0}=17.5$ 
& $\sqrt{s_0}=35$ & $\sqrt{s_0}=70$ & $\sqrt{s_0}=17.5$ \\
\hline
  6   & 0.566 & 0.08 & -0.07 & 0.940 & -0.04 & 0.1 \\
  6.5 & 0.514 & 0.07 & -0.07 & 0.919 & -0.04 & 0.1 \\
  7   & 0.465 & 0.07 & -0.06 & 0.902 & -0.04 & 0.1 \\
  7.5 & 0.421 & 0.06 & -0.06 & 0.893 & -0.05 & 0.1 \\
  8   & 0.381 & 0.06 & -0.05 & 0.901 & -0.05 & 0.1 \\
  8.5 & 0.344 & 0.05 & -0.05 & 0.895 & -0.05 & 0.1 \\
  9   & 0.311 & 0.05 & -0.04 & 0.898 & -0.05 & 0.09 \\
  9.5 & 0.280 & 0.04 & -0.04 & 0.908 & -0.05 & 0.05 \\
 10   & 0.253 & 0.04 & -0.03 & 0.908 & -0.03 & -0.04
\end{tabular}
\caption{tabled values for Fig.~\ref{fig:c1c03535_s0}.}
  \label{tab:c1c03535_s0}
\end{table}

\begin{table}[h]
  \centering
 \tiny\begin{tabular}{l|l|l|l|l|l|l}
&\multicolumn{3}{|c}{ NLL}&\multicolumn{3}{|c}{ NLL+}\\
$Y$& { PDF0} & { PDF+} & { PDF-}
& { PDF0} & { PDF+} & { PDF-}\\
\hline
  6   & 0.851 & -0.006 & 0.002 & 0.940 & -0.007 & 0.003 \\
  6.5 & 0.824 & -0.006 & 0.003 & 0.919 & -0.007 & 0.004 \\
  7   & 0.801 & -0.005 & 0.004 & 0.902 & -0.007 & 0.005 \\
  7.5 & 0.783 & -0.005 & 0.003 & 0.893 & -0.006 & 0.004 \\
  8   & 0.777 & -0.005 & 0.003 & 0.901 & -0.007 & 0.005 \\
  8.5 & 0.761 & -0.005 & 0.004 & 0.895 & -0.007 & 0.005 \\
  9   & 0.751 & -0.005 & 0.003 & 0.898 & -0.007 & 0.004 \\
  9.5 & 0.749 & -0.006 & 0.003 & 0.908 & -0.008 & 0.004 \\
 10   & 0.753 & -0.007 & 0.004 & 0.908 & -0.006 & 0.003
\end{tabular}
\caption{tabled values for Fig.~\ref{fig:c1c03535_pdf}.}
  \label{tab:c1c03535_pdf}
\end{table}

\begin{table}[h]
  \centering
  \tiny\begin{tabular}{l|l|l|l|l|l|l}
&\multicolumn{3}{|c}{ LL}&\multicolumn{3}{|c}{ NLL}\\
$Y$& { MC0} & { MC+} & { MC-}
& { MC0} & { MC+} & { MC-}\\
\hline
  6   & 0.258 & 0.003 & -0.003 & 0.512 & 0.006 & -0.006 \\
  6.5 & 0.195 & 0.001 & -0.001 & 0.471 & 0.004 & -0.004 \\
  7   & 0.147 & 0.0006 & -0.0006 & 0.434 & 0.003 & -0.003 \\
  7.5 & 0.111 & 0.0004 & -0.0004 & 0.404 & 0.002 & -0.002 \\
  8   & 0.0836 & 0.0002 & -0.0002 & 0.383 & 0.002 & -0.002 \\
  8.5 & 0.0633 & 0.0002 & -0.0002 & 0.358 & 0.003 & -0.003 \\
  9   & 0.0481 & 0.0002 & -0.0002 & 0.337 & 0.003 & -0.003 \\
  9.5 & 0.0365 & 0.0002 & -0.0002 & 0.324 & 0.005 & -0.005 \\
 10   & 0.0278 & 0.0002 & -0.0002 & 0.317 & 0.009 & -0.009\\
\hline
&\multicolumn{3}{|c}{ LL+}&\multicolumn{3}{|c}{ NLL+}\\
$Y$& { MC0} & { MC+} & { MC-}
& { MC0} & { MC+} & { MC-}\\
\hline
  6   & 0.317 & 0.002 & -0.002 & 0.566 & 0.008 & -0.008 \\
  6.5 & 0.264 & 0.001 & -0.001 & 0.525 & 0.005 & -0.005 \\
  7   & 0.220 & 0.0008 & -0.0008 & 0.489 & 0.004 & -0.004 \\
  7.5 & 0.183 & 0.0005 & -0.0005 & 0.461 & 0.003 & -0.003 \\
  8   & 0.152 & 0.0004 & -0.0004 & 0.445 & 0.004 & -0.004 \\
  8.5 & 0.126 & 0.0004 & -0.0004 & 0.421 & 0.004 & -0.004 \\
  9   & 0.105 & 0.0004 & -0.0004 & 0.403 & 0.006 & -0.006 \\
  9.5 & 0.0870 & 0.0004 & -0.0004 & 0.393 & 0.008 & -0.008 \\
 10   & 0.0721 & 0.0005 & -0.0005 & 0.382 & 0.02 & -0.02
\end{tabular}
\caption{tabled values for Fig.~\ref{fig:c2c03535}.}
  \label{tab:c2c03535}
\end{table}

\begin{table}[h]
  \centering
  \tiny\begin{tabular}{l|l|l|l|l|l|l}
&\multicolumn{3}{|c}{ LL}&\multicolumn{3}{|c}{ NLL}\\
$Y$ & $\mu=35$ & $\mu=70$ & $\mu=17.5$ 
    & $\mu=35$ & $\mu=70$ & $\mu=17.5$ \\
\hline
  6   & 0.258 & 0.1 & -0.1 & 0.512 & -0.02 & 0.03 \\
  6.5 & 0.195 & 0.1 & -0.08 & 0.471 & -0.03 & 0.04 \\
  7   & 0.147 & 0.08 & -0.06 & 0.434 & -0.03 & 0.05 \\
  7.5 & 0.111 & 0.07 & -0.05 & 0.404 & -0.04 & 0.06 \\
  8   & 0.0836 & 0.05 & -0.04 & 0.383 & -0.04 & 0.08 \\
  8.5 & 0.0633 & 0.04 & -0.03 & 0.358 & -0.05 & 0.08 \\
  9   & 0.0481 & 0.03 & -0.02 & 0.337 & -0.05 & 0.09 \\
  9.5 & 0.0365 & 0.03 & -0.02 & 0.324 & -0.05 & 0.09 \\
 10   & 0.0278 & 0.02 & -0.01 & 0.317 & -0.06 & 0.07\\
\hline
&\multicolumn{3}{|c}{ LL+}&\multicolumn{3}{|c}{ NLL+}\\
$Y$ & $\mu=35$ & $\mu=70$ & $\mu=17.5$ 
    & $\mu=35$ & $\mu=70$ & $\mu=17.5$ \\
\hline
  6   & 0.317 & 0.03 & -0.02 & 0.566 & -0.02 & 0.04 \\
  6.5 & 0.264 & 0.03 & -0.02 & 0.525 & -0.02 & 0.05 \\
  7   & 0.220 & 0.02 & -0.02 & 0.489 & -0.03 & 0.06 \\
  7.5 & 0.183 & 0.02 & -0.02 & 0.461 & -0.04 & 0.08 \\
  8   & 0.152 & 0.02 & -0.01 & 0.445 & -0.05 & 0.1 \\
  8.5 & 0.126 & 0.02 & -0.01 & 0.421 & -0.06 & 0.09 \\
  9   & 0.105 & 0.01 & -0.01 & 0.403 & -0.06 & 0.07 \\
  9.5 & 0.0870 & 0.01 & -0.008 & 0.393 & -0.07 & 0.01 \\
 10   & 0.0721 & 0.01 & -0.007 & 0.382 & -0.07 & -0.09
\end{tabular}
\caption{tabled values for Fig.~\ref{fig:c2c03535_mu}.}
  \label{tab:c2c03535_mu}
\end{table}

\begin{table}[h]
  \centering
  \tiny\begin{tabular}{l|l|l|l|l|l|l}
&\multicolumn{3}{|c}{ LL}&\multicolumn{3}{|c}{ NLL}\\
$Y$ & $\sqrt{s_0}=35$ & $\sqrt{s_0}=70$ & $\sqrt{s_0}=17.5 $
    & $\sqrt{s_0}=35$ & $\sqrt{s_0}=70$ & $\sqrt{s_0}=17.5$ \\
\hline
  6   & 0.258 & 0.1 & -0.08 & 0.512 & -0.002 & 0.08 \\
  6.5 & 0.195 & 0.09 & -0.06 & 0.471 & -0.005 & 0.07 \\
  7   & 0.147 & 0.07 & -0.05 & 0.434 & -0.005 & 0.07 \\
  7.5 & 0.111 & 0.05 & -0.04 & 0.404 & -0.005 & 0.06 \\
  8   & 0.0836 & 0.04 & -0.03 & 0.383 & -0.004 & 0.06 \\
  8.5 & 0.0633 & 0.03 & -0.02 & 0.358 & -0.003 & 0.05 \\
  9   & 0.0481 & 0.02 & -0.02 & 0.337 & -0.002 & 0.05 \\
  9.5 & 0.0365 & 0.02 & -0.01 & 0.324 & -0.001 & 0.04 \\
 10   & 0.0278 & 0.01 & -0.009 & 0.317 & 0.001 & 0.03\\
\hline
&\multicolumn{3}{|c}{ LL+}&\multicolumn{3}{|c}{ NLL+}\\
$Y$& $\sqrt{s_0}=35$ & $\sqrt{s_0}=70$ & $\sqrt{s_0}=17.5$ 
& $\sqrt{s_0}=35$ & $\sqrt{s_0}=70$ & $\sqrt{s_0}=17.5$ \\
\hline
  6   & 0.317 & 0.09 & -0.07 & 0.566 & -0.009 & 0.09 \\
  6.5 & 0.264 & 0.07 & -0.06 & 0.525 & -0.01 & 0.09 \\
  7   & 0.220 & 0.06 & -0.05 & 0.489 & -0.01 & 0.09 \\
  7.5 & 0.183 & 0.05 & -0.04 & 0.461 & -0.01 & 0.08 \\
  8   & 0.152 & 0.04 & -0.03 & 0.445 & -0.02 & 0.08 \\
  8.5 & 0.126 & 0.04 & -0.03 & 0.421 & -0.02 & 0.07 \\
  9   & 0.105 & 0.03 & -0.02 & 0.403 & -0.01 & 0.06 \\
  9.5 & 0.0870 & 0.03 & -0.02 & 0.393 & -0.01 & 0.03 \\
 10   & 0.0721 & 0.02 & -0.02 & 0.382 & -0.001 & -0.007
\end{tabular}
\caption{tabled values for Fig.~\ref{fig:c2c03535_s0}.}
  \label{tab:c2c03535_s0}
\end{table}

\clearpage

\begin{table}[h]
  \centering
  \tiny\begin{tabular}{l|l|l|l|l|l|l}
&\multicolumn{3}{|c}{ LL}&\multicolumn{3}{|c}{ NLL}\\
$Y$& { MC0} & { MC+} & { MC-}
& { MC0} & { MC+} & { MC-}\\
\hline
  6   & 0.686 & 0.009 & -0.009 & 0.602 & 0.006 & -0.006 \\
  6.5 & 0.632 & 0.005 & -0.005 & 0.571 & 0.004 & -0.004 \\
  7   & 0.582 & 0.003 & -0.003 & 0.542 & 0.003 & -0.003 \\
  7.5 & 0.536 & 0.002 & -0.002 & 0.516 & 0.002 & -0.002 \\
  8   & 0.493 & 0.002 & -0.002 & 0.493 & 0.002 & -0.002 \\
  8.5 & 0.454 & 0.002 & -0.002 & 0.470 & 0.002 & -0.002 \\
  9   & 0.418 & 0.002 & -0.002 & 0.449 & 0.002 & -0.002 \\
  9.5 & 0.385 & 0.002 & -0.002 & 0.432 & 0.003 & -0.003 \\
 10   & 0.354 & 0.003 & -0.003 & 0.421 & 0.005 & -0.005\\
\hline
&\multicolumn{3}{|c}{ LL+}\\
$Y$& { MC0} & { MC+} & { MC-}\\
\hline
  6   & 0.560 & 0.005 & -0.005\\
  6.5 & 0.515 & 0.003 & -0.003\\
  7   & 0.473 & 0.002 & -0.002\\
  7.5 & 0.435 & 0.002 & -0.002\\
  8   & 0.400 & 0.002 & -0.002\\
  8.5 & 0.368 & 0.002 & -0.002\\
  9   & 0.338 & 0.002 & -0.002\\
  9.5 & 0.310 & 0.002 & -0.002\\
 10   & 0.285 & 0.003 & -0.003
\end{tabular}
\caption{tabled values for Fig.~\ref{fig:c2c13535}.}
  \label{tab:c2c13535}
\end{table}

\begin{table}[h]
  \centering
  \tiny\begin{tabular}{l|l|l|l|l|l|l}
&\multicolumn{3}{|c}{ LL}&\multicolumn{3}{|c}{ NLL}\\
$Y$ & $\mu=35$ & $\mu=70$ & $\mu=17.5$ 
    & $\mu=35$ & $\mu=70$ & $\mu=17.5$ \\
\hline
  6   & 0.686 & 0.07 & -0.09 & 0.602 & 0.02 & -0.007 \\
  6.5 & 0.632 & 0.08 & -0.09 & 0.571 & 0.02 & -0.004 \\
  7   & 0.582 & 0.08 & -0.09 & 0.542 & 0.01 & 0.0009 \\
  7.5 & 0.536 & 0.08 & -0.09 & 0.516 & 0.01 & 0.007 \\
  8   & 0.493 & 0.08 & -0.08 & 0.493 & 0.007 & 0.01 \\
  8.5 & 0.454 & 0.08 & -0.08 & 0.470 & 0.004 & 0.02 \\
  9   & 0.418 & 0.07 & -0.08 & 0.449 & 0.0007 & 0.03 \\
  9.5 & 0.385 & 0.07 & -0.08 & 0.432 & -0.003 & 0.04 \\
 10   & 0.354 & 0.07 & -0.07 & 0.421 & -0.009 & 0.05\\
\hline
&\multicolumn{3}{|c}{ LL+}\\
$Y$ & $\mu=35$ & $\mu=70$ & $\mu=17.5$ \\
\hline
  6   & 0.560 & 0.03 & -0.02\\
  6.5 & 0.515 & 0.03 & -0.02\\
  7   & 0.473 & 0.03 & -0.02\\
  7.5 & 0.435 & 0.03 & -0.02\\
  8   & 0.400 & 0.02 & -0.02\\
  8.5 & 0.368 & 0.02 & -0.02\\
  9   & 0.338 & 0.02 & -0.01\\
  9.5 & 0.310 & 0.02 & -0.01\\
 10   & 0.285 & 0.02 & -0.01
\end{tabular}
\caption{tabled values for Fig.~\ref{fig:c2c13535_mu}.}
  \label{tab:c2c13535_mu}
\end{table}

\begin{table}[h]
  \centering
  \tiny\begin{tabular}{l|l|l|l|l|l|l}
&\multicolumn{3}{|c}{ LL}&\multicolumn{3}{|c}{ NLL}\\
$Y$ & $\sqrt{s_0}=35$ & $\sqrt{s_0}=70$ & $\sqrt{s_0}=17.5 $
    & $\sqrt{s_0}=35$ & $\sqrt{s_0}=70$ & $\sqrt{s_0}=17.5$ \\
\hline
  6   & 0.686 & 0.08 & -0.07 & 0.602 & 0.01 & 0.03 \\
  6.5 & 0.632 & 0.08 & -0.07 & 0.571 & 0.01 & 0.02 \\
  7   & 0.582 & 0.07 & -0.06 & 0.542 & 0.01 & 0.02 \\
  7.5 & 0.536 & 0.07 & -0.06 & 0.516 & 0.01 & 0.02 \\
  8   & 0.493 & 0.06 & -0.05 & 0.493 & 0.01 & 0.02 \\
  8.5 & 0.454 & 0.06 & -0.05 & 0.470 & 0.01 & 0.02 \\
  9   & 0.418 & 0.05 & -0.05 & 0.449 & 0.01 & 0.01 \\
  9.5 & 0.385 & 0.05 & -0.04 & 0.432 & 0.01 & 0.01 \\
 10   & 0.354 & 0.04 & -0.04 & 0.421 & 0.01 & 0.01\\
\hline
&\multicolumn{3}{|c}{ LL+}\\
$Y$& $\sqrt{s_0}=35$ & $\sqrt{s_0}=70$ & $\sqrt{s_0}=17.5$ \\
\hline
  6   & 0.560 & 0.07 & -0.06\\
  6.5 & 0.515 & 0.06 & -0.06\\
  7   & 0.473 & 0.06 & -0.05\\
  7.5 & 0.435 & 0.05 & -0.0\\
  8   & 0.400 & 0.05 & -0.04\\
  8.5 & 0.368 & 0.05 & -0.04\\
  9   & 0.338 & 0.04 & -0.04\\
  9.5 & 0.310 & 0.04 & -0.03\\
 10   & 0.285 & 0.04 & -0.03
\end{tabular}
\caption{tabled values for Fig.~\ref{fig:c2c13535_s0}.}
  \label{tab:c2c13535_s0}
\end{table}

\begin{table}[h]
  \centering
  \tiny\begin{tabular}{l|l|l|l|l|l|l}
    & \multicolumn{3}{|c}{ LL} & \multicolumn{3}{|c}{ NLL}\\
$Y$ & MC0  &  MC+ &  MC-       &  MC0  &  MC+ &  MC- \\ 
\hline 
  6   & 0.287 & 0.002 & -0.002 & 0.258 & 0.002 & -0.002 \\
  6.5 & 0.316 & 0.002 & -0.002 & 0.310 & 0.002 & -0.002 \\
  7   & 0.257 & 0.0009 & -0.0009 & 0.268 & 0.001 & -0.001 \\
  7.5 & 0.176 & 0.0005 & -0.0005 & 0.192 & 0.0007 & -0.0007 \\
  8   & 0.100 & 0.0003 & -0.0003 & 0.112 & 0.0004 & -0.0004 \\
  8.5 & 0.0420 & 0.0001 & -0.0001 & 0.0480 & 0.0002 & -0.0002 \\
  9   & 0.0154 & 0.00005 & -0.00005 & 0.0178 & 0.00007 & -0.00007 \\
  9.5 & 0.00450 & 0.00002 & -0.00002 & 0.00516 & 0.00003 & -0.00003 \\
 10   & 0.000810 & 0.000005 & -0.000005 & 0.000897 & 0.000008 & -0.000008\\
\hline
    & \multicolumn{3}{|c}{ LL+}\\
$Y$ & MC0  &  MC+ &  MC-       \\
\hline
  6   & 0.240 & 0.001 & -0.001\\
  6.5 & 0.282 & 0.001 & -0.001\\
  7   & 0.239 & 0.0008 & -0.0008\\
  7.5 & 0.170 & 0.0005 & -0.0005\\
  8   & 0.0996 & 0.0003 & -0.0003\\
  8.5 & 0.0426 & 0.0001 & -0.0001\\
  9   & 0.0159 & 0.00005 & -0.00005\\
  9.5 & 0.00472 & 0.00002 & -0.00002\\
 10   & 0.000861 & 0.000005 & -0.000005
  \end{tabular}
  \caption{tabled values for Fig.~\ref{fig:c13535}, \ref{fig:c13535rel_pdf_cuba}.}
  \label{tab:c13535}
\end{table}

\begin{table}[h]
  \centering
  \tiny\begin{tabular}{l|l|l|l|l|l|l}
&\multicolumn{3}{|c}{ LL}&\multicolumn{3}{|c}{ NLL}\\
$Y$ & $\mu=35$ & $\mu=70$ & $\mu=17.5$ 
    & $\mu=35$ & $\mu=70$ & $\mu=17.5$ \\
\hline
  6   & 0.287 & -0.01 & 0.02 & 0.258 & -0.008 & 0.001 \\
  6.5 & 0.316 & -0.03 & 0.04 & 0.310 & -0.01 & 0.004 \\
  7   & 0.257 & -0.03 & 0.05 & 0.268 & -0.01 & 0.006 \\
  7.5 & 0.176 & -0.03 & 0.04 & 0.192 & -0.01 & 0.005 \\
  8   & 0.100 & -0.02 & 0.03 & 0.112 & -0.009 & 0.003 \\
  8.5 & 0.0420 & -0.01 & 0.02 & 0.0480 & -0.004 & 0.001 \\
  9   & 0.0154 & -0.004 & 0.007 & 0.0178 & -0.002 & 0.0004 \\
  9.5 & 0.00450 & -0.001 & 0.002 & 0.00516 & -0.0005 & 0.00006 \\
 10   & 0.000810 & -0.0003 & 0.0005 & 0.000897 & -0.0001 & -0.000007\\
\hline
&\multicolumn{3}{|c}{ LL+}\\
$Y$ & $\mu=35$ & $\mu=70$ & $\mu=17.5$\\
\hline
  6   & 0.240 & -0.05 & 0.08\\
  6.5 & 0.282 & -0.07 & 0.1\\
  7   & 0.239 & -0.06 & 0.1\\
  7.5 & 0.170 & -0.05 & 0.08\\
  8   & 0.0996 & -0.03 & 0.06\\
  8.5 & 0.0426 & -0.01 & 0.03\\
  9   & 0.0159 & -0.006 & 0.01\\
  9.5 & 0.00472 & -0.002 & 0.003\\
 10   & 0.000861 & -0.0003 & 0.0007\\
\end{tabular}
\caption{tabled values for left figure of Fig.~\ref{fig:c13535rel_mu_s0}.}
  \label{tab:c13535_mu}
\end{table}

\begin{table}[h]
  \centering
  \tiny\begin{tabular}{l|l|l|l|l|l|l}
&\multicolumn{3}{|c}{ LL}&\multicolumn{3}{|c}{ NLL}\\
$Y$ & $\sqrt{s_0}=35$ & $\sqrt{s_0}=70$ & $\sqrt{s_0}=17.5$ 
    & $\sqrt{s_0}=35$ & $\sqrt{s_0}=70$ & $\sqrt{s_0}=17.5$ \\
\hline
 6 & 0.287 & 0.09 & -0.05 & 0.258 & 0.004 & 0.02 \\
 6.5 & 0.316 & 0.07 & -0.05 & 0.310 & 0.002 & 0.03 \\
 7 & 0.257 & 0.05 & -0.03 & 0.268 & 0.0008 & 0.02 \\
 7.5 & 0.176 & 0.03 & -0.02 & 0.192 & 0.0005 & 0.01 \\
 8 & 0.100 & 0.01 & -0.009 & 0.112 & 0.0003 & 0.007 \\
 8.5 & 0.0420 & 0.005 & -0.003 & 0.0480 & 0.0002 & 0.003 \\
 9 & 0.0154 & 0.001 & -0.001 & 0.0178 & 0.00007 & 0.001 \\
 9.5 & 0.00450 & 0.0004 & -0.0003 & 0.00516 & 0.00002 & 0.0003 \\
 10 & 0.000810 & 0.00006 & -0.00005 & 0.000897 & 0.000003 & 0.00005\\
\hline
&\multicolumn{3}{|c}{ LL+}\\
$Y$ & $\sqrt{s_0}=35$ & $\sqrt{s_0}=70$ & $\sqrt{s_0}=17.5$\\
\hline
 6 & 0.240 & 0.04 & -0.03\\
 6.5 & 0.282 & 0.04 & -0.03\\
 7 & 0.239 & 0.03 & -0.02\\
 7.5 & 0.170 & 0.02 & -0.01\\
 8 & 0.0996 & 0.008 & -0.007\\
 8.5 & 0.0426 & 0.003 & -0.003\\
 9 & 0.0159 & 0.001 & -0.0009\\
 9.5 & 0.00472 & 0.0003 & -0.0002\\
 10 & 0.000861 & 0.00005 & -0.00004
\end{tabular}
\caption{tabled values for right figure of Fig.~\ref{fig:c13535rel_mu_s0}.}
  \label{tab:c13535_s0}
\end{table}

\begin{table}[h]
  \centering
  \tiny\begin{tabular}{l|l|l|l|l|l|l}
&\multicolumn{3}{|c}{ LL}&\multicolumn{3}{|c}{ NLL}\\
$Y$& { PDF0} & { PDF+} & { PDF-}
& { PDF0} & { PDF+} & { PDF-}\\
\hline
  6   & 0.287 & 0.02 & -0.02 & 0.258 & 0.01 & -0.02 \\
  6.5 & 0.316 & 0.02 & -0.02 & 0.310 & 0.02 & -0.02 \\
  7   & 0.257 & 0.02 & -0.02 & 0.268 & 0.02 & -0.02 \\
  7.5 & 0.176 & 0.02 & -0.02 & 0.192 & 0.02 & -0.02 \\
  8   & 0.100 & 0.01 & -0.01 & 0.112 & 0.01 & -0.01 \\
  8.5 & 0.0420 & 0.006 & -0.005 & 0.0480 & 0.007 & -0.006 \\
  9   & 0.0154 & 0.003 & -0.002 & 0.0178 & 0.003 & -0.002 \\
  9.5 & 0.00450 & 0.0009 & -0.0007 & 0.00516 & 0.001 & -0.0008 \\
 10   & 0.000810 & 0.0002 & -0.0001 & 0.000897 & 0.0002 & -0.0002\\
\hline
&\multicolumn{3}{|c}{ LL+}\\
$Y$& { PDF0} & { PDF+} & { PDF-}\\
\hline
  6   & 0.240 & 0.01 & -0.02 \\
  6.5 & 0.282 & 0.02 & -0.02 \\
  7   & 0.239 & 0.02 & -0.02 \\
  7.5 & 0.170 & 0.02 & -0.02 \\
  8   & 0.0996 & 0.01 & -0.01 \\
  8.5 & 0.0426 & 0.006 & -0.005 \\
  9   & 0.0159 & 0.003 & -0.002 \\
  9.5 & 0.00472 & 0.0009 & -0.0008 \\
 10   & 0.000861 & 0.0002 & -0.0002
\end{tabular}
\caption{tabled values for left figure of Fig.~\ref{fig:c13535rel_pdf_cuba}.}
  \label{tab:c13535_pdf}
\end{table}

\begin{table}[h]
  \centering
  \tiny\begin{tabular}{l|l|l|l|l|l|l}
    & \multicolumn{3}{|c}{ LL} & \multicolumn{3}{|c}{ NLL}\\
$Y$ & MC0  &  MC+ &  MC-       &  MC0  &  MC+ &  MC- \\ 
\hline 
  6   & 0.197 & 0.002 & -0.002 & 0.155 & 0.001 & -0.001 \\
  6.5 & 0.200 & 0.001 & -0.001 & 0.177 & 0.0008 & -0.0008 \\
  7   & 0.149 & 0.0006 & -0.0006 & 0.145 & 0.0005 & -0.0005 \\
  7.5 & 0.0942 & 0.0003 & -0.0003 & 0.0991 & 0.0003 & -0.0003 \\
  8   & 0.0495 & 0.0001 & -0.0001 & 0.0554 & 0.0001 & -0.0001 \\
  8.5 & 0.0191 & 0.00006 & -0.00006 & 0.0226 & 0.00006 & -0.00006 \\
  9   & 0.00644 & 0.00002 & -0.00002 & 0.00798 & 0.00002 & -0.00002 \\
  9.5 & 0.00173 & 0.000007 & -0.000007 & 0.00223 & 0.000008 & -0.000008\\
 10 & 0.000287 & 0.000002 & -0.000002 & 0.000377 & 0.000002 & -0.000002\\
\hline
    & \multicolumn{3}{|c}{ LL+}\\
$Y$ & MC0  &  MC+ &  MC-\\
\hline
  6   & 0.134 & 0.0009 & -0.0009\\
  6.5 & 0.145 & 0.0006 & -0.0006\\
  7   & 0.113 & 0.0004 & -0.0004\\
  7.5 & 0.0738 & 0.0002 & -0.0002\\
  8   & 0.0398 & 0.0001 & -0.0001\\
  8.5 & 0.0156 & 0.00005 & -0.00005\\
  9   & 0.00537 & 0.00002 & -0.00002\\
  9.5 & 0.00146 & 0.000006 & -0.000006\\
 10 & 0.000245 & 0.00002 & -0.00002
  \end{tabular}
  \caption{tabled values for Fig.~\ref{fig:c23535}, \ref{fig:c23535rel_pdf_cuba}.}
  \label{tab:c23535}
\end{table}

\begin{table}[h]
  \centering
  \tiny\begin{tabular}{l|l|l|l|l|l|l}
&\multicolumn{3}{|c}{ LL}&\multicolumn{3}{|c}{ NLL}\\
$Y$ & $\mu=35$ & $\mu=70$ & $\mu=17.5$ 
    & $\mu=35$ & $\mu=70$ & $\mu=17.5$ \\
\hline
  6   & 0.197 & 0.01 & -0.01 & 0.155 & 0.0002 & -0.001 \\
  6.5 & 0.200 & 0.006 & -0.004 & 0.177 & -0.003 & 0.001 \\
  7   & 0.149 & -0.001 & 0.002 & 0.145 & -0.004 & 0.003 \\
  7.5 & 0.0942 & -0.004 & 0.005 & 0.0991 & -0.005 & 0.004 \\
  8   & 0.0495 & -0.004 & 0.004 & 0.0554 & -0.004 & 0.003 \\
  8.5 & 0.0191 & -0.002 & 0.002 & 0.0226 & -0.002 & 0.002 \\
  9   & 0.00644 & -0.0008 & 0.001 & 0.00798 & -0.0008 & 0.0007 \\
  9.5 & 0.00173 & -0.0003 & 0.0003 & 0.00223 & -0.0002 & 0.0002 \\
 10   & 0.000287 & -0.00005 & 0.00007 & 0.000377 & -0.00005 & 0.00004\\
\hline
&\multicolumn{3}{|c}{ LL+}\\
$Y$ & $\mu=35$ & $\mu=70$ & $\mu=17.5$\\
\hline
  6   & 0.134 & -0.02 & 0.04\\
  6.5 & 0.145 & -0.03 & 0.05\\
  7   & 0.113 & -0.03 & 0.04\\
  7.5 & 0.0738 & -0.02 & 0.03\\
  8   & 0.0398 & -0.01 & 0.02\\
  8.5 & 0.0156 & -0.004 & 0.009\\
  9   & 0.00537 & -0.002 & 0.003\\
  9.5 & 0.00146 & -0.0005 & 0.001\\
 10   & 0.000245 & -0.00009 & 0.0002
\end{tabular}
\caption{tabled values for left figure of Fig.~\ref{fig:c23535rel_mu_s0}.}
  \label{tab:c23535_mu}
\end{table}

\begin{table}[h]
  \centering
  \tiny\begin{tabular}{l|l|l|l|l|l|l}
&\multicolumn{3}{|c}{ LL}&\multicolumn{3}{|c}{ NLL}\\
$Y$ & $\sqrt{s_0}=35$ & $\sqrt{s_0}=70$ & $\sqrt{s_0}=17.5$ 
    & $\sqrt{s_0}=35$ & $\sqrt{s_0}=70$ & $\sqrt{s_0}=17.5$ \\
\hline
  6   & 0.197 & 0.09 & -0.05 & 0.155 & 0.006 & 0.02 \\
  6.5 & 0.200 & 0.08 & -0.05 & 0.177 & 0.004 & 0.02 \\
  7   & 0.149 & 0.05 & -0.03 & 0.145 & 0.003 & 0.02 \\
  7.5 & 0.0942 & 0.03 & -0.02 & 0.0991 & 0.002 & 0.01 \\
  8   & 0.0495 & 0.01 & -0.009 & 0.0554 & 0.001 & 0.006 \\
  8.5 & 0.0191 & 0.005 & -0.003 & 0.0226 & 0.0007 & 0.002 \\
  9   & 0.00644 & 0.001 & -0.001 & 0.00798 & 0.0003 & 0.0007 \\
  9.5 & 0.00173 & 0.0004 & -0.0003 & 0.00223 & 0.00008 & 0.0002 \\
 10   & 0.000287 & 0.00006 & -0.00005 & 0.000377 & 0.00001 & 0.00003\\
\hline
&\multicolumn{3}{|c}{ LL+}\\
$Y$ & $\sqrt{s_0}=35$ & $\sqrt{s_0}=70$ & $\sqrt{s_0}=17.5$\\
\hline
  6   & 0.134 & 0.04 & -0.03\\
  6.5 & 0.145 & 0.04 & -0.03\\
  7   & 0.113 & 0.03 & -0.02\\
  7.5 & 0.0738 & 0.02 & -0.01\\
  8   & 0.0398 & 0.009 & -0.007\\
  8.5 & 0.0156 & 0.003 & -0.003\\
  9   & 0.00537 & 0.001 & -0.0009\\
  9.5 & 0.00146 & 0.0003 & -0.0002\\
 10   & 0.000245 & 0.00005 & -0.00004
\end{tabular}
\caption{tabled values for right figure of Fig.~\ref{fig:c23535rel_mu_s0}.}
  \label{tab:c23535_s0}
\end{table}

\begin{table}[h]
  \centering
  \tiny\begin{tabular}{l|l|l|l|l|l|l}
&\multicolumn{3}{|c}{ LL}&\multicolumn{3}{|c}{ NLL}\\
$Y$& { PDF0} & { PDF+} & { PDF-}
& { PDF0} & { PDF+} & { PDF-}\\
\hline
 6 & 0.197 & 0.01 & -0.01 & 0.155 & 0.007 & -0.01 \\
 6.5 & 0.200 & 0.01 & -0.01 & 0.177 & 0.01 & -0.01 \\
 7 & 0.149 & 0.01 & -0.01 & 0.145 & 0.01 & -0.01 \\
 7.5 & 0.0942 & 0.009 & -0.009 & 0.0991 & 0.009 & -0.009 \\
 8 & 0.0495 & 0.006 & -0.006 & 0.0554 & 0.006 & -0.006 \\
 8.5 & 0.0191 & 0.003 & -0.002 & 0.0226 & 0.003 & -0.003 \\
 9 & 0.00644 & 0.001 & -0.0009 & 0.00798 & 0.001 & -0.001 \\
 9.5 & 0.00173 & 0.0003 & -0.0003 & 0.00223 & 0.0004 & -0.0003 \\
 10 & 0.000287 & 0.00007 & -0.00005 & 0.000377 & 0.00009 & -0.00007\\
\hline
&\multicolumn{3}{|c}{ LL+}\\
$Y$& { PDF0} & { PDF+} & { PDF-}\\
\hline
 6 & 0.134 & 0.04 & -0.03 \\
 6.5 & 0.145 & 0.04 & -0.03 \\
 7 & 0.113 & 0.03 & -0.02 \\
 7.5 & 0.0738 & 0.02 & -0.01\\
 8 & 0.0398 & 0.009 & -0.007 \\
 8.5 & 0.0156 & 0.003 & -0.003 \\
 9 & 0.00537 & 0.001 & -0.0009 \\
 9.5 & 0.00146 & 0.0003 & -0.0002 \\
 10 & 0.000245 & 0.00005 & -0.00004
\end{tabular}
\caption{tabled values for left figure of Fig.~\ref{fig:c23535rel_pdf_cuba}.}
  \label{tab:c23535_pdf}
\end{table}

\begin{table}[h]
  \centering
  \tiny\begin{tabular}{l|l|l|l|l}
& LL &\multicolumn{3}{|c}{ NLL}\\
$Y$& 
& MC0  &  MC+ &  MC- \\ 
\hline 
 6 & 0.162 &  0.0699 & 0.0006 & -0.0006 \\
 6.5 & 0.200 & 0.0806 & 0.0005 & -0.0005 \\
 7 & 0.179 &  0.0651 & 0.0003 & -0.0003 \\
 7.5 & 0.128 &  0.0405 & 0.0002 & -0.0002 \\
 8 & 0.0689 &  0.0182 & 0.0001 & -0.0001 \\
 8.5 & 0.0288 & 0.00638 & 0.00005 & -0.00005 \\
 9 & 0.00973 &  0.00173 & 0.00002 & -0.00002 \\
 9.5 & 0.00227 &  0.000305 & 0.00007 & -0.000007\\
 10 & 0.000258 &  0.0000267 & 0.000001 & -0.000001\\
\hline
& LL+ &\multicolumn{3}{|c}{NLL+}\\
$Y$& 
& MC0  &  MC+ &  MC- \\ 
\hline
 6 & 0.0940 & 0.0643 & 0.0006 & -0.0006 \\
 6.5 & 0.113 & 0.0736 & 0.0005 & -0.0005 \\
 7 & 0.0966 & 0.0589 & 0.0004 & -0.0004 \\
 7.5 & 0.0654 &  0.0362 & 0.0002 & -0.0002 \\
 8 & 0.0332 &  0.0160 & 0.0001 & -0.0001 \\
 8.5 & 0.0131 &  0.00553 & 0.00006 & -0.00006 \\
 9 & 0.00413 & 0.00148 & 0.00003 & -0.00003 \\
 9.5 & 0.000900 & 0.000260 & 0.000009 & -0.000009\\
 10 & 0.0000959  & 0.0000253 & 0.000002 & -0.000002
  \end{tabular}
  \caption{tabled values for Fig.~\ref{fig:c05050}.}
  \label{tab:c05050}
\end{table}

\begin{table}[h]
  \centering
  \tiny\begin{tabular}{l|l|l|l|l|l|l}
&\multicolumn{3}{|c}{ LL}&\multicolumn{3}{|c}{ NLL}\\
$Y$ & $\mu=50$ & $\mu=100$ & $\mu=25$ 
    & $\mu=50$ & $\mu=100$ & $\mu=25$ \\
\hline
 6 & 0.162 & -0.04 & 0.08 & 0.0699 & 0.002 & 0.01 \\
 6.5 & 0.200 & -0.06 & 0.1 & 0.0806 & 0.003 & 0.003 \\
 7 & 0.179 & -0.06 & 0.1 & 0.0651 & 0.002 & -0.001 \\
 7.5 & 0.128 & -0.05 & 0.1 & 0.0405 & 0.002 & -0.003 \\
 8 & 0.0689 & -0.03 & 0.07 & 0.0182 & 0.001 & -0.002 \\
 8.5 & 0.0288 & -0.01 & 0.03 & 0.00638 & 0.0004 & -0.0009 \\
 9 & 0.00973 & -0.005 & 0.01 & 0.00173 & 0.0001 & -0.0002 \\
 9.5 & 0.00227 & -0.001 & 0.003 & 0.000305 & 0.00003 & -0.00003 \\
 10 & 0.000258 & -0.0001 & 0.0004 & 0.0000267 & 0.0000009 & 0.000006\\
\hline
&\multicolumn{3}{|c}{ LL+}&\multicolumn{3}{|c}{ NLL+}\\
$Y$ & $\mu=50$ & $\mu=100$ & $\mu=25$ 
    & $\mu=50$ & $\mu=100$ & $\mu=25$ \\
\hline
 6 & 0.0940 & -0.02 & 0.04 & 0.0643 & 0.002 & 0.01 \\
 6.5 & 0.113 & -0.03 & 0.05 & 0.0736 & 0.002 & 0.003 \\
 7 & 0.0966 & -0.03 & 0.05 & 0.0589 & 0.001 & -0.002 \\
 7.5 & 0.0654 & -0.02 & 0.04 & 0.0362 & 0.001 & -0.003 \\
 8 & 0.0332 & -0.01 & 0.02 & 0.0160 & 0.0009 & -0.002 \\
 8.5 & 0.0131 & -0.005 & 0.009 & 0.00553 & 0.0004 & -0.0006 \\
 9 & 0.00413 & -0.002 & 0.003 & 0.00148 & 0.0001 & 0.00001 \\
 9.5 & 0.000900 & -0.0004 & 0.0008 & 0.000260 & 0.00002 & 0.0001 \\
 10 & 0.0000959 & -0.00004 & 0.00009 & 0.0000253 & -0.000002 &    0.00004
\end{tabular}
\caption{tabled values for left figure of Fig.~\ref{fig:c05050rel_mu_s0}.}
  \label{tab:c05050_mu}
\end{table}

\begin{table}[h]
  \centering
  \tiny\begin{tabular}{l|l|l|l|l|l|l}
&\multicolumn{3}{|c}{ LL}&\multicolumn{3}{|c}{ NLL}\\
$Y$ & $\sqrt{s_0}=50$ & $\sqrt{s_0}=100$ & $\sqrt{s_0}=25$ 
    & $\sqrt{s_0}=50$ & $\sqrt{s_0}=100$ & $\sqrt{s_0}=25$ \\
\hline
 6 & 0.162 & 0.005 & 0.009 & 0.0699 & 0.002 & -0.0001 \\
 6.5 & 0.200 & -0.006 & 0.02 & 0.0806 & 0.002 & -0.001 \\
 7 & 0.179 & -0.01 & 0.02 & 0.0651 & 0.002 & -0.002 \\
 7.5 & 0.128 & -0.01 & 0.02 & 0.0405 & 0.002 & -0.001 \\
 8 & 0.0689 & -0.008 & 0.01 & 0.0182 & 0.001 & -0.0006 \\
 8.5 & 0.0288 & -0.004 & 0.005 & 0.00638 & 0.0004 & -0.0002 \\
 9 & 0.00973 & -0.001 & 0.002 & 0.00173 & 0.0001 & -0.00003 \\
 9.5 & 0.00227 & -0.0003 & 0.0004 & 0.000305 & 0.00001 & 0.000008\\
 10 & 0.000258 & -0.00004 & 0.00005 & 0.0000267 & -0.0000007 & 0.000003\\
\hline
&\multicolumn{3}{|c}{ LL+}&\multicolumn{3}{|c}{ NLL+}\\
$Y$ & $\sqrt{s_0}=50$ & $\sqrt{s_0}=100$ & $\sqrt{s_0}=25$ 
    & $\sqrt{s_0}=50$ & $\sqrt{s_0}=100$ & $\sqrt{s_0}=25$ \\
\hline
 6 & 0.0940 & 0.003 & 0.0007 & 0.0643 & 0.003 & -0.0005 \\
 6.5 & 0.113 & 0.0002 & 0.003 & 0.0736 & 0.003 & -0.002 \\
 7 & 0.0966 & -0.002 & 0.004 & 0.0589 & 0.003 & -0.002 \\
 7.5 & 0.0654 & -0.002 & 0.004 & 0.0362 & 0.003 & -0.001 \\
 8 & 0.0332 & -0.002 & 0.002 & 0.0160 & 0.001 & -0.0006 \\
 8.5 & 0.0131 & -0.0008 & 0.001 & 0.00553 & 0.0005 & -0.0001 \\
 9 & 0.00413 & -0.0003 & 0.0003 & 0.00148 & 0.0001 & 0.00003 \\
 9.5 & 0.000900 & -0.00007 & 0.00008 & 0.000260 & 0.00001 & 0.00003 \\
 10 & 0.0000959 & -0.000007 & 0.000009  & 0.0000253 & -0.000003 & 0.000009
\end{tabular}
\caption{tabled values for right figure of Fig.~\ref{fig:c05050rel_mu_s0}.}
  \label{tab:c05050_s0}
\end{table}

\begin{table}[h]
  \centering
  \tiny\begin{tabular}{l|l|l|l|l|l|l}
&\multicolumn{3}{|c}{ LL}&\multicolumn{3}{|c}{ NLL}\\
$Y$& { MC0} & { MC+} & { MC-}
& { MC0} & { MC+} & { MC-}\\
\hline
 6 & 0.435 & 0.004 & -0.004 & 0.886 & 0.009 & -0.009 \\
 6.5 & 0.360 & 0.002 & -0.002 & 0.865 & 0.006 & -0.006 \\
 7 & 0.298 & 0.001 & -0.001 & 0.849 & 0.005 & -0.005 \\
 7.5 & 0.246 & 0.0008 & -0.0008 & 0.842 & 0.005 & -0.005 \\
 8 & 0.204 & 0.0006 & -0.0006 & 0.849 & 0.006 & -0.006 \\
 8.5 & 0.170 & 0.0005 & -0.0005 & 0.850 & 0.008 & -0.008 \\
 9 & 0.141 & 0.0005 & -0.0005 & 0.866 & 0.01 & -0.01 \\
 9.5 & 0.118 & 0.0006 & -0.0006 & 0.899 & 0.02 & -0.02 \\
 10 & 0.0984 & 0.0008 & -0.0008 & 0.878 & 0.05 & -0.05\\
\hline
&\multicolumn{3}{|c}{ LL+}&\multicolumn{3}{|c}{ NLL+}\\
$Y$& { MC0} & { MC+} & { MC-}
& { MC0} & { MC+} & { MC-}\\
\hline
 6 & 0.597 & 0.003 & -0.003 & 0.963 & 0.01 & -0.01 \\
 6.5 & 0.545 & 0.002 & -0.002 & 0.948 & 0.008 & -0.008 \\
 7 & 0.496 & 0.002 & -0.002 & 0.938 & 0.007 & -0.007 \\
 7.5 & 0.451 & 0.001 & -0.001 & 0.942 & 0.007 & -0.007 \\
 8 & 0.409 & 0.001 & -0.001 & 0.965 & 0.008 & -0.008 \\
 8.5 & 0.371 & 0.001 & -0.001 & 0.980 & 0.01 & -0.01 \\
 9 & 0.336 & 0.001 & -0.001 & 1.01 & 0.02 & -0.02 \\
 9.5 & 0.305 & 0.001 & -0.001 & 1.05 & 0.04 & -0.04 \\
 10 & 0.276 & 0.002 & -0.002 & 0.925 & 0.07 & -0.07
\end{tabular}
\caption{tabled values for Fig.~\ref{fig:c1c05050}.}
  \label{tab:c1c05050}
\end{table}

\begin{table}[h]
  \centering
  \tiny\begin{tabular}{l|l|l|l|l|l|l}
&\multicolumn{3}{|c}{ LL}&\multicolumn{3}{|c}{ NLL}\\
$Y$ & $\mu=50$ & $\mu=100$ & $\mu=25$ 
    & $\mu=50$ & $\mu=100$ & $\mu=25$ \\
\hline
 6 & 0.435 & 0.1 & -0.1 & 0.886 & -0.06 & 0.02 \\
 6.5 & 0.360 & 0.1 & -0.1 & 0.865 & -0.07 & 0.07 \\
 7 & 0.298 & 0.09 & -0.09 & 0.849 & -0.08 & 0.09 \\
 7.5 & 0.246 & 0.08 & -0.08 & 0.842 & -0.09 & 0.1 \\
 8 & 0.204 & 0.07 & -0.07 & 0.849 & -0.1 & 0.2 \\
 8.5 & 0.170 & 0.07 & -0.06 & 0.850 & -0.1 & 0.1 \\
 9 & 0.141 & 0.06 & -0.05 & 0.866 & -0.1 & 0.1 \\
 9.5 & 0.118 & 0.05 & -0.04 & 0.899 & -0.1 & 0.05 \\
 10 & 0.0984 & 0.04 & -0.04 & 0.878 & -0.1 & -0.2\\
\hline
&\multicolumn{3}{|c}{ LL+}&\multicolumn{3}{|c}{ NLL+}\\
$Y$ & $\mu=50$ & $\mu=100$ & $\mu=25$ 
    & $\mu=50$ & $\mu=100$ & $\mu=25$ \\
\hline
 6 & 0.597 & 0.02 & -0.02 & 0.963 & -0.05 & 0.02 \\
 6.5 & 0.545 & 0.02 & -0.02 & 0.948 & -0.07 & 0.08 \\
 7 & 0.496 & 0.02 & -0.02 & 0.938 & -0.08 & 0.1 \\
 7.5 & 0.451 & 0.02 & -0.02 & 0.942 & -0.1 & 0.2 \\
 8 & 0.409 & 0.02 & -0.02 & 0.965 & -0.1 & 0.2 \\
 8.5 & 0.371 & 0.02 & -0.02 & 0.980 & -0.1 & 0.1 \\
 9 & 0.336 & 0.02 & -0.02 & 1.01 & -0.2 & -0.03 \\
 9.5 & 0.305 & 0.02 & -0.01 & 1.05 & -0.2 & -0.3 \\
 10 & 0.276 & 0.02 & -0.01 & 0.925 & -0.003 & -0.6
\end{tabular}
\caption{tabled values for Fig.~\ref{fig:c1c05050_mu}.}
  \label{tab:c1c05050_mu}
\end{table}

\begin{table}[h]
  \centering
  \tiny\begin{tabular}{l|l|l|l|l|l|l}
&\multicolumn{3}{|c}{ LL}&\multicolumn{3}{|c}{ NLL}\\
$Y$ & $\sqrt{s_0}=50$ & $\sqrt{s_0}=100$ & $\sqrt{s_0}=25 $
    & $\sqrt{s_0}=50$ & $\sqrt{s_0}=100$ & $\sqrt{s_0}=25$ \\
\hline
 6 & 0.435 & 0.1 & -0.1 & 0.886 & -0.1 & 0.08 \\
 6.5 & 0.360 & 0.1 & -0.08 & 0.865 & -0.1 & 0.08 \\
 7 & 0.298 & 0.09 & -0.07 & 0.849 & -0.1 & 0.09 \\
 7.5 & 0.246 & 0.07 & -0.06 & 0.842 & -0.1 & 0.09 \\
 8 & 0.204 & 0.06 & -0.05 & 0.849 & -0.1 & 0.08 \\
 8.5 & 0.170 & 0.05 & -0.04 & 0.850 & -0.1 & 0.08 \\
 9 & 0.141 & 0.04 & -0.03 & 0.866 & -0.1 & 0.06 \\
 9.5 & 0.118 & 0.03 & -0.03 & 0.899 & -0.1 & 0.03 \\
 10 & 0.0984 & 0.03 & -0.02 & 0.878 & -0.05 & -0.05\\
\hline
&\multicolumn{3}{|c}{ LL+}&\multicolumn{3}{|c}{ NLL+}\\
$Y$& $\sqrt{s_0}=50$ & $\sqrt{s_0}=100$ & $\sqrt{s_0}=25$ 
   & $\sqrt{s_0}=50$ & $\sqrt{s_0}=100$ & $\sqrt{s_0}=25$ \\
\hline
 6 & 0.597 & 0.08 & -0.07 & 0.963 & -0.1 & 0.09 \\
 6.5 & 0.545 & 0.07 & -0.07 & 0.948 & -0.1 & 0.1 \\
 7 & 0.496 & 0.07 & -0.06 & 0.938 & -0.1 & 0.1 \\
 7.5 & 0.451 & 0.06 & -0.06 & 0.942 & -0.1 & 0.1 \\
 8 & 0.409 & 0.06 & -0.05 & 0.965 & -0.2 & 0.1 \\
 8.5 & 0.371 & 0.05 & -0.05 & 0.980 & -0.2 & 0.08 \\
 9 & 0.336 & 0.05 & -0.04 & 1.01 & -0.2 & 0.04 \\
 9.5 & 0.305 & 0.04 & -0.04 & 1.05 & -0.1 & -0.07 \\
 10 & 0.276 & 0.04 & -0.04 & 0.925 & 0.03 & -0.2
\end{tabular}
\caption{tabled values for Fig.~\ref{fig:c1c05050_s0}.}
  \label{tab:c1c05050_s0}
\end{table}

\clearpage 

\begin{table}[h]
  \centering
  \tiny\begin{tabular}{l|l|l|l|l|l|l}
&\multicolumn{3}{|c}{ LL}&\multicolumn{3}{|c}{ NLL}\\
$Y$& { MC0} & { MC+} & { MC-}
& { MC0} & { MC+} & { MC-}\\
\hline
 6 & 0.316 & 0.003 & -0.003 & 0.576 & 0.006 & -0.006 \\
 6.5 & 0.242 & 0.001 & -0.001 & 0.537 & 0.004 & -0.004 \\
 7 & 0.185 & 0.0007 & -0.0007 & 0.504 & 0.003 & -0.003 \\
 7.5 & 0.142 & 0.0005 & -0.0005 & 0.481 & 0.003 & -0.003 \\
 8 & 0.109 & 0.0003 & -0.0003 & 0.469 & 0.003 & -0.003 \\
 8.5 & 0.0838 & 0.0003 & -0.0003 & 0.453 & 0.004 & -0.004 \\
 9 & 0.0645 & 0.0002 & -0.0002 & 0.450 & 0.006 & -0.006 \\
 9.5 & 0.0498 & 0.0002 & -0.0002 & 0.464 & 0.01 & -0.01 \\
 10 & 0.0384 & 0.0003 & -0.0003 & 0.459 & 0.02 & -0.02\\
\hline
&\multicolumn{3}{|c}{ LL+}&\multicolumn{3}{|c}{ NLL+}\\
$Y$& { MC0} & { MC+} & { MC-}
& { MC0} & { MC+} & { MC-}\\
\hline
 6 & 0.358 & 0.002 & -0.002 & 0.626 & 0.007 & -0.007 \\
 6.5 & 0.302 & 0.001 & -0.001 & 0.589 & 0.005 & -0.005 \\
 7 & 0.254 & 0.0008 & -0.0008 & 0.557 & 0.004 & -0.004 \\
 7.5 & 0.213 & 0.0006 & -0.0006 & 0.537 & 0.004 & -0.004 \\
 8 & 0.179 & 0.0005 & -0.0005 & 0.533 & 0.004 & -0.004 \\
 8.5 & 0.150 & 0.0005 & -0.0005 & 0.523 & 0.006 & -0.006 \\
 9 & 0.126 & 0.0005 & -0.0005 & 0.527 & 0.009 & -0.009 \\
 9.5 & 0.105 & 0.0005 & -0.0005 & 0.543 & 0.02 & -0.02 \\
 10 & 0.0880 & 0.0007 & -0.0007 & 0.484 & 0.04 & -0.04
\end{tabular}
\caption{tabled values for Fig.~\ref{fig:c2c05050}.}
  \label{tab:c2c05050}
\end{table}

\begin{table}[h]
  \centering
  \tiny\begin{tabular}{l|l|l|l|l|l|l}
&\multicolumn{3}{|c}{ LL}&\multicolumn{3}{|c}{ NLL}\\
$Y$ & $\mu=50$ & $\mu=100$ & $\mu=25$ 
    & $\mu=50$ & $\mu=100$ & $\mu=25$ \\
\hline
 6 & 0.316 & 0.1 & -0.1 & 0.576 & -0.03 & -0.09 \\
 6.5 & 0.242 & 0.1 & -0.09 & 0.537 & -0.03 & -0.01 \\
 7 & 0.185 & 0.09 & -0.07 & 0.504 & -0.04 & 0.03 \\
 7.5 & 0.142 & 0.07 & -0.06 & 0.481 & -0.05 & 0.06 \\
 8 & 0.109 & 0.06 & -0.05 & 0.469 & -0.06 & 0.1 \\
 8.5 & 0.0838 & 0.05 & -0.04 & 0.453 & -0.07 & 0.1 \\
 9 & 0.0645 & 0.04 & -0.03 & 0.450 & -0.08 & 0.1 \\
 9.5 & 0.0498 & 0.03 & -0.02 & 0.464 & -0.09 & 0.09 \\
 10 & 0.0384 & 0.03 & -0.02 & 0.459 & -0.08 & -0.05\\
\hline
&\multicolumn{3}{|c}{ LL+}&\multicolumn{3}{|c}{ NLL+}\\
$Y$ & $\mu=50$ & $\mu=100$ & $\mu=25$ 
    & $\mu=50$ & $\mu=100$ & $\mu=25$ \\
\hline
 6 & 0.358 & 0.03 & -0.02 & 0.626 & -0.02 & -0.1 \\
 6.5 & 0.302 & 0.03 & -0.02 & 0.589 & -0.03 & -0.01 \\
 7 & 0.254 & 0.02 & -0.02 & 0.557 & -0.04 & 0.04 \\
 7.5 & 0.213 & 0.02 & -0.02 & 0.537 & -0.05 & 0.08 \\
 8 & 0.179 & 0.02 & -0.01 & 0.533 & -0.07 & 0.1 \\
 8.5 & 0.150 & 0.02 & -0.01 & 0.523 & -0.08 & 0.1 \\
 9 & 0.126 & 0.01 & -0.01 & 0.527 & -0.09 & 0.04 \\
 9.5 & 0.105 & 0.01 & -0.009 & 0.543 & -0.1 & -0.1 \\
 10 & 0.0880 & 0.01 & -0.007 & 0.484 & -0.03 & -0.3
\end{tabular}
\caption{tabled values for Fig.~\ref{fig:c2c05050_mu}.}
  \label{tab:c2c05050_mu}
\end{table}

\begin{table}[h]
  \centering
  \tiny\begin{tabular}{l|l|l|l|l|l|l}
&\multicolumn{3}{|c}{ LL}&\multicolumn{3}{|c}{ NLL}\\
$Y$ & $\sqrt{s_0}=50$ & $\sqrt{s_0}=100$ & $\sqrt{s_0}=17.5 $
    & $\sqrt{s_0}=50$ & $\sqrt{s_0}=100$ & $\sqrt{s_0}=25$ \\
\hline
 6 & 0.316 & 0.1 & -0.1 & 0.576 & -0.08 & 0.08 \\
 6.5 & 0.242 & 0.1 & -0.07 & 0.537 & -0.09 & 0.07 \\
 7 & 0.185 & 0.08 & -0.06 & 0.504 & -0.09 & 0.07 \\
 7.5 & 0.142 & 0.06 & -0.04 & 0.481 & -0.09 & 0.07 \\
 8 & 0.109 & 0.05 & -0.03 & 0.469 & -0.08 & 0.03 \\
 8.5 & 0.0838 & 0.04 & -0.03 & 0.453 & -0.08 & 0.01 \\
 9 & 0.0645 & 0.03 & -0.02 & 0.450 & -0.08 & -0.03 \\
 9.5 & 0.0498 & 0.02 & -0.01 & 0.464 & -0.07 & -0.1 \\
 10 & 0.0384 & 0.02 & -0.01 & 0.459 & -0.04 & -0.3\\
\hline
&\multicolumn{3}{|c}{ LL+}&\multicolumn{3}{|c}{ NLL+}\\
$Y$& $\sqrt{s_0}=50$ & $\sqrt{s_0}=100$ & $\sqrt{s_0}=25$ 
& $\sqrt{s_0}=50$ & $\sqrt{s_0}=100$ & $\sqrt{s_0}=25$ \\
\hline
 6 & 0.358 & 0.09 & -0.08 & 0.626 & -0.1 & 0.09 \\
 6.5 & 0.302 & 0.08 & -0.06 & 0.589 & -0.1 & 0.09 \\
 7 & 0.254 & 0.07 & -0.05 & 0.557 & -0.1 & 0.08 \\
 7.5 & 0.213 & 0.06 & -0.05 & 0.537 & -0.1 & 0.08 \\
 8 & 0.179 & 0.05 & -0.04 & 0.533 & -0.1 & 0.04 \\
 8.5 & 0.150 & 0.04 & -0.03 & 0.523 & -0.1 & 0.01 \\
 9 & 0.126 & 0.03 & -0.03 & 0.527 & -0.1 & -0.05 \\
 9.5 & 0.105 & 0.03 & -0.02 & 0.543 & -0.08 & -0.2 \\
 10 & 0.0880 & 0.02 & -0.02 & 0.484 & -0.005 & -0.3
\end{tabular}
\caption{tabled values for Fig.~\ref{fig:c2c05050_s0}.}
  \label{tab:c2c05050_s0}
\end{table}

\begin{table}[h]
  \centering
  \tiny\begin{tabular}{l|l|l|l|l|l|l}
&\multicolumn{3}{|c}{ LL}&\multicolumn{3}{|c}{ NLL}\\
$Y$& { MC0} & { MC+} & { MC-}
& { MC0} & { MC+} & { MC-}\\
\hline
 6 & 0.726 & 0.01 & -0.01 & 0.650 & 0.006 & -0.006 \\
 6.5 & 0.673 & 0.005 & -0.005 & 0.621 & 0.004 & -0.004 \\
 7 & 0.623 & 0.003 & -0.003 & 0.594 & 0.003 & -0.003 \\
 7.5 & 0.576 & 0.003 & -0.003 & 0.571 & 0.002 & -0.002 \\
 8 & 0.533 & 0.002 & -0.002 & 0.552 & 0.002 & -0.002 \\
 8.5 & 0.493 & 0.002 & -0.002 & 0.533 & 0.003 & -0.003 \\
 9 & 0.456 & 0.002 & -0.002 & 0.520 & 0.003 & -0.003 \\
 9.5 & 0.422 & 0.003 & -0.003 & 0.516 & 0.004 & -0.004 \\
 10 & 0.390 & 0.004 & -0.004 & 0.523 & 0.008 & -0.008\\
\hline
&\multicolumn{3}{|c}{ LL+}\\
$Y$& { MC0} & { MC+} & { MC-}\\
\hline
 6 & 0.599 & 0.005 & -0.005 \\
 6.5 & 0.554 & 0.003 & -0.003 \\
 7 & 0.512 & 0.002 & -0.002 \\
 7.5 & 0.473 & 0.002 & -0.002 \\
 8 & 0.437 & 0.002 & -0.002 \\
 8.5 & 0.404 & 0.002 & -0.002 \\
 9 & 0.373 & 0.002 & -0.002 \\
 9.5 & 0.345 & 0.002 & -0.002 \\
 10 & 0.319 & 0.003 & -0.003
\end{tabular}
\caption{tabled values for Fig.~\ref{fig:c2c15050}.}
  \label{tab:c2c15050}
\end{table}

\begin{table}[h]
  \centering
  \tiny\begin{tabular}{l|l|l|l|l|l|l}
&\multicolumn{3}{|c}{ LL}&\multicolumn{3}{|c}{ NLL}\\
$Y$ & $\mu=50$ & $\mu=100$ & $\mu=25$ 
    & $\mu=50$ & $\mu=100$ & $\mu=25$ \\
\hline
 6 & 0.726 & 0.07 & -0.08 & 0.650 & 0.01 & -0.1 \\
 6.5 & 0.673 & 0.07 & -0.08 & 0.621 & 0.01 & -0.06 \\
 7 & 0.623 & 0.07 & -0.08 & 0.594 & 0.009 & -0.03 \\
 7.5 & 0.576 & 0.07 & -0.08 & 0.571 & 0.005 & -0.006 \\
 8 & 0.533 & 0.07 & -0.08 & 0.552 & 0.0004 & 0.02 \\
 8.5 & 0.493 & 0.07 & -0.08 & 0.533 & -0.004 & 0.04 \\
 9 & 0.456 & 0.07 & -0.08 & 0.520 & -0.01 & 0.05 \\
 9.5 & 0.422 & 0.07 & -0.07 & 0.516 & -0.02 & 0.07 \\
 10 & 0.390 & 0.07 & -0.07 & 0.523 & -0.03 & 0.1\\
\hline
&\multicolumn{3}{|c}{ LL+}\\
$Y$ & $\mu=50$ & $\mu=100$ & $\mu=25$ \\
\hline
 6 & 0.599 & 0.03 & -0.02 \\
 6.5 & 0.554 & 0.03 & -0.02 \\
 7 & 0.512 & 0.02 & -0.02 \\
 7.5 & 0.473 & 0.02 & -0.02 \\
 8 & 0.437 & 0.02 & -0.02 \\
 8.5 & 0.404 & 0.02 & -0.01 \\
 9 & 0.373 & 0.02 & -0.01 \\
 9.5 & 0.345 & 0.02 & -0.01 \\
 10 & 0.319 & 0.02 & -0.01
\end{tabular}
\caption{tabled values for Fig.~\ref{fig:c2c15050_mu}.}
  \label{tab:c2c15050_mu}
\end{table}

\begin{table}[h]
  \centering
  \tiny\begin{tabular}{l|l|l|l|l|l|l}
&\multicolumn{3}{|c}{ LL}&\multicolumn{3}{|c}{ NLL}\\
$Y$ & $\sqrt{s_0}=50$ & $\sqrt{s_0}=100$ & $\sqrt{s_0}=17.5 $
    & $\sqrt{s_0}=50$ & $\sqrt{s_0}=100$ & $\sqrt{s_0}=25$ \\
\hline
 6 & 0.726 & 0.07 & -0.07 & 0.650 & -0.03 & 0.02 \\
 6.5 & 0.673 & 0.07 & -0.07 & 0.621 & -0.04 & 0.02 \\
 7 & 0.623 & 0.07 & -0.06 & 0.594 & -0.04 & 0.02 \\
 7.5 & 0.576 & 0.07 & -0.06 & 0.571 & -0.03 & 0.02 \\
 8 & 0.533 & 0.06 & -0.05 & 0.552 & -0.03 & -0.01 \\
 8.5 & 0.493 & 0.06 & -0.05 & 0.533 & -0.03 & -0.03 \\
 9 & 0.456 & 0.05 & -0.05 & 0.520 & -0.03 & -0.07 \\
 9.5 & 0.422 & 0.05 & -0.04 & 0.516 & -0.02 & -0.2 \\
 10 & 0.390 & 0.04 & -0.04 & 0.523 & -0.02 & -0.3\\
\hline
&\multicolumn{3}{|c}{ LL+}\\
$Y$& $\sqrt{s_0}=50$ & $\sqrt{s_0}=100$ & $\sqrt{s_0}=25$ \\
\hline
 6 & 0.599 & 0.07 & -0.06 \\
 6.5 & 0.554 & 0.06 & -0.06 \\
 7 & 0.512 & 0.06 & -0.05 \\
 7.5 & 0.473 & 0.05 & -0.05 \\
 8 & 0.437 & 0.05 & -0.05 \\
 8.5 & 0.404 & 0.05 & -0.04 \\
 9 & 0.373 & 0.04 & -0.04 \\
 9.5 & 0.345 & 0.04 & -0.04 \\
 10 & 0.319 & 0.04 & -0.03
\end{tabular}
\caption{tabled values for Fig.~\ref{fig:c2c15050_s0}.}
  \label{tab:c2c15050_s0}
\end{table}

\begin{table}
  \centering
  \tiny\begin{tabular}{l|l|l|l|l|l|l}
    & \multicolumn{3}{|c}{ LL} & \multicolumn{3}{|c}{ NLL}\\
$Y$ & MC0  &  MC+ &  MC-       &  MC0  &  MC+ &  MC- \\ 
\hline 
 6 & 0.0352 & 0.0003 & -0.0003 & 0.0310 & 0.0002 & -0.0002 \\
 6.5 & 0.0360 & 0.0002 & -0.0002 & 0.0349 & 0.0002 & -0.0002 \\
 7 & 0.0266 & 0.0001 & -0.0001 & 0.0276 & 0.0001 & -0.0001 \\
 7.5 & 0.0157 & 0.00005 & -0.00005 & 0.0170 & 0.00005 & -0.00005 \\
 8 & 0.00704 & 0.00002 & -0.00002 & 0.00773 & 0.00003 & -0.00003 \\
 8.5 & 0.00245 & 0.000008 & -0.000008 & 0.00271 & 0.00001 & -0.00001 \\
 9 & 0.000688 & 0.000003 & -0.000003 & 0.000748 & 0.000004 & -0.000004\\
 9.5 & 0.000134 & 0.0000006 & -0.0000006 & 0.000137 & 0.000001 & -0.000001\\
 10 & 0.0000127 & 0.0000001 & -0.0000001 & 0.0000117 & 0.0000002 & -0.0000002\\
\hline
    & \multicolumn{3}{|c}{ LL+}\\
$Y$ & MC0  &  MC+ &  MC-       \\
\hline
 6 & 0.0281 & 0.0002 & -0.0002 \\
 6.5 & 0.0308 & 0.0001 & -0.0001 \\
 7 & 0.0239 & 0.00007 & -0.00007 \\
 7.5 & 0.0147 & 0.00004 & -0.00004 \\
 8 & 0.00680 & 0.00002 & -0.00002 \\
 8.5 & 0.00242 & 0.000008 & -0.000008 \\
 9 & 0.000695 & 0.000003& -0.000003\\
 9.5 & 0.000137 & 0.0000007 & -0.0000007\\
 10 & 0.0000132 & 0.0000001 & -0.0000001
  \end{tabular}
  \caption{tabled values for Fig.~\ref{fig:c15050}.}
  \label{tab:c15050}
\end{table}

\begin{table}[h]
  \centering
  \tiny\begin{tabular}{l|l|l|l|l|l|l}
&\multicolumn{3}{|c}{ LL}&\multicolumn{3}{|c}{ NLL}\\
$Y$ & $\mu=50$ & $\mu=100$ & $\mu=25$ 
    & $\mu=50$ & $\mu=100$ & $\mu=25$ \\
\hline
 6 & 0.0352 & -0.001 & 0.003 & 0.0310 & -0.001 & 0.007 \\
 6.5 & 0.0360 & -0.003 & 0.006 & 0.0349 & -0.002 & 0.004 \\
 7 & 0.0266 & -0.004 & 0.006 & 0.0276 & -0.002 & 0.002 \\
 7.5 & 0.0157 & -0.003 & 0.004 & 0.0170 & -0.001 & 0.001 \\
 8 & 0.00704 & -0.001 & 0.002 & 0.00773 & -0.0006 & 0.0002 \\
 8.5 & 0.00245 & -0.0006 & 0.001 & 0.00271 & -0.0002 & -0.000007 \\
 9 & 0.000688 & -0.0002 & 0.0003 & 0.000748 & -0.00007 & -0.00001 \\
 9.5 & 0.000134 & -0.00004 & 0.00007 & 0.000137 & -0.00001 & -0.000007\\
 10 & 0.0000127 & -0.000004 & 0.000008 & 0.0000117 & -0.0000009 & -0.000001\\
\hline
&\multicolumn{3}{|c}{ LL+}\\
$Y$ & $\mu=50$ & $\mu=100$ & $\mu=25$\\
\hline
 6 & 0.0281 & -0.006 & 0.01 \\
 6.5 & 0.0308 & -0.008 & 0.01 \\
 7 & 0.0239 & -0.007 & 0.01 \\
 7.5 & 0.0147 & -0.004 & 0.008 \\
 8 & 0.00680 & -0.002 & 0.004 \\
 8.5 & 0.00242 & -0.0008 & 0.002 \\
 9 & 0.000695 & -0.0002 & 0.0005 \\
 9.5 & 0.000137 & -0.00005 & 0.0001 \\
 10 & 0.0000132 & -0.000005 & 0.00001
\end{tabular}
\caption{tabled values for left figure of Fig.~\ref{fig:c15050rel_mu_s0}.}
  \label{tab:c15050_mu}
\end{table}

\begin{table}[h]
  \centering
  \tiny\begin{tabular}{l|l|l|l|l|l|l}
&\multicolumn{3}{|c}{ LL}&\multicolumn{3}{|c}{ NLL}\\
$Y$ & $\sqrt{s_0}=50$ & $\sqrt{s_0}=100$ & $\sqrt{s_0}=25$ 
    & $\sqrt{s_0}=50$ & $\sqrt{s_0}=100$ & $\sqrt{s_0}=25$ \\
\hline
 6 & 0.0352 & 0.01 & -0.007 & 0.0310 & -0.002 & 0.003 \\
 6.5 & 0.0360 & 0.009 & -0.006 & 0.0349 & -0.003 & 0.003 \\
 7 & 0.0266 & 0.005 & -0.004 & 0.0276 & -0.003 & 0.002 \\
 7.5 & 0.0157 & 0.003 & -0.002 & 0.0170 & -0.001 & 0.001 \\
 8 & 0.00704 & 0.001 & -0.0007 & 0.00773 & -0.0006 & 0.0005 \\
 8.5 & 0.00245 & 0.0003 & -0.0002 & 0.00271 & -0.0002 & 0.0002 \\
 9 & 0.000688 & 0.00007 & -0.00005 & 0.000748 & -0.00006 & 0.00004 \\
 9.5 & 0.000134 & 0.00001 & -0.000009 & 0.000137 & -0.00001 & 0.000008\\
 10 & 0.0000127 & 0.000001 & -0.0000008 & 0.0000117 & -0.0000009 & 0.0000007\\
\hline
&\multicolumn{3}{|c}{ LL+}\\
$Y$ & $\sqrt{s_0}=50$ & $\sqrt{s_0}=100$ & $\sqrt{s_0}=25$\\
\hline
 6 & 0.0281 & 0.005 & -0.003 \\
 6.5 & 0.0308 & 0.004 & -0.003 \\
 7 & 0.0239 & 0.003 & -0.002 \\
 7.5 & 0.0147 & 0.001 & -0.001 \\
 8 & 0.00680 & 0.0006 & -0.0005 \\
 8.5 & 0.00242 & 0.0002 & -0.0002 \\
 9 & 0.000695 & 0.00005 & -0.00004 \\
 9.5 & 0.000137 & 0.000009 & -0.000007\\
 10 & 0.0000132 & 0.0000008 & -0.0000006
\end{tabular}
\caption{tabled values for right figure of Fig.~\ref{fig:c15050rel_mu_s0}.}
  \label{tab:c15050_s0}
\end{table}

\begin{table}[h]
  \centering
  \tiny\begin{tabular}{l|l|l|l|l|l|l}
    & \multicolumn{3}{|c}{ LL} & \multicolumn{3}{|c}{ NLL}\\
$Y$ & MC0  &  MC+ &  MC-       &  MC0  &  MC+ &  MC- \\ 
\hline 
 6 & 0.0256 & 0.0003 & -0.0003 & 0.0201 & 0.0001 & -0.0001 \\
 6.5 & 0.0243 & 0.0001 & -0.0001 & 0.0217 & 0.00008 & -0.00008 \\
 7 & 0.0166 & 0.00007 & -0.00007 & 0.0164 & 0.00005 & -0.00005 \\
 7.5 & 0.00907 & 0.00003 & -0.00003 & 0.00972 & 0.00002 & -0.00002 \\
 8 & 0.00375 & 0.00001 & -0.00001 & 0.00427 & 0.00001 & -0.00001 \\
 8.5 & 0.00121 & 0.000004 & -0.000004 & 0.00144 & 0.000004 & -0.000004\\
 9 & 0.000314 & 0.000001 & -0.000001 & 0.000389 & 0.000001 & -0.000001\\
 9.5 & 0.0000564 & 0.0000003 & -0.0000003 & 0.0000707 & 0.0000003 & -0.0000003\\
 10 & 0.00000496 &  0.00000004 & - 0.00000004 &   0.00000612 & 0.00000005 & -0.00000005\\
\hline
    & \multicolumn{3}{|c}{ LL+}\\
$Y$ & MC0  &  MC+ &  MC-\\
\hline
 6 & 0.0168 & 0.00009 & -0.00009 \\
 6.5 & 0.0171 & 0.00006 & -0.00006 \\
 7 & 0.0123 & 0.00004 & -0.00004 \\
 7.5 & 0.00698 & 0.00002 & -0.00002 \\
 8 & 0.00297 & 0.000008 & -0.000008\\
 8.5 & 0.000979 & 0.000003 & -0.000003\\
 9 & 0.000260 & 0.000001 & -0.000001\\
 9.5 & 0.0000473 & 0.0000002 & -0.0000002\\
 10 & 0.00000422 & 0.00000003 & -0.00000007
  \end{tabular}
  \caption{tabled values for Fig.~\ref{fig:c25050}.}
  \label{tab:c25050}
\end{table}

\begin{table}[h]
  \centering
  \tiny\begin{tabular}{l|l|l|l|l|l|l}
&\multicolumn{3}{|c}{ LL}&\multicolumn{3}{|c}{ NLL}\\
$Y$ & $\mu=50$ & $\mu=100$ & $\mu=25$ 
    & $\mu=50$ & $\mu=100$ & $\mu=25$ \\
\hline
 6 & 0.0256 & 0.001 & -0.0007 & 0.0201 & -0.0003 & 0.00003 \\
 6.5 & 0.0243 & 0.0001 & 0.0003 & 0.0217 & -0.0007 & 0.0003 \\
 7 & 0.0166 & -0.0005 & 0.0009 & 0.0164 & -0.0008 & 0.0005 \\
 7.5 & 0.00907 & -0.0006 & 0.0008 & 0.00972 & -0.0006 & 0.0005 \\
 8 & 0.00375 & -0.0004 & 0.0005 & 0.00427 & -0.0003 & 0.0003 \\
 8.5 & 0.00121 & -0.0002 & 0.0002 & 0.00144 & -0.0001 & 0.0001 \\
 9 & 0.000314 & -0.00005 & 0.00007 & 0.000389 & -0.00004 & 0.00003 \\
 9.5 & 0.0000564 & -0.00001 & 0.00002 & 0.0000707 & -0.000009 & 0.000006\\
 10 & 0.00000496 & -0.000001 & 0.000001 & 0.00000612 & -0.0000009 & 0.0000005\\
\hline
&\multicolumn{3}{|c}{ LL+}\\
$Y$ & $\mu=50$ & $\mu=100$ & $\mu=25$\\
\hline
 6 & 0.0168 & -0.003 & 0.005 \\
 6.5 & 0.0171 & -0.004 & 0.006 \\
 7 & 0.0123 & -0.003 & 0.005 \\
 7.5 & 0.00698 & -0.002 & 0.003 \\
 8 & 0.00297 & -0.0008 & 0.002 \\
 8.5 & 0.000979 & -0.0003 & 0.0006 \\
 9 & 0.000260 & -0.00008 & 0.0002 \\
 9.5 & 0.0000473 & -0.00002 & 0.00003 \\
 10 & 0.00000422 & -0.000002 & 0.000003
\end{tabular}
\caption{tabled values for left figure of Fig.~\ref{fig:c25050rel_mu_s0}.}
  \label{tab:c25050_mu}
\end{table}

\begin{table}[h]
  \centering
  \tiny\begin{tabular}{l|l|l|l|l|l|l}
&\multicolumn{3}{|c}{ LL}&\multicolumn{3}{|c}{ NLL}\\
$Y$ & $\sqrt{s_0}=50$ & $\sqrt{s_0}=100$ & $\sqrt{s_0}=25$ 
    & $\sqrt{s_0}=50$ & $\sqrt{s_0}=100$ & $\sqrt{s_0}=25$ \\
\hline
 6 & 0.0256 & 0.01 & -0.007 & 0.0201 & -0.002 & 0.003 \\
 6.5 & 0.0243 & 0.01 & -0.006 & 0.0217 & -0.003 & 0.003 \\
 7 & 0.0166 & 0.006 & -0.004 & 0.0164 & -0.002 & 0.002 \\
 7.5 & 0.00907 & 0.003 & -0.002 & 0.00972 & -0.001 & 0.001 \\
 8 & 0.00375 & 0.001 & -0.0007 & 0.00427 & -0.0006 & 0.0002 \\
 8.5 & 0.00121 & 0.0003 & -0.0002 & 0.00144 & -0.0002 & 0.0000006\\
 9 & 0.000314 & 0.00007 & -0.00005 & 0.000389 & -0.00005 & -0.00003 \\
 9.5 & 0.0000564 & 0.00001 & -0.000009 & 0.0000707 & -0.000008& -0.00002 \\
 10 & 0.00000496 & 0.000001 & -0.0000008 & 0.00000612 & -0.0000007 & -0.000003\\
\hline
&\multicolumn{3}{|c}{ LL+}\\
$Y$ & $\sqrt{s_0}=50$ & $\sqrt{s_0}=100$ & $\sqrt{s_0}=25$\\
\hline
 6 & 0.0168 & 0.005 & -0.003 \\
 6.5 & 0.0171 & 0.005 & -0.003 \\
 7 & 0.0123 & 0.003 & -0.002 \\
 7.5 & 0.00698 & 0.002 & -0.001 \\
 8 & 0.00297 & 0.0006 & -0.0005 \\
 8.5 & 0.000979 & 0.0002 & -0.0002 \\
 9 & 0.000260 & 0.00005 & -0.00004 \\
 9.5 & 0.0000473 & 0.000009 & -0.000007 \\
 10 & 0.00000422 & 0.0000008 & -0.0000006
\end{tabular}%
\caption{tabled values for Fig.~\ref{fig:c1c03550_mu}.}
  \label{tab:c1c03550_mu}
\end{table}

\clearpage 

\begin{table}[h]
  \centering
  \tiny%
\caption{tabled values for right figure of Fig.~\ref{fig:c23550rel_mu_s0}.}
  \label{tab:c23550_s0}
\end{table}

\clearpage

\providecommand{\href}[2]{#2}\begingroup\raggedright\endgroup

\end{document}